\documentclass[fleqn,usenatbib]{mnras}

\usepackage{newtxtext,newtxmath}

\usepackage[T1]{fontenc}
\usepackage{lineno}

\DeclareRobustCommand{\VAN}[3]{#2}
\let\VANthebibliography\thebibliography
\def\thebibliography{\DeclareRobustCommand{\VAN}[3]{##3}\VANthebibliography}

\usepackage{graphicx}	
\usepackage{amsmath}	

\usepackage{subfigure}
\usepackage{booktabs}
\usepackage{multirow}
\usepackage{array}

\usepackage{xcolor}

\usepackage{ulem}

\title[SMBBHs: Continuum Spectral Features and Periodic Variabilities]{A Population Study for Searching Supermassive Binary Black Holes in Active Galactic Nuclei: Continuum Spectral Features and Periodic Variabilities}

\author[Li et al.]{
Zekun Li,$^{1,2}$
Changshuo Yan,$^{1,2}$\thanks{yancs@nao.cas.cn}
and Youjun Lu$^{1,2}$\thanks{luyj@nao.cas.cn}
\\
$^{1}$National Astronomical Observatories, Chinese Academy of Sciences, 20A Datun Road, Beĳing 100101, China\\
$^{2}$School of Astronomy and Space Sciences, University of Chinese Academy of Sciences, 19A Yuquan Road, Beĳing 100049, China
}

\date{Accepted XXX. Received YYY; in original form ZZZ}

\pubyear{\the\year{}}

\begin{document}

\label{firstpage}
\pagerange{\pageref{firstpage}--\pageref{lastpage}}
\maketitle

\begin{abstract}
Active sub-pc supermassive binary black holes (SMBBHs) are expected to exhibit various electromagnetic signatures due to unique dynamical and geometric structures of their accretion, but observational identification of them remains challenging. In this paper, we adopt semianalytic models to investigate both deficits in spectral-energy-distributions (SEDs) of these systems induced by gaps/holes in their accretion disks and periodic variations in their light curves induced by either orbital-modulated Doppler boosting or accretion rate variation of each SMBH component. We construct a population model to generate SMBBHs across cosmic time by considering their orbital evolution and associated accretion and radiation processes. By estimating the continuum emission from each mock system and its variation, we investigate the detection of SMBBHs via either SED-deficit signature or light curve periodicity under reasonably given criteria. We find that all-sky surveys with filters similar to those of the China Space Station Telescope or Rubin/LSST could identify up to approximately $3\times 10^2$ SMBBHs via SED-deficit features and/or $10^3$ SMBBHs via periodicity ($\lesssim5$\,yr), although both selections may suffer from a high rate of false positives. A few to a dozen SED-deficit SMBBHs may be detected by future pulsar timing arrays with signal-to-noise ratio $\gtrsim3$, enabling multi-messenger observations. Only $\sim20\%-53\%$ of SED-deficit selected SMBBHs may also be detected via periodic variations, and $\sim7\%-9\%$ of periodic variation selected SMBBHs may be detected via SED-deficit signatures. The false positives for those SMBBHs selected jointly by both methods are negligible, highlighting the importance of searching for active SMBBH systems using joint methods. 
\end{abstract}

\begin{keywords}
accretion: accretion disks -- black hole physics -- galaxies: evolution -- galaxies: kinematics and dynamics -- quasars: supermassive black holes
\end{keywords}



\section{Introduction} 
\label{sec:intro}

Supermassive binary black holes (SMBBHs) in galactic centers are natural products of galaxy mergers within the framework of $\Lambda$CDM cosmology \citep[e.g.,][]{1980Natur.287..307B,2002MNRAS.331..935Y}. Following a galaxy merger, the two central supermassive black holes (SMBHs) evolve from kiloparsec (kpc) to sub-parsec separations and become a closely bound binary due to dynamical friction and interactions with surrounding stars. In gas-rich mergers, significant amounts of gas can accumulate in the vicinity of the central SMBBHs, triggering nuclear activity. It appears as dual active galactic nuclei (AGNs) or offset AGNs when the separation of the two BHs is on the kpc scale \citep[e.g.,][]{2010ApJ...708..427L, 2011ApJ...738...92Y, 2021ApJ...916..110L, 2023MNRAS.522.1895C} and a single AGN with inner structure cultivated by the central SMBBH when the separation is on sub-parsec scales or smaller \citep[e.g.,][]{1996ApJ...467L..77A, 2012ApJ...761...90G, 2023ARA&A..61..517L}. While numerous dual AGNs at kiloparsec (kpc) scales have been observationally identified using various techniques \citep[e.g.,][]{2009ApJ...702L..82C, 2011ApJ...735...48S, 2012ApJS..201...31G, 2021A&A...646A.153S, 2022NatAs...6.1185M, 2023Natur.616...45C}, direct evidence for AGNs hosting sub-parsec SMBBHs remains rare \citep[e.g.,][]{2022LRR....25....3B,2023arXiv231016896D,2025CQGra..42k3001L}. 

Indirect methods, such as AGN periodic variation (attributed to orbital modulations; e.g., \citealt{2015Natur.525..351D, 2015MNRAS.453.1562G, 2016MNRAS.463.2145C, 2019MNRAS.485.1579K, 2025ApJ...978...86L, 2025A&A...699A..55F, 2025arXiv250516884H,2025arXiv250821510C}) and double-peaked or asymmetric broad emission lines (BELs), have been adopted to select a larger number of SMBBH candidates \citep[e.g.,][]{2012ApJS..201...23E, 2013ApJ...775...49S, 2021ApJ...910..101J}. However, the confirmation of these candidates as SMBBHs remains challenging, e.g., given the high false positive probabilities for periodically variable candidates \citep[e.g.,][]{2018ApJ...856...42S, 2023arXiv231016896D,2025arXiv250910601E} and alternative interpretations for the double-peaked or asymmetric BELs \citep[e.g.,][]{1989ApJ...344..115C}. Nevertheless, candidates like PG1302-102 have been studied using multiple methods \citep[e.g.,][]{2018ApJ...859L..12L, 2020MNRAS.496.1683X, 2021A&A...645A..15S} for further validation, though were not confirmed yet. These considerations underscore the growing importance of developing various detection methods and applying combined diagnostic techniques to enhance reliability. 

Population models of SMBBHs \citep[e.g.,][]{2009ApJ...700.1952H}, which incorporate detailed accretion and radiation processes, enable systematic investigations on the detectability of SMBBHs using various signatures. For instance, \citet{2019MNRAS.485.1579K} conducted a comprehensive study on periodic variable SMBBHs, revealing significant discrepancies in candidate number from observations, which supports that the false positive rate may be indeed high. Additionally, co-evolution analysis of SMBBHs can bridge electromagnetic (EM) observations with gravitational wave (GW) detections \citep[e.g.,][]{2017MNRAS.464.3131K, 2020ApJ...897...86C, 2021MNRAS.506.2408X, 2025ApJ...987..106C, 2025arXiv250610846X}, which is of great importance in the near future for investigating individual SMBBHs that will be revealed by pulsar timing arrays (PTAs) \citep[e.g.,][]{2017NatAs...1..886M, 2018MNRAS.477..964K, 2023ApJ...955..132C, 2024ApJ...974..261C}. This also motivates studies to overcome the challenges in EM observations of sub-parsec SMBBHs through systematic modeling of SMBBHs.

The spectral energy distribution (SED) features of AGNs hosting SMBBHs have been proposed as a probe of sub-parsec SMBBHs \citep[e.g.,][]{2012MNRAS.420..705T,2012MNRAS.420..860S,2014ApJ...785..115R,2022LRR....25....3B,2019ApJ...879..110K,2023arXiv231016896D}. Both analytical analysis \citep[e.g.,][]{2012MNRAS.427.2660K, 2012ApJ...761...90G} and hydrodynamic simulations \citep[e.g.,][]{1996ApJ...467L..77A,2002ApJ...567L...9A,2007MNRAS.379..956D,2013MNRAS.436.2997D,2014ApJ...783..134F,2020ApJ...901...25D,2023ARA&A..61..517L,2024arXiv240514843G} show that the disk accretion onto a sub-pc SMBBH may have a unique triple-accretion disk structure with a hole/gap between them. As a result, the continuum emission from such a triple-disk-SMBBH system may be short of emission at some wavelengths corresponding to the gap location (e.g., the optical-to-UV bands) comparing with that from a single-BH accretion disk. Such a SED-deficit has been used as a signature for searching SMBBHs among AGNs.  \citet{2015ApJ...809..117Y} first analyzed the distinct optical-to-UV spectrum of Mrk\,231 and demonstrated that it could be well-explained by an SMBBH scenario. \citet{2019ApJ...879..110K} estimated that approximately $100$ SMBBH systems (assuming a flux limit of $m_{\mathrm{V}} = 18$) have the potential to exhibit SED-deficit (or ``notch'' defined therein) features.
It is important to note here, such SED-deficit signatures may also be mimicked by dust reddening or obscuration effects in single AGNs \citep[e.g.,][]{2016ApJ...829....4L,2016ApJ...825...42V,2022ApJ...927....3F}.

If the SED-deficit selected SMBBH candidates are true SMBBHs, at least some of them should also exhibit periodic variations or vice versa. For Mrk\,231, observations do suggest it exhibits periodic variation \citep[][]{2018MNRAS.480.5504Y, 2020MNRAS.494.4069K}. For $138$ periodic variation selected SMBBH candidates, however, \citet{2020MNRAS.492.2910G} performed a systematic search for SED-deficit features and did not find clear evidence of SED-deficit in the IR–optical–UV SEDs, which may suggest a high false positive rate of periodic variation selected SMBBH candidates as mentioned above \citep[see][]{2016MNRAS.461.3145V, 2018ApJ...856...42S, 2020MNRAS.492.2910G}. This raises two important questions: (1) How many AGNs hosting SMBBHs can, in principle, be distinguished from normal AGNs via either SED-deficit or periodic variation? (2) How many such SMBBHs can be jointly identified through both SED-deficit and periodic variation, thereby providing mutual confirmation to overcome the high false positive rate by only using one of them? In this paper, we aim to address these questions by constructing a statistical framework for the cosmic SMBBH population, and exploring future detection prospects by detailed SED-deficit and periodicity modeling. 

Recent advancements in GW astronomy have further emphasized the importance of studying SMBBHs. The detection of the stochastic gravitational wave background (GWB) signal at the nanohertz band by several PTAs \citep{2023ApJ...951L...8A, 2023A&A...678A..50E, 2023ApJ...951L...6R, 2023RAA....23g5024X, 2025MNRAS.536.1489M} has revealed possible deviations from that predicted by population synthesis models of SMBBHs \citep[e.g.,][]{2023ApJ...952L..37A}. These findings highlight the need for further investigations into SMBBH evolution and merging models \citep[e.g.,][]{2023ApJ...955..132C, 2024ApJ...965..164G, 2024ApJ...974..261C, 2024PhRvD.109b1302E, 2024A&A...691A.270E, 2025PhRvD.111b3043S, 2025arXiv250109786S}, emphasizing the complementary role of electromagnetic observations in enhancing our understanding of these sub-parsec SMBBH systems. In this work, we also examine the dependence of our predictions on the SMBBH modeling by taking into account the constraint given by the GWB signal. 

The paper is organized as follows. In Section~\ref{sec:method}, we introduce our population model for cosmic SMBBHs formed from galaxy mergers that includes their orbital evolution in gas-rich environments. We generate mock samples of active SMBBHs, and further consider the accretion and radiation processes of these systems and calculate their SEDs. In the meantime, we also estimate the periodic variation of these systems by considering both orbital-modulated Doppler boosting effect and periodic rate variation of accretion onto each BH component. Applying reasonably quantified selection criteria, we adopt either the SED-deficit or periodic variation method to search for SMBBHs among the mock sample, and also obtain the mock sample for SMBBHs that can be selected by both methods. We present our results on the detectability of sub-parsec SMBBHs under various model assumptions and analyze the property distributions of the selected SMBBHs and consider the detection of gravitational waves from these SMBBHs by PTAs in Section~\ref{sec:results}. Conclusions and discussions are given in Section~\ref{sec:discussion}.

Throughout the paper, we adopt the flat $\Lambda$CDM cosmology with $\Omega_{\mathrm{m}}= 0.3$, $\Omega_{\Lambda}= 0.7$, and $H_0=70$\,km\,s$^{-1}$\,Mpc$^{-1}$ \citep[e.g.,][]{2016A&A...594A..13P}.

\section{Method} 
\label{sec:method}

In this section, we first introduce the population synthesis model for generating mock samples of active SMBBHs, then introduce the model for estimating the SEDs and periodic variation light curves of the mock SMBBHs, and set the criteria for selecting SMBBHs from the mock sample via either the SED-deficit or periodic variation method. 

\subsection{Population Synthesis Model for Active SMBBHs} 
\label{subsec:mocksamples}

We adopt a widely used SMBBH sampling procedure \citep[e.g.,][]{2017MNRAS.464.3131K, 2020ApJ...897...86C, 2020ApJ...889...79Y} and incorporate reasonable adjustments inspired by recent observations, such as the observational evidence for the nanohertz GWB signal. For a simplified galaxy merger model, the number of merging galaxies per unit time $dt$ and per redshift interval can be expressed as
\begin{equation}
\begin{aligned}
\frac{d^4 N_{\mathrm{gal}}(z,M_{*},q_{\mathrm{gal}})}{dz dM_{*} dq_{\mathrm{gal}} dt}=& \frac{d^2N_{\mathrm{mrg}}(M_{\mathrm{tot}}(M_{*}),q_{\mathrm{gal}},z)}{dq_{\mathrm{gal}} dt}\\
&\times\Phi(M_{*},z)\times\frac{dV}{dz},
\label{eq:dNdt}
\end{aligned}
\end{equation}
where $V$ is the comoving volume, $M_{\mathrm{tot}}(M_{*})\simeq(1+q_{\mathrm{gal}})M_{*}$ is the mass of the merger remnant galaxy (assuming the total mass change in the merger is negligible). $N_{\mathrm{mrg}}$ represents the merger rate per galaxy (MRPG) at redshift $z$, in the mass range from $M_{\mathrm{tot}}$ to $M_{\mathrm{tot}}+dM_{\mathrm{tot}}$ and the galaxy mass ratio range from $q_{\mathrm{gal}}$ to $q_{\mathrm{gal}}+dq_{\mathrm{gal}}$, and $\Phi (M_{*},z)$ denotes the galaxy stellar mass function (GSMF). As the main concern in this paper is about the active SMBBHs, we therefore focus on wet mergers, i.e., mergers with gas-rich progenitor galaxies (see Section~\ref{subsubsec:galmerger}), of which the central BHs are linked to their properties following a procedure similar to that described in \citet{2011ApJ...738...92Y}.

\subsubsection{From galaxy merger to active SMBBHs} 
\label{subsubsec:galmerger}

To calculate the properties of host galaxy mergers, we adopt the MRPG model from \citet{2015MNRAS.449...49R}, which was derived from galaxy merger trees in the Illustris simulations. The Illustris simulations trace galaxy-galaxy mergers from the early universe to the present time ($z=0$) and follow merging SMBHs down to the sub-kpc (about 0.1 kpc) scale. The MRPG model derived from the Illustris simulations is widely used in studies of the co-evolution of SMBHs with their host galaxies \citep[c.f.,][]{2017MNRAS.464.3131K, 2020ApJ...897...86C}. 

For the GSMF, we utilize observational datasets \citep[][]{2010A&A...523A..13P} and adopt the best-fitting double ($z\lesssim 3$) and single ($z\gtrsim 3$) Schechter function from \citet{2023A&A...677A.184W}, while employing linear interpolation for the piecewise function to suppress artificial discontinuities. These adjustments are minor but ensure smoother results and better consistency with previous studies \citep[][]{2017MNRAS.471.3098L,2019MNRAS.488.3143B}. 

The presence of accretion disk features of active SMBBHs depends on the gas environment, which is related to the morphology of the progenitor galaxies. Dry mergers by two elliptical galaxies are unlikely to produce SMBBHs with strong nuclear activity \citep[e.g.,][]{2011ApJ...738...92Y}. To model SMBBHs in gas-rich mergers, we assign morphology types to each galaxy following \citet{2011ApJ...738...92Y}. Assuming that the morphology types of merging galaxies are independent and maintain the same proportions as those in the general galaxy population \citep{2005AJ....130.1516D, 2014MNRAS.444.2200D}, we randomly sample galaxy types based on their total fractions \citep{2010MNRAS.404.2087B, 2011ApJ...738...92Y}. This factor of gas supply during galaxy mergers has not been adequately addressed in previous studies of SMBBH detection using EM features originating from accretion disks.

While galaxy morphology introduces various factors-such as differences in the GSMF for red and blue galaxies, which are further influenced by the galaxy classification methodologies \citep{2010MNRAS.404.2087B, 2010A&A...523A..13P, 2017MNRAS.471.3098L}, the primary factor arising from different types of galaxies is the difference in the SMBH mass-host galaxy mass relation. We focus on this issue and first relate galaxy mass $M_{*}$ to bulge mass $M_{\mathrm{bulge}}$ for different galaxy types \citep[][see also \citealt{2009ApJ...696..411W} and \citealt{2015MNRAS.447.2772R}]{2011ApJ...738...92Y} as
\begin{equation}
\bar M_{\mathrm{bulge}}\left(M_*\right)= \begin{cases} M_*, & \text { for elliptical, } \\
0.7 M_*, & \text { for S0, } \\
0.2 M_*, & \text { for Sab, } \\
0.05M_*, & \text { for Scd/Irr. }\end{cases} 
\label{eq:bulgemass}
\end{equation}
Here we assume that $M_{\mathrm{bulge}}(M_{*})$ follows a log-normal distribution with a constant scatter $\Delta \mathrm{log} M_{\mathrm{bulge}} = 0.1$ dex for each galaxy type with a cut of $M_{\mathrm{bulge}}<M_*$.

We then proceed to generate the masses of the central SMBHs. The SMBH-bulge mass scaling relation is another complex and uncertain aspect of galaxy property studies \citep[e.g.,][]{2013ARA&A..51..511K}, particularly for AGNs \citep[][]{2020NatAs...4..282S, 2023NatAs...7.1376Z, 2024MNRAS.527.4690L}. We adopt the following form for the $M_{\mathrm{BH}}-M_{\mathrm{bulge}}$ relation at any given redshift $z$ as
\begin{equation}
\log M_{\mathrm{BH}}=\tilde{\gamma}+\tilde{\omega} \log (1+z)+\tilde{\alpha} \log M_{\mathrm{bulge}, 11}, 
\label{eq:massrelation}
\end{equation}
with an intrinsic scatter of $\tilde{\epsilon}$. Here $\tilde{\gamma}$, $\tilde{\omega}$, and $\tilde{\alpha}$ are constant parameters, and $M_{\mathrm{bulge}, 11}$ represents the bulge mass in unit of $10^{11} M_{\odot}$. For SMBHs in the local universe ($z=0$), we adopt the empirical relation given by \citet{2013ARA&A..51..511K} (hereafter KH13), i.e., $\tilde{\gamma}=8.69$, $\tilde{\omega}=0$, and $\tilde{\alpha}=1.17$, with an intrinsic scatter of $\tilde{\epsilon}=0.29$. In this paper, we consider two cases for the $M_{\mathrm{BH}}-M_{\mathrm{bulge}}$ relation. The first case is that the BH mass-bulge mass scaling relation is assumed not to evolve with redshift and the relation at high redshift is the same as that obtained for those observed BHs in the local universe (KH13). The second one is that the relation evolves with redshift as inferred from high redshift observations of SMBHs \citep[e.g.,][]{2024A&A...691A.145M,2024ApJ...964...39G,2024Natur.633..318C} and from PTA observations of the GWB assumed to be from SMBBHs \citep[e.g.,][]{2023ApJ...952L..37A, 2024ApJ...965..164G, 2023ApJ...955..132C, 2024ApJ...974..261C}. We adopt $(\tilde{\gamma},\tilde{\omega},\tilde{\alpha},\tilde{\epsilon}) = (8.69,1.99,1.00,0.31)$ for the redshift-dependent model given by \citet[][hereafter CYL23]{2023ApJ...955..132C}, where they obtained the constraint on the $M_{\rm BH}-M_{\mathrm{bulge}}$ relation by matching the observed GWB signal with that predicted from the population synthesis model of SMBBHs. Below we denote our SMBBH population synthesis model adopting the redshift-independent $M_{\rm BH}-M_{\mathrm{bulge}}$ relation (the first case) as BBH-nonevol, and the model adopting the redshift-dependent $M_{\rm BH}-M_{\mathrm{bulge}}$ relation (the second case) as BBH-zevol. We adopt BBH-zevol as our fiducial model, while using BBH-nonevol as a comparative no-evolution baseline.

We may derive the merger rate of SMBBHs, $d^4 N(z,M_{\mathrm{BH}},q)/dz d\log M_{\mathrm{BH}} d\log qdt$, from the merger rate of galaxies $d^4 N_{\mathrm{gal}}(z,M_{*},q_{\mathrm{gal}})/dz dM_* dq_{\mathrm{gal}} dt$ through the following procedure with the above model assumptions. First, we sample all galaxy mergers using Equation~(1) within the redshift range of $z\in(0, 5)$, as both the number of periodically variable SMBBHs and SED-deficit (induced by accretion disk truncation) SMBBHs decrease rapidly beyond $z\gtrsim2$ due to both the evolution and redshift effects. Second, we assign galaxy morphology types and exclude gas-poor mergers by two elliptical galaxies. According to the scaling relations given in Equations~\eqref{eq:bulgemass} and \eqref{eq:massrelation}, we can randomly assign mass to the central SMBH in a galaxy with a given morphology type and total stellar mass of $M_*$. Therefore, the merger rate of SMBBHs can be obtained as 
\begin{equation}
\begin{aligned}
\frac{d^4 N(z,M_{\mathrm{BH}},q)}{dz d\log M_{\mathrm{BH}} d\log q dt }=&
\iint P(\log M_{\mathrm{BH}}\mid \log \bar{ M}_{\mathrm{BH}}(M_{*},z)) \times\\
& P(\log (q M_{\mathrm{BH}}) \mid \log (\bar{M}_{\mathrm{BH}} (q_{\mathrm{gal}} M_{*},z))) \\
&\times \frac{d^4 N_{\mathrm{gal}}(z,M_*,q_{\mathrm{gal}})}{dz dM_{*} dq_{\mathrm{gal}} dt} dM_{\mathrm{*}}  dq_{\mathrm{gal}}. 
\label{eq:dN-dNgal}
\end{aligned}
\end{equation}
Here $q M_{\mathrm{BH}}$ and $q_{\mathrm{gal}} M_{*}$ refer to the mass of the secondary SMBH (when the primary SMBH mass is fixed at $M_{\mathrm{BH}}$) and that of its pre-merger host galaxy. According to this formula, we can then derive the SMBBH distribution across cosmic time by considering the orbital evolution of SMBBHs, especially in the disk accretion stages, with which the residence time of each SMBBH system at any given semimajor axis interval and thus the probability for the SMBBH with that semimajor axis can be estimated. 

\subsubsection{Orbital evolution of active SMBBHs} 
\label{subsubsec:evolution}

SMBBHs form in gas-rich environments and are driven toward coalescence through several different mechanisms, including dynamical friction (DF), stellar scattering, gaseous disk interactions, and GW emission \citep{1980Natur.287..307B, 2002MNRAS.331..935Y, 2017MNRAS.464.3131K, 2019MNRAS.487.4985B, 2020MNRAS.495.4681I, 2020ApJ...897...86C, 2022LRR....25....3B, 2023MNRAS.519.5031S}. For simplicity, we assume that the binary orbit is well circularized during the stages of interest, though recent studies have sparked numerous discussions regarding eccentricity (see Section~\ref{sec:discussion}). Extreme mass ratio ($q \ll 0.01$) SMBBHs are unlikely to merge within a Hubble time \citep{2002MNRAS.331..935Y}, and are therefore not considered in this paper. Below, we summarize the evolutionary processes of SMBBH systems with total mass $M_\mathrm{BBH}$, mass ratio $q$, separation $a$, and $e=0$.

The Chandrasekhar dynamical friction dominates the orbital decay of SMBBHs at separations on kpc-scale, driving each SMBH to sink toward the center of the remnant galaxies through interactions with stars and gaseous materials. The orbital decay timescale depends on the properties of host galaxy bulges, such as the stellar density distribution and stellar velocity dispersion $\sigma$ \citep[][]{2002MNRAS.331..935Y, 2012ApJ...745...83A, 2020ApJ...897...86C}. The DF merger timescale for the secondary SMBH with mass $M_2$ on a circular orbit can be roughly estimated using Equation~(1) in \citet{2002MNRAS.331..935Y} as
\begin{equation}
\begin{aligned}
t_{\mathrm{df}} \sim \frac{2 \times 10^6}{\log N_*}\left(\frac{\sigma}{100 \mathrm{~km} \mathrm{~s}^{-1}}\right)\left(\frac{r_{\mathrm{c}}}{100 \mathrm{pc}}\right)^2\left(\frac{10^8 M_{\odot}}{M_2}\right) \mathrm{yr},
\end{aligned}
\end{equation}
where $r_{\mathrm{c}}$ and $\sigma$ are the radius and velocity dispersion of the core, respectively, and $N_*$ is the number of stars contained in the core \citep[for more detailed calculations, see][]{2020ApJ...897...86C}. 

A bound SMBBH forms when the combined mass of gas and stars within the binary orbit is comparable to the total mass of the two BHs. Once the SMBBH becomes bound, three-body interactions between the SMBBH and low angular momentum stars passing by its vicinity extract energy and angular momentum from the binary system and lead to further orbital decay. The separation at which stellar scattering starts to dominate over the dynamical friction and drives the SMBBH inwards is given by \citep{1996NewA....1...35Q,2002MNRAS.331..935Y}
\begin{equation}
a_{\mathrm{h}}=\frac{Gm_{2}}{4\sigma^2} \simeq 1.08\mathrm{pc}\left(\frac{m_{2}}{10^7M_{\odot}}\right)\left(\frac{100\mathrm{~km} \mathrm{~s}^{-1}}{\sigma}\right)^2.
\end{equation}
The SMBBH orbital evolution during the so called hard binary stage with $a\le a_{\rm h}$ is strongly influenced by the supply of interacting stars, which may be governed by two-body relaxation and other dynamical processes that can be calculated following the procedure outlined in the literature for different dynamical processes, for example, see \citet{2002MNRAS.331..935Y} and \citet{2020ApJ...897...86C}. One can thus obtain the orbital decay of this stage. For gas-rich mergers, viscous disk interactions may dominate the orbital decay during the later phase of the hard binary stage, especially in cases of loss cone star depletion due to insufficient refilling of low angular momentum stars to the vicinity of SMBBHs. For simplicity, we assume a sufficient supply of loss cone stars (full loss cone, neglecting loss cone depletion) when estimating the timescale of the hard binary stage, i.e.,
\begin{equation}
\begin{aligned} 
t_{\mathrm{h}}\approx& 14 \mathrm{Myr}\left(\frac{\sigma}{100 \mathrm{~km} \mathrm{~s}^{-1}}\right)\\
&\times\left(\frac{10^3 M_{\odot} \mathrm{pc}^{-3}}{\rho}\right)\left(\frac{0.1 \mathrm{pc}}{a}\right)\left(\frac{16}{H}\right),
\end{aligned}
\end{equation}
where the constant $H\approx16$ is the typical hardening rate for a hard binary \citep{1996NewA....1...35Q}, and $\rho$ represents the galaxy core density. 

Viscous disk driven SMBBH evolution is of primary interest in this work, as the spectral features arise from dynamical interactions between SMBBHs and accretion disks and the occurrence rate of such SMBBH systems depends on the orbital decay timescale during this stage \citep[e.g.,][]{2002ApJ...567L...9A, 2012MNRAS.420..705T, 2012MNRAS.427.2660K}. \citet{2009ApJ...700.1952H} first adopted the steady-state standard thin accretion disk model \citep{1973A&A....24..337S} to analytically study the orbital decay of an SMBBH embedded in a circumbinary disk, providing detailed calculations of the coalescence timescale. We follow \citet{2009ApJ...700.1952H} by adopting the steady thin disk accretion model to estimate the SMBBH evolution timescale during the viscous disk driven stage via the Type 2 migration (orbital decay $\dot{a}_{\mathrm{sec}}$), analogous to that for planetary migration, i.e.,
\begin{equation}
\begin{aligned} 
t_{\mathrm{visc}}=t_{\mathrm{sec}}=-\frac{a}{\dot{a}_{\mathrm{sec}}}=\left(\frac{q_s}{8 f_{\mathrm{Edd}}} \frac{t_{\mathrm{Edd}}}{t_\nu}\right)^k t_\nu.
\end{aligned}
\end{equation}
In the above equation, $t_\nu =-a / \dot{a}_{\nu}=2 \pi R_0^2 \Sigma_0 / \dot{M}$ is the viscous diffusion time dependent on the secondary's location $R_0$, total accretion rate $\dot{M}$, and the disk surface density $\Sigma_0$ at that location, $q_s$ is the normalized symmetric mass ratio, $t_{\mathrm{Edd}}$ is the characteristic growth timescale associated with Eddington accretion $\dot{M}_{\mathrm{Edd}}$, and $f_{\mathrm{Edd}}= \dot{M}/ \dot{M}_{\mathrm{Edd}}$ is the dimensionless accretion rate. By combining this with the thin disk solutions for three regions, divided by distinct disk properties dominated by different pressures (radiation or gas) and opacities (electron scattering or free-free absorption), we are able to obtain the corresponding SMBBH evolution timescales (see Eqs. 26a, b, c in \citealt{2009ApJ...700.1952H}). 

The accretion rate determines the disk properties, thereby affecting both binary evolution and radiation processes. We assume that the accretion rate distribution for the outer circumbinary disks of SMBBH systems does not differ much from that for single SMBH AGNs. We adopt the Eddington ratio distribution obtained from Illustris, which has been shown to be in agreement with observations \citep{2012ApJ...746..169S,2015MNRAS.452..575S}. The Illustris results indicate that the distribution evolves with redshift but exhibits no significant dependence on SMBH mass. We fit the redshift-dependent distribution $f_{\mathrm{Edd},z} \sim \mathrm{LogNormal}(\bar f_{\mathrm{Edd}}(z),\Delta \log f_{\mathrm{Edd}})$ from \citet{2015MNRAS.452..575S}, truncated at the maximum accretion rate limit $f_{\mathrm{Edd,max}}=1$ to be consistent with that in the adopted galaxy merger model \citep{2015MNRAS.452..575S}, and use it as the accretion rate model in our calculations. For comparison, we also define a control sample with a fixed average accretion rate distribution that does not vary with redshift and is identical to that observed at the AGN abundance peak ($z \sim 2$) \citep{2012ApJ...746..169S}, while maintaining the same Gaussian scatter, $\Delta \log f_{\mathrm{Edd}} = 0.8$. Therefore, we randomly assign an accretion rate to the outer circumbinary disk of each SMBBH system according to these two distributions of Eddington ratios. 

The final stage of SMBBH orbital evolution is driven by GW radiation and the corresponding orbital decay timescale is \citep{1964PhRv..136.1224P}
\begin{equation}
t_{\mathrm{GW}}(a, e)=\frac{5}{64} \frac{c^5 a^4\left(1-e^2\right)^{7 / 2}}{G^3 \mu_{12} M^2_{ \mathrm{BBH}}\left(1+73 e^2 / 24+37 e^4 / 96\right)}.
\end{equation}
In this paper, we set $e=0$ as the SMBBH orbit is likely well circularized at this final stage. In this stage, the GW radiation drives the SMBBH rapidly towards the innermost stable circular orbit (ISCO) and merging within it \citep[e.g.,][]{2002MNRAS.331..935Y, 2022LRR....25....3B}. As mentioned in \citet{2009ApJ...700.1952H} and supported by hydrodynamic simulations \citep[e.g.,][]{2024arXiv240514843G}, the circumbinary disk can still diffuse inward and follow the binary evolution at the beginning of the GW-driven stage.

We consider the competing effects of different mechanisms simultaneously. For instance, for SMBBHs at separations where stellar scattering, viscous disk interactions and GW emission all contribute to orbital decay, the evolution timescale at $a<a_{\rm h}$ can be expressed as
\begin{equation}
\frac{1}{t_{\mathrm{evol}}(a)} \simeq \frac{1}{t_{\mathrm{GW}}(a)} +\frac{1}{t_{\mathrm{visc}}(a)} +\frac{1}{t_{\mathrm{h}}(a)}.
\end{equation}

\begin{figure}
\centering
\includegraphics[width=1\linewidth]{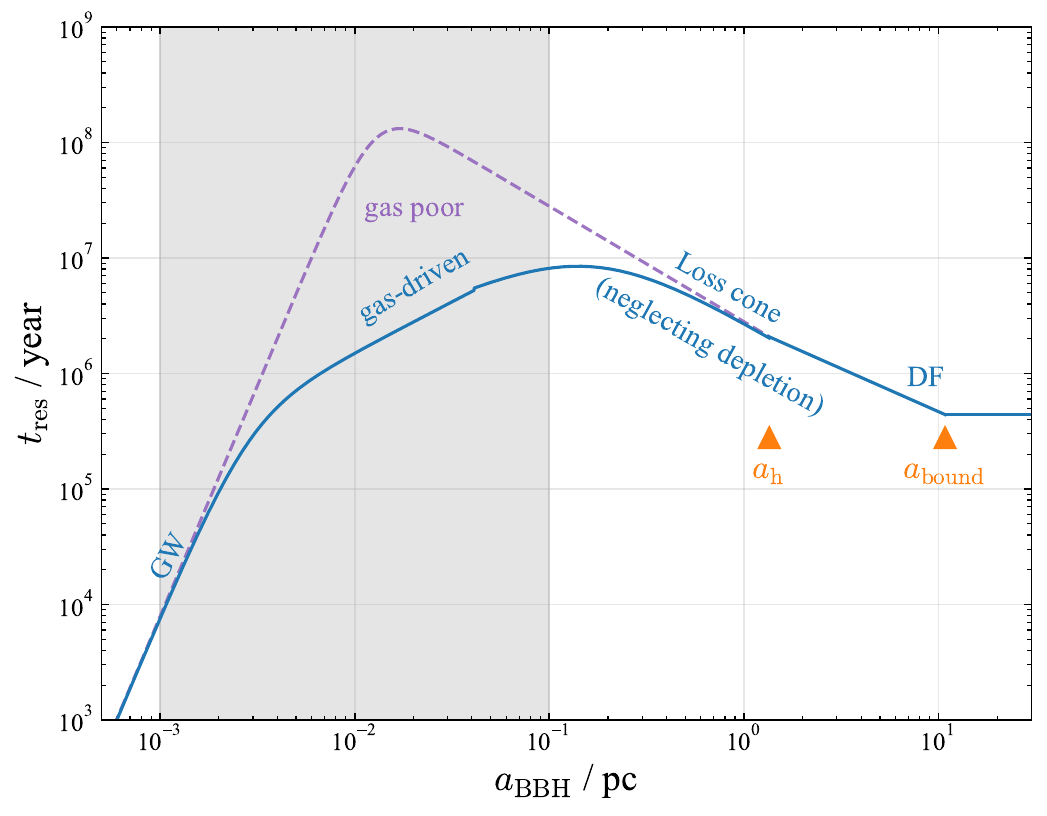}
\caption{
An example for SMBBH evolution timescale as a function of binary separation (for a representative SMBBH system with $q=1$, $M_{\mathrm{BBH}}=10^8M_{\odot}$, and $f_{\mathrm{Edd}}=0.5$). The blue curve shows the evolution timescale for gas-rich systems, where distinct mechanisms dominate at different separation ranges. Compared to SMBBH evolution in gas-poor environments (purple dashed line), interactions with accretion disks shorten the overall evolution timescale. The shaded region represents roughly the semi-major axis range of our SMBBH candidates with SED deficits, see Section~\ref{subsubsec:evolution}. Here, $a_{\mathrm{b}}$ denotes the separation at which the binary becomes gravitationally bound, while $a_{\mathrm{h}}$ represents the semimajor axis at which the binary enters the hard-binary regime, where its orbital evolution is dominated by interactions with surrounding stars. 
}
\label{fig:f1}
\end{figure}

Figure~\ref{fig:f1} shows the orbital evolution timescale for an example SMBBH system for illustration. The host galaxy has a velocity dispersion of $\sigma \simeq 200 \mathrm{km~s^{-1}}$ and the properties of the SMBBH system are set as $M_{\rm BBH}=10^8M_\odot$; $q=1$; and $f_{\rm Edd}=0.5$. As seen from this figure, the SMBBH system may quickly evolve to within sub-pc scale driven by the disk viscosity and the GW radiation, where it may be revealed via its spectral or variation signatures (the grey shadow region). The probability of such an SMBBH system at a given semimajor axis range is determined by its residence time (see Eq.~\ref{eq:dNdlogr}).

We note that the binary evolution timescale prior to the disk-driven stage, i.e., the dynamical friction and hard binary stages, can be calculated more precisely, for example, as performed in \citet{2020ApJ...897...86C}. For simplicity and efficiency, we ignore the evolutionary time period before the end of dynamical friction and assume full loss cone. This introduces only minor biases in our final results as the evolution time period before dynamical friction is typical of $\sim 10^9$\,yr and relatively short compared with the age of the universe. 

After calculating the residence time at each evolutionary stage, we convert the distribution of SMBBH merger rate obtained in Section~\ref{subsubsec:galmerger}, to the comoving number density of SMBBHs per unit logarithmic semi-major axis $\Delta \mathrm{log}a$ as
\begin{equation}
\begin{aligned}
&\frac{d^4 \mathcal{N}(z, M_{\mathrm{BH}}, q, a, f_{\rm Edd})} 
{dz d\log M_{\mathrm{BH}} d\log q d\log a df_{\rm Edd}} \\
& = \ln 10 \frac{a}{\dot{a}} P(f_{\mathrm{Edd}}) \frac{d^4 N(z, M_{\mathrm{BH}}, q)}
{dz d\log M_{\mathrm{BH}} d\log q dt}.
\label{eq:dNdlogr}
\end{aligned}
\end{equation}
where $P(f_{\mathrm{Edd}})$ is the Eddington ratio distribution of the accretion onto the circumbinary disks around SMBBHs before their merger and is assumed to be independent of $M_{\rm BH}$ and $q$, $\dot{a}$ is a function of $M_{\rm BH}$, $q$, and $f_{\rm Edd}$ as described above determined by the orbital evolution timescale. 

We then generate mock SMBBHs according to the above Equation, with different parameters covering redshift $z$, primary BH mass $M_{\mathrm{BH}}$, mass ratio $q$, semi-major axis $a$, Eddington ratio $f_{\mathrm{Edd}}$, and the inclination angle $i$ uniformly distributed in cos$i$. As to be seen below, the potential detectable SMBBHs come mainly from redshift $z<5$ and distribute in mass range $10^8M_\odot <M_{\mathrm{BH}}<10^{10}M_\odot$. We perform the above sampling procedure by employing the Monte Carlo method in the parameter space $z\in(0,5)$, $q\in(0.01,1)$, $f_{\mathrm{Edd}}\in(0.01,1)$, $M_{\mathrm{BH}}\in(10^6M_{\odot},10^{11}M_{\odot})$, and $a\in(10R_{\mathrm{g}},10^4R_{\mathrm{g}})$.
Here $R_{\mathrm{g}}$ is defined as $R_{\mathrm{g}}=GM_{ \mathrm{BBH}}/c^2$. Meanwhile, we keep the sampling number the same as the overall counts of SMBBHs calculated across cosmic volume within our parameter space. We repeat this calculation for each model. 

\subsection{Electromagnetic features of SMBBH in AGNs}
\label{subsec:features}

In this subsection, we present the accretion and radiation models adopted for SMBBHs. We focus on the SED and continuum spectrum of the unique triple-disk system \citep[][]{ 2012ApJ...761...90G,2012MNRAS.420..860S,2014ApJ...785..115R,2015MNRAS.446L..36F,2015ApJ...809..117Y,2019ApJ...879..110K}, with the goal of quantifying spectral diagnostics for identifying SMBBHs using the SED-deficit, as will be discussed in Section~\ref{subsubsec:diagnostics}. For comparison, we also describe the traditional approach to SMBBH detection based on periodic variability \citep[][]{2015MNRAS.453.1562G,2019MNRAS.485.1579K}. 

\subsubsection{SMBBH accretion and radiation} 
\label{subsubsec:accretion model}

\begin{figure*}
\includegraphics[width=1\linewidth]{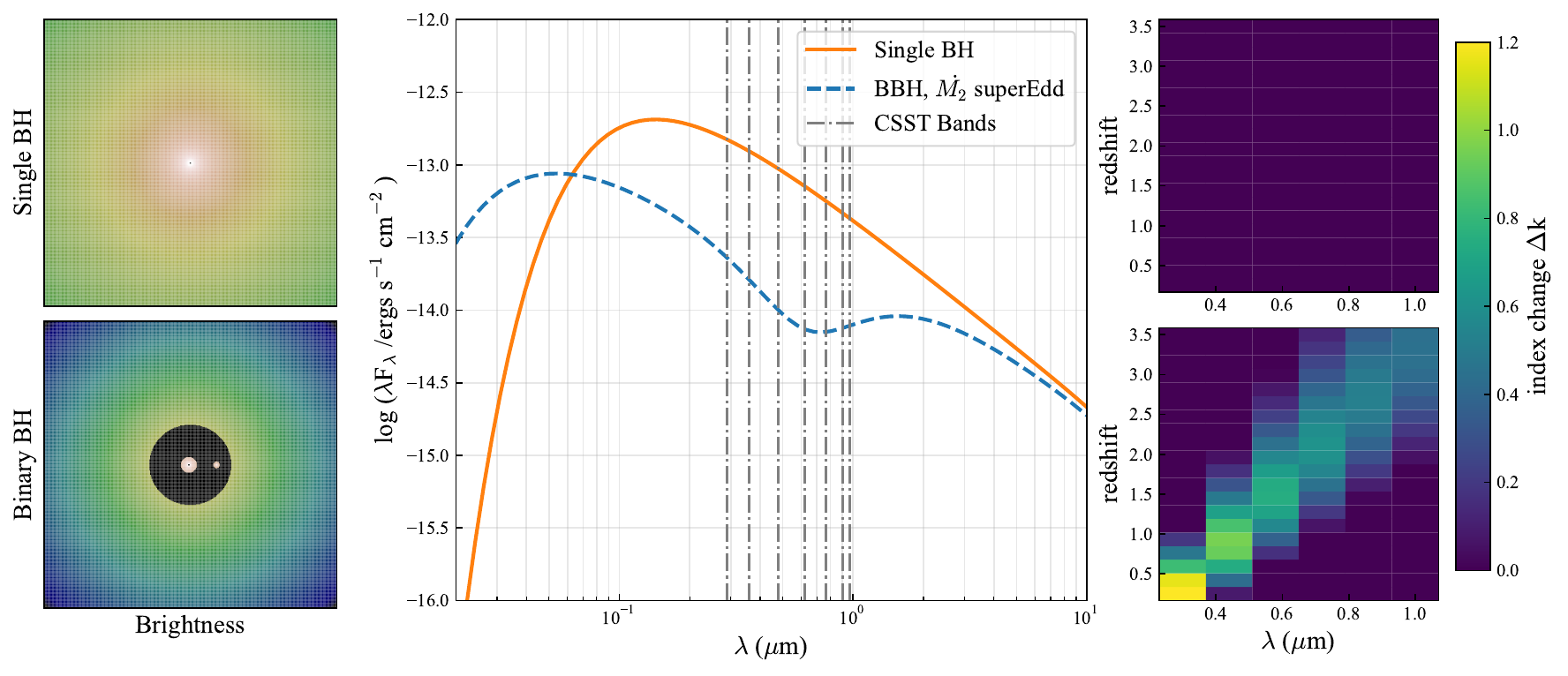}
\caption{
Spectral diagnostics for SMBBHs. {\it Left panel:} Brightness distribution examples for single BH and SMBBH systems. {\it Middle panel:} SED of an example SMBBH system ($q=0.057,a=164 R_{\mathrm{g}}$) chosen from our deficit SMBBHs (blue dashed line). The SED of a single BH system with the same redshift $z = 1.75$, the central SMBH mass $M_{\mathrm{BH,1}}=10^{8.8}M_{\odot}$, and accretion rate $f_{\mathrm{Edd}} = 0.71$ is shown for comparison (orange solid line). The central wavelengths of CSST bands are plotted with grey dash dot line. {\it Right panel:} The SED power index change between two adjacent bands for this example SMBBH system (lower panel) is shown for various redshifts and wave bands. Here, $\Delta k$ quantifies the change in the spectral power-law slope between adjacent bands. For a single-BH accretion disk whose optical-UV SED is well approximated by a single power law, the spectral slope remains nearly constant across bands, resulting in $\Delta k \approx 0$ (upper panel). For the example SMBBH system shown here, the dominant spectral-index changes associated with the SED deficit correspond to an increase in the local slope across adjacent bands, yielding predominantly positive $\Delta k$ values. By properly choosing $\Delta _{\mathrm{min}}$ as the identifying threshold for slope change, SMBBH spectrum can be effectively distinguished from single BH (upper panel) systems.
}
\label{fig:f2}
\end{figure*}

To obtain the emission from the accretion onto SMBBHs, it is necessary to derive the structure and rate of the accretion flow, as well as the inner and outer edges of the flow, with which one can obtain the detailed disk parameters, including local density, temperature, gas and radiation pressure. As suggested by numerous numerical simulations and analytical analysis \citep{1996ApJ...467L..77A, 2007MNRAS.379..956D, 2008ApJ...672...83M, 2012MNRAS.427.2660K, 2013MNRAS.436.2997D, 2014ApJ...783..134F, 2016MNRAS.455.1989G, 2020ApJ...901...25D, 2023ARA&A..61..517L}, for SMBBHs resulting from gas-rich mergers in the later stages, their accretion can be divided into three parts: two mini-disks around the primary and secondary BH components, and a circumbinary disk surrounding the central system. The rotating secondary BH component creates a cavity (or a gap for binaries with small mass ratio) between the circumbinary disk and the two mini-disks. The existence of accretion streams and a central cavity have also been shown by numerical simulations. The accretion streams will bridge the inner edge of the circumbinary disk with the mini-disks, yet remain weakly heated \citep{2016ApJ...832...22S,2019ApJ...879..110K}, contributing minimally to the system's overall radiation. 

In addition to the accretion rate model introduced in Section~\ref{subsubsec:evolution}, another crucial factor is how the accreting material is supplied to the two mini-disks. Both analytical and numerical studies indicate that the secondary SMBH accretes more mass than the primary SMBH. A recent study by \citet{2020ApJ...901...25D} provides a fit of the primary and secondary accretion rates ($\dot{M}_{1}$ and $\dot{M}_{2}$) for this mass ratio dependence as $\dot{M}_{1}=\dot{M}_{2}(0.1+0.9q)$, which is applicable for $q > 0.01$ and has been adopted in recent studies of SMBBH systems \citep[e.g.,][]{2023MNRAS.519.2083I}. For such a setting, the accretion onto the secondary BH component can be super-Eddington when $q$ is small, even if the accretion of the outer circumbinary disk is sub-Eddington in the standard thin disk regime; and the accretion rate onto the primary BH component can be substantially smaller than that of the circumbinary disk, which could be radiatively inefficient and the radiation from it is significantly suppressed. Below we introduce the models for calculating both the radiation from the standard thin disk and the super-Eddington disk, respectively.

In the standard thin disk regime ($\alpha$ disk), the disk temperature profile is described as
\begin{equation}
T_{\mathrm{eff}}(R)=\left[\frac{3 G M_\bullet \dot{M}}{8 \pi \sigma_{\mathrm{B}}R^3}\left(1-\sqrt{\frac{R_{\mathrm{in}}}{R}}\right)\right]^{1 / 4}, 
\end{equation}
where $\sigma_{\mathrm{B}}$ is the Stefan–Boltzmann constant, $R$ is the distance to the mass center, and $M_{\bullet}$ is the central BH mass. Assuming a radiative efficiency of $\epsilon \sim 0.1$ \citep[e.g.,][]{2002MNRAS.335..965Y, 2008ApJ...689..732Y}, the inner boundary of the mini-disks $R_{\mathrm{in}}$ is approximated as $3.5R_{\mathrm{g}}$ for Kerr SMBHs. 

For disks with rate exceeding the Eddington limit, the disk temperature profile may be described as \citep{2006ApJ...648..523W}
\begin{equation}
\begin{aligned}
T_{\text{eff}}(R) &= \left[ \frac{8}{\kappa_{\text{es}} I_4 a} \sqrt{\frac{B \Gamma_{\Omega}}{(2N + 3) \xi}} \right]^{1/4} f^{1/8} R^{1/4} \Omega_{K}^{1/2},\\
&\approx 2.52 \times 10^7~\text{K} f^{1/8} \left( \frac{M_\bullet}{10 M_{\odot}} \right)^{-1/4} \left( \frac{R}{R_{\text{g}}} \right)^{-1/2}.
\end{aligned}
\end{equation}
Here the parameter $f$ is a function of the accretion rate, $\kappa_{\text{es}}$ denotes the electron scattering opacity, $a$ and $I_4$ are the radiation and numerical constants, respectively, $B = (1 - R_{\mathrm{in}}/R)$ is the boundary term, $\xi$ is a dimensionless parameter set to $1.5$, $\Gamma_{\Omega}$ is a coefficient of order unity, $N$ is the polytropic index, and $\Omega_{K}$ represents the Keplerian angular frequency. This formulation corresponds to the slim disk approximation for super-Eddington accretion flows \citep{2006ApJ...648..523W}.

The accretion flow is likely to become radiatively inefficient when the accretion rate falls below roughly $10\%$ of the Eddington limit. Indeed, for certain choices of viscosity parameter $\alpha$ and other flow parameters, the transition to advection-dominated accretion flow (ADAF) may occur at $f_{\rm{Edd}}\lesssim0.1$ \citep[e.g.,][]{2001MNRAS.324..119Y}. In this regime, advection begins to play a significant role in the energy transport \citep[e.g.,][]{1995ApJ...452..710N,2014ARA&A..52..529Y}, particularly given that the majority of accretion energy is dissipated in the corona rather than the disk at the inner region \citep[e.g.,][]{2000A&A...361..175M,2008MmSAI..79..134M,2009MNRAS.394..207C,2017MNRAS.467..898Q,2022iSci...25j3544L}, which can further diminish the disk accretion and result in relatively negligible emission. Therefore, we assume that a mini-disk becomes inefficient in radiation once its accretion rate falls below $0.1$. We also repeat our calculations allowing for unrestricted primary emission to test the robustness of our conclusions in Section~\ref{sec:discussion}. We note that this assumption barely affects the circumbinary disk emission, since the majority ($\gtrsim90\%$) of our target systems have accretion rates above this threshold and only the emission from the disk at large radii is considered. The outer circumbinary disk is still a standard thin disk \citep[see][]{2000A&A...361..175M}.

The emission from each disk is treated as blackbody radiation, governed by the effective temperature $T_{\mathrm{eff}}$:
\begin{equation}
F_{\mathrm{\lambda}}=\frac{1}{ D_{\mathrm{L}}^2}\int_{R_{\text {in }}}^{R_{\text {out }}} \frac{2 h c^2 \cos i / \lambda^5}{\exp \left[h c / \lambda k_{\mathrm{B}} T_{\text {eff }}(R)\right]-1} 2\pi R dR, 
\end{equation}
where $h$ is the Planck constant, $k_{\mathrm{B}}$ is the Boltzmann constant, and $i$ is the inclination angle defined earlier, with $i = 0$ corresponding to a face-on accretion disk. 

The unique three-component structure of SMBBH accretion disks may introduce differences in the SED compared to the single BH case (see Figure~\ref{fig:f2} for an example), consistent with earlier studies \citep[e.g.,][]{2012ApJ...761...90G, 2012MNRAS.420..860S, 2019MNRAS.485.1579K, 2025ApJ...995...68T, 2026arXiv260102288V}. Although the exact amount of missing (or additional) energy depends on the specific SMBBH parameters, our overall results are not significantly affected. For simplicity, we do not consider misaligned disks, which may occur under specific conditions \citep{2023ARA&A..61..517L}. The outer boundaries of the mini-disks are approximated by the mean Roche radius \citep[][]{1983ApJ...268..368E}
\begin{equation}
R_{\mathrm{RL}}(x)=0.49 a x^{2 / 3} /\left[0.6 x^{2 / 3}+\ln \left(1+x^{1 / 3}\right)\right],
\end{equation}
where $a$ still represents the binary separation, x equals $q$ or $1/q$ for the secondary and primary SMBHs, respectively. Assuming that each mini-disk fills half of the Roche lobe \citep[e.g.,][]{2024arXiv240514843G}, we set the outer radius as $R_{\mathrm{out}}(x)=0.5R_{\mathrm{RL}}(x)$. The inner edge of the circumbinary disk is approximated as $R_{\mathrm{c}} \approx a/(1+q)+2R_{\mathrm{H}}$, motivated by hydrodynamic simulations \citep{2017MNRAS.466.1170M,2024ApJ...970..156D}, where $R_{\mathrm{H}}\approx 0.69 q^{1/3}a$ is the Hill radius. By combining contributions from all three components, we compute the spectral energy distribution (SED) for each SMBBH. Figure~\ref{fig:f2} shows an example for such a system (see the disk structure illustrated in the left panels). 

In addition, we explore the effect of X-ray reprocessing/irradiation on the generated SEDs, since it may partially refill the notch feature \citep[e.g.,][]{2015MNRAS.446L..36F,2019NewAR..8601525D}. In SMBBH systems, the X-ray emission is expected to arise mainly from AGN-like coronas associated with the two mini-disks, as well as from shocks produced by the impact of the accretion streams from the circumbinary disk on the outer edges of the mini-disks \citep[e.g.,][]{2014ApJ...785..115R,2017ApJ...835..199R,2018MNRAS.476.2249T,2018ApJ...865..140D,2022ApJ...928..137G,2025MNRAS.543.2670K}. In reprocessed/irradiated disks, high-energy photons from the inner accretion flow are absorbed by the outer disk and re-emitted as blackbody radiation, altering the observed spectral energy distribution. Motivated by recent calculations of irradiated circumbinary disks \citep[e.g.,][]{2024ApJ...975..141L,2025arXiv250614141B}, we adopt the reprocessing formalism of \citet{2024ApJ...975..141L} as a geometrical prescription, but replace the central irradiation source by the  coronal and shock X-ray luminosities appropriate for SMBHBs. 

We estimate the total high-energy luminosity $L_{\rm X,\rm tot}$ as $L_{\rm X,\rm tot}=L_{\rm X,\rm cor}+L_{\rm X,\rm shock}$. The coronal component $L_{\rm X,\rm cor}$ is computed from the intrinsic rest-frame $2-10\,{\rm keV}$ luminosity of each mini-disk using the Eddington-ratio-dependent bolometric correction of \citet{2020A&A...636A..73D}, as adopted in recent SMBBH X-ray reflection modeling by \citet{2025ApJ...989..190M},
$L_{2-10,i}=\frac{f_{{\rm Edd},i}L_{{\rm Edd},i}}{K_{\rm X}(f_{{\rm Edd},i})}$, where $K_{\rm X}(f_{{\rm Edd},i})=7.51\left[1+\left(\frac{f_{{\rm Edd},i}}{0.05}\right)^{0.61}\right]$, $i=1,2$.
The component Eddington ratios $f_{{\rm Edd},i}$ are obtained using the same mini-disk accretion allocation as in our fiducial model (dependent on mass ratio). We convert the $2-10\,{\rm keV}$ luminosity to a broad-band coronal luminosity through $L_{X,\rm cor}=\eta_{\rm X}\sum_i L_{2-10,i}$, with fiducial $\eta_{\rm X}=3$, corresponding approximately to integrating a coronal spectrum with index $\Gamma=2$, consistent with \citet{2025ApJ...989..190M}, over $0.5-100$ keV. We next estimate $L_{X,\rm shock}$ directly from the stream-impact hot-spot luminosity formula of \citet{2014ApJ...785..115R}. The X-ray irradiation mainly alters the temperature profile in the outer disk where the original temperature is relatively low \citep{2024ApJ...975..141L}. The irradiation heating of the circumbinary disk is then included as $F_{\rm irr}(r)=\frac{C_{\rm irr}L_{\rm X,\rm tot}}{4\pi r^2},$ where $C_{\rm irr}\sim 2A\,d(H/r)/d\ln r$ is an effective irradiation efficiency on the order of $10^{-3}$, $H$ is the disk scale height, and $A\sim 0.1$ \citep{2024ApJ...975..141L} is the absorption factor.
This updates the circumbinary disk temperature to $T_{\rm irr,CBD}(r)=\left[T_{\rm eff,CBD}^4(r)+\frac{C_{\rm irr}L_{\rm X,\rm tot}}{4\pi\sigma r^2}\right]^{1/4}$. The subsequent multi-temperature blackbody integration is unchanged. This approximate treatment serves as a conservative test of whether X-ray reprocessing can refill the SED deficit.

\subsubsection{SED-deficit diagnostics} 
\label{subsubsec:diagnostics}

The SED of an SMBH accretion disk typically follows a power-law relation in the optical-to-UV bands, $\lambda F_{\lambda} \propto \lambda^{k}$, with $k=-\frac{4}{3}$. However, the cavity or gap opened by the secondary SMBH can introduce deviations from this relation. \citet{2012ApJ...761...90G} modeled the SED of an SMBBH accretion disk with a broken power-law, quantifying the prominence of the SED deficit through changes in the power-law index (See Section 5 therein). Based on their calculations and the observed SED deficit in Mrk 231 \citep{2015ApJ...809..117Y}, we propose a diagnostic approach for identifying SED deficits according to the changes of the spectral index. The power-law index $k_n$ between two adjacent bands $\lambda_n$ and $\lambda_{n+1}$ ($\lambda_{n+1}>\lambda_n$) can be defined as
\begin{equation}
\begin{aligned}
k_{n} \equiv & \frac{
\ln \left[ \lambda_{n+1} F_\lambda(n+1) \right] - \ln \left[ \lambda_n F_\lambda(n) \right]
}{
\ln \lambda_{n+1} - \ln \lambda_n
}, \\
&  (n = 1,2,\dots,N-1),  
\end{aligned}    
\end{equation}
Here, $N$ denotes the number of photometric bands under consideration. Among all adjacent bands within the detector’s wavelength coverage, the most significant spectral index change can be defined as
\begin{equation}
\Delta k_{\max} \equiv \bigl\{\Delta k \bigr\}_{\max}= \max \left\{ k_{i+1} - k_{i} \,\middle|\, i = 1, 2, \ldots, N - 2 \right\}. 
\label{eq:delta_k}
\end{equation}
The parameter $\Delta k$ effectively captures the second-order derivative properties of the SED profile, and the maximum value $\Delta k_{\max}$ highlights the most prominent SED-deficit feature, analogous to the broken power-law fitting approach used in \citet{2012ApJ...761...90G}. By analyzing the SEDs of the mock SMBBHs and incorporating the quantitative results from \citet{2012ApJ...761...90G}, we establish a threshold criterion of $\Delta _{\mathrm{min}}=1.0$ for detecting SED deficits, which is empirical but effective in distinguishing SMBBHs for the purpose of theoretical investigations (as discussed later). 

To capture spectral index changes across a broader wavelength intervals, we also consider the index changes between non-adjacent bands, denoted as $\Delta k^{\mathrm{all}}$, the maximum of which is
\begin{equation}
\Delta k_{\max}^{\mathrm{all}} \equiv \max \left\{ k_{i} - k_{j} \,\middle|\, i > j;\ i, j = 1, 2, \ldots, N - 1 \right\}.  
\end{equation}
We compute the distribution of these index changes and adopt a threshold of $\Delta _{\mathrm{min}}^{\mathrm{all}} = 1.2$, chosen to maintain a confidence level comparable to the adjacent band threshold $\Delta _{\mathrm{min}}=1.0$ (as will be discussed later in Section~\ref{subsec:false alarm}) incorporating noise modeling. An SMBBH is considered identifiable if either of the following criteria is satisfied, i.e., $\Delta k_{\max}>\Delta _{\mathrm{min}}$ or $\Delta k_{\max}^{\mathrm{all}}>\Delta _{\mathrm{min}}^{\mathrm{all}}$. 

The change of spectral index as the expected SED deficit caused by SMBBH triple-disk emission is generally continuous, which is different from those caused by other factors such as random variations in single disk emission. For this reason, we also introduce a cutoff on $\Delta k$ to control the SMBBH selection, i.e., only those systems with $\bigl\{\Delta k \bigr\}_{\min} > -0.1$ are considered further as SMBBH candidates rather than false positive single AGNs (see Section~\ref{subsec:false alarm}). 

These SED-deficit features are sensitive to SMBBH parameters, particularly the redshift. The right panel of Figure~\ref{fig:f2} demonstrates how the location of the deficit wavelength shifts with redshift $z$, as diagnosed by the index change $\Delta k$. Unlike the rough estimation of SED deficits based solely on the characteristic energy of missing photons, this routine excludes sources indistinguishable from normal AGNs, thereby reducing potential overestimation. 

Utilizing the criteria above, we identify SED-deficit SMBBHs among the mock samples. We adopt the seven photometric bands of the China Space Station Telescope (CSST) with parameters given in \citet{Zhan2021} (see Fig.~\ref{fig:f2}). To ensure the detection of all bands, the magnitude limits of which are $(25.4,25.4,26.3,26.0,25.9,25.3,24.5)$\,mag for the $(NUV, u, g, r, i, z, y)$ bands with central wavelengths at $(0.295, 0.36, 0.475, 0.62, 0.764, 0.905, 0.965)$\,$\mu\mathrm{m}$, we assume a minimum sensitivity of $\sim 24$ magnitudes across all bands in the following analysis to control the false positives. The central wavelength of the $z$ and $y$ bands are quite close to each other, and the $y$-band photometric uncertainties are relatively large, increasing the likelihood of false alarms. Therefore, we exclude the index change between these two bands from our analysis. Other facilities with comparable detection capabilities, such as Rubin Observatory Legacy Survey of Space and Time (LSST) \citep[e.g.,][]{2019ApJ...873..111I,2019MNRAS.485.1579K,2025arXiv250610846X,2025arXiv250821510C}, are expected to offer similar prospects. We also test our selection criteria against single AGNs exhibiting various noise factors, and estimate the corresponding false alarm rate of our diagnostic in Section~\ref{subsec:false alarm}. Note that our control of $\bigl\{\Delta k \bigr\}_{\min} > -0.1$ also excludes a significant fraction (about $70\%$) of genuine SMBBHs identified by the deficit threshold ($\Delta k_{\max}>\Delta _{\mathrm{min}}$ or $\Delta k_{\max}^{\mathrm{all}}>\Delta _{\mathrm{min}}^{\mathrm{all}}$) after incorporating noise.

\subsubsection{Periodic variation} 
\label{subsubsec:variation}

In this section, we describe the selection of periodically variable SMBBHs in our mock samples, similarly to that done in \citet{2019MNRAS.485.1579K}. We adopt a V-band variability threshold of $\delta m_{\mathrm{V}} \geq \delta m_{\mathrm{V,min}}=0.05$, numerically analogous to the minimum variability floor $\delta_{\mathrm{F,min}}=0.05$ used by \citet{2019MNRAS.485.1579K}. Similarly, we account for the influence of the signal-to-noise ratio (SNR) on the variability threshold. For both Doppler modulated variable SMBBHs (hereafter DopVar) and hydrodynamical accretion modulated variable SMBBHs (hereafter HydroVar), we assume that the variability period corresponds directly to the SMBBH orbital period, i.e., $P_\mathrm{var}=P_\mathrm{orb}$. 

For Doppler-induced variations, the periodicity arises from the periodical change of the line-of-sight velocity $v_{\parallel}=v_{\mathrm{orb}}\sin i$ due to SMBBH rotation. The observed flux from each mini-disk of an SMBBH system can be expressed as (see \citealt{2015Natur.525..351D})
\begin{equation}
F_{\nu}=D^3 F'_{\nu '}, 
\end{equation}
with the Doppler factor defined as
$D=\left(1-v^2_{\mathrm{orb}}/c^2\right)^{\frac{1}{2}}/(1-v_{\parallel}/c)$,
where $v_{\rm orb}$ is the orbital velocity of the BH in the center of the mini-disk, $v_\parallel$ is its line-of-sight velocity, the rest-frame frequency $\nu'$ is shifted to the observer's frame frequency as $\nu=D\nu'$, and $F'_{\nu'}$ is the intrinsic flux. In our calculations, the frequency shift caused by Doppler boosting is included together with the K-correction. We thus calculate the Doppler variation for each source in the SMBBH sample in terms of apparent magnitude $\delta m_{V}^{\rm D}$, and compare it with the threshold $\delta m_{\mathrm{V,min}}$ to determine whether the variation can be detected.

Periodic modulation of accretion streams onto the secondary BH is another possible origin for variability, as revealed by hydrodynamical simulations, \citep{1996ApJ...467L..77A, 2013MNRAS.436.2997D, 2014ApJ...783..134F, 2023ARA&A..61..517L, 2025ApJ...984..144T, 2024APh...15402892C}. Although some studies suggest that such variations may occur at periods longer than the orbital period (e.g., $5P_\mathrm{orb}$) for large mass ratio SMBBHs, the SMBBHs to our interests are barely affected, as they tend to have small mass ratios. Recent radiative magnetohydrodynamic (RMHD) simulations \citep{2025ApJ...991...71C} reveal that the periodic radiation may vary for different viewing angles, leaving more doubts about this issue. We assume the variability period to be equal to the SMBBH orbital period, for simplicity. The flux change caused by this modulation onto the secondary BH can be expressed as 
\begin{equation}
\Delta F_{\nu,2}^{h} = F_{\nu,2}(\chi f_{\mathrm{Edd},2})-F_{\nu,2}(f_{\mathrm{Edd},2}), 
\label{eq:Hydrovar}
\end{equation} 
where the modulation factor for the accretion rate is set to $\chi = 1.5$ \citep{2019MNRAS.485.1579K}. We note that the modulation factor may depend on binary parameters such as mass ratio \citep[][]{2014ApJ...783..134F,2019MNRAS.485.1579K}, but we adopt a fixed fiducial value for consistency and as a baseline simplification. Similar to the Doppler variable case, we calculate the hydrodynamic variation for each SMBBH in the mock sample in terms of apparent magnitude $\delta m_{\mathrm{V}}^h$, and compare it with the criterion to identify HydroVar candidates. Since accretion modulation remains an active topic in hydrodynamic simulations \citep[e.g.,][]{2023arXiv231016896D}, we test an alternative modulation pattern for comparison. By assuming a modulation factor $\chi = 1.5$ in Equation~\eqref{eq:Hydrovar} to be synchronous for both mini-disks, while still adopting the accretion allocation from \citet{2020ApJ...901...25D}, the predicted number of detectable HydroVar SMBBHs increases by around $20\%$.

We do not perform a survey-specific simulation. Instead, following the population-level detectability prescription of \citet{2019MNRAS.485.1579K}, we adopt an observer-frame period cut of $P_{\rm var}\lesssim5\,{\rm yr}$ together with the variability amplitude threshold described above. For an LSST-like observing baseline of $\sim10$yr, this period cut covers at least two observed cycles. Longer period systems are more strongly affected by baseline-related selection biases, as the observing baseline limits the recoverable periods (see \citealt[][their Fig.~A1]{2019MNRAS.485.1579K}). We emphasize that this is an idealized selection criterion. Although the cadence of LSST-like wide-field surveys is much shorter than the year-scale periods considered here, irregular sampling, seasonal gaps, stochastic AGN variability, and survey-specific window functions can reduce the period-recovery efficiency \citep[e.g.,][]{2016MNRAS.461.3145V, 2018ApJ...856...42S}. Therefore, the predicted numbers of periodically variable SMBBHs in this work are idealized detectability estimates rather than completeness-corrected survey forecasts.

\subsubsection{Joint selection} 
\label{subsubsec:joint}

We investigate the potential for jointly detecting SMBBHs using both periodic variations and SED deficits. This exploration provides a cross check for the selection of SMBBHs, partially mitigating the inherent limitations of each method, i.e., the false positives in periodic variability searches and the SED complexities for identifying deficits, thereby offering more reliable detection than either diagnostic alone. This defines a dual requirement: these joint candidates must be close enough to meet the variability criterion, while also displaying identifiable SED deficits within the detector's wavelength range. Our theoretical estimations aim to inspire such observational efforts \citep[e.g.,][]{2020MNRAS.492.2910G}, by analyzing the fraction and properties of these joint candidates. 

\section{Results}
\label{sec:results}

\subsection{SED-Deficit SMBBHs}
\label{subsec:detecability}

\begin{figure*}
\subfigure[BBH-nonevol model]{
\includegraphics[width=0.45\textwidth]{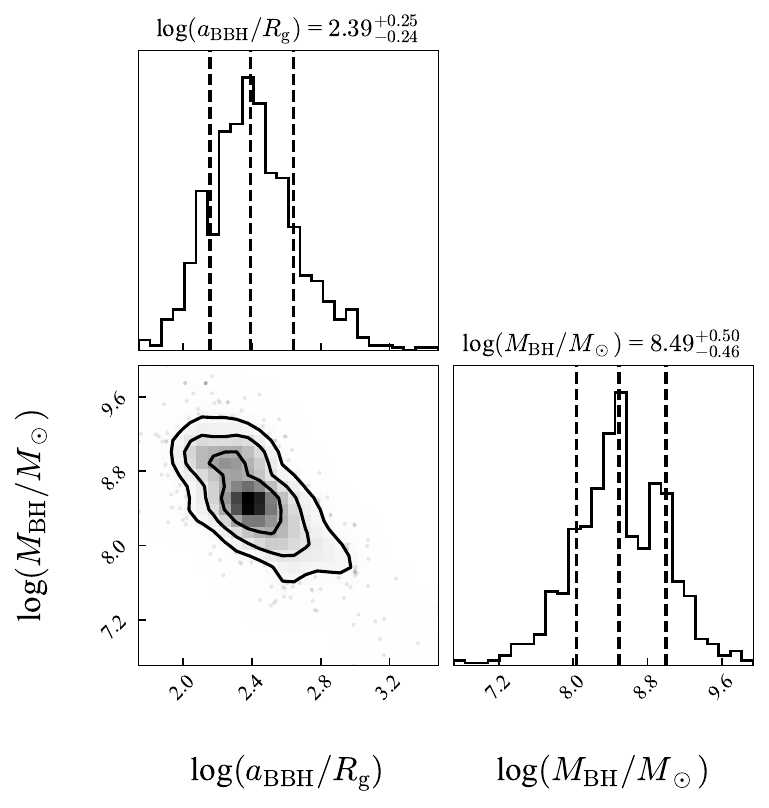} }
\subfigure[BBH-zevol model]{
\includegraphics[width=0.45\textwidth]{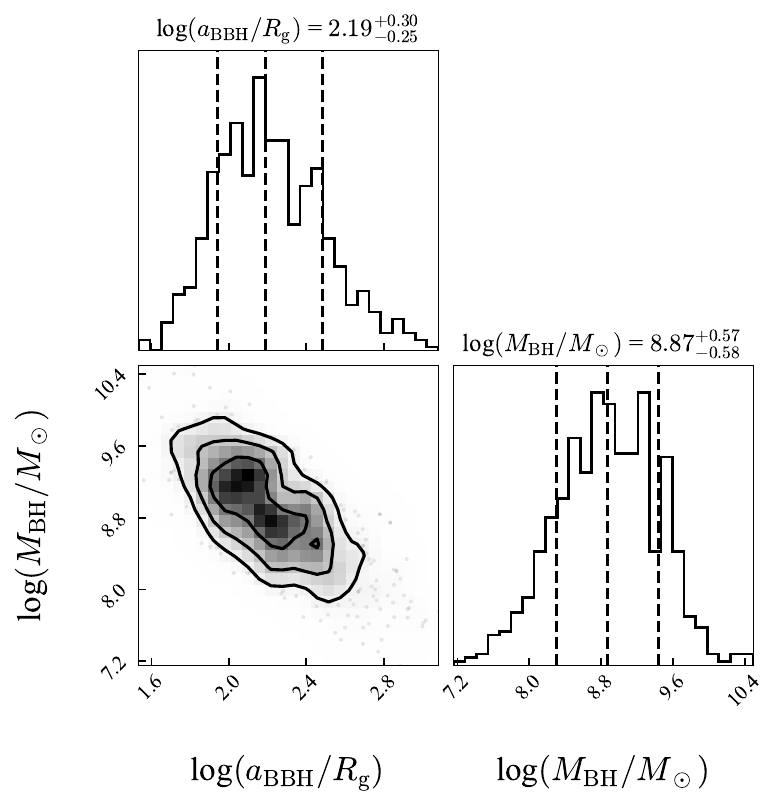} }
\caption{
Distributions of the primary mass and semi-major axis of those mock SMBBHs selected via the SED-deficit features. The left and right panels show the results obtained from the BBH-nonevol and BBH-zevol models, respectively.
} 
\label{fig:f3}
\end{figure*}

\begin{figure}
\centering
\includegraphics[width=0.9\linewidth]{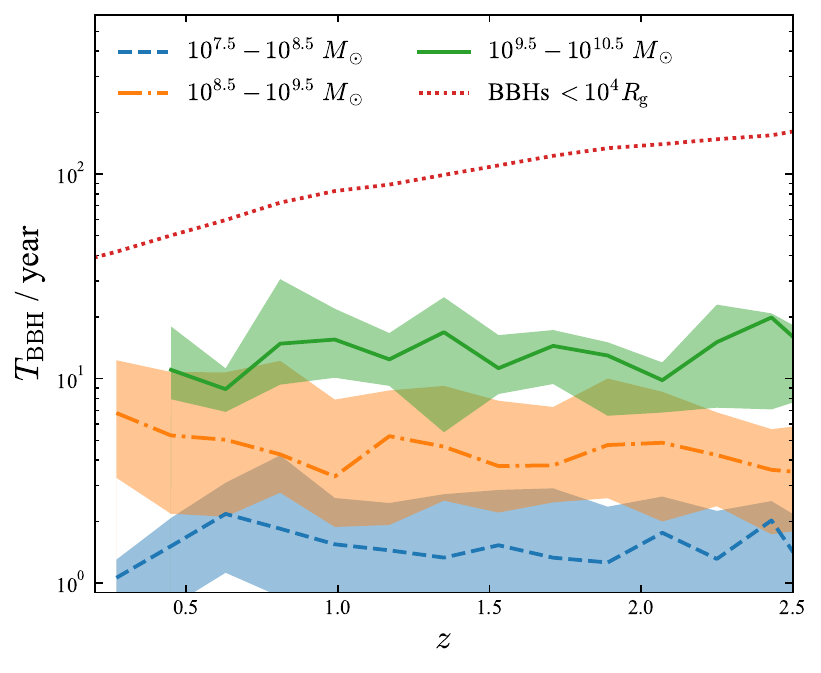}
\caption{
Orbital periods (in the observer's frame) of mock SMBBHs selected via SED-deficits obtained from the BBH-zevol model (see Section~\ref{subsec:detecability}). The blue, orange, and green lines indicate the medians of the orbital periods for the SMBBHs at different redshift bins with masses in the range of $10^{7.5}-10^{8.5} M_{\odot}$, $10^{8.5}-10^{9.5} M_{\odot}$, and $10^{9.5}-10^{10.5} M_{\odot}$, respectively, and the corresponding shaded regions mark the $16$-th to $84$-th percentile of the orbital periods. The dotted line shows the median of the orbital periods at each redshift for those SMBBHs with semi-major axes smaller than $10^4R_{\rm g}$.
} 
\label{fig:f4}
\end{figure}

\begin{figure*}
\centering
\includegraphics[width=0.9\linewidth]{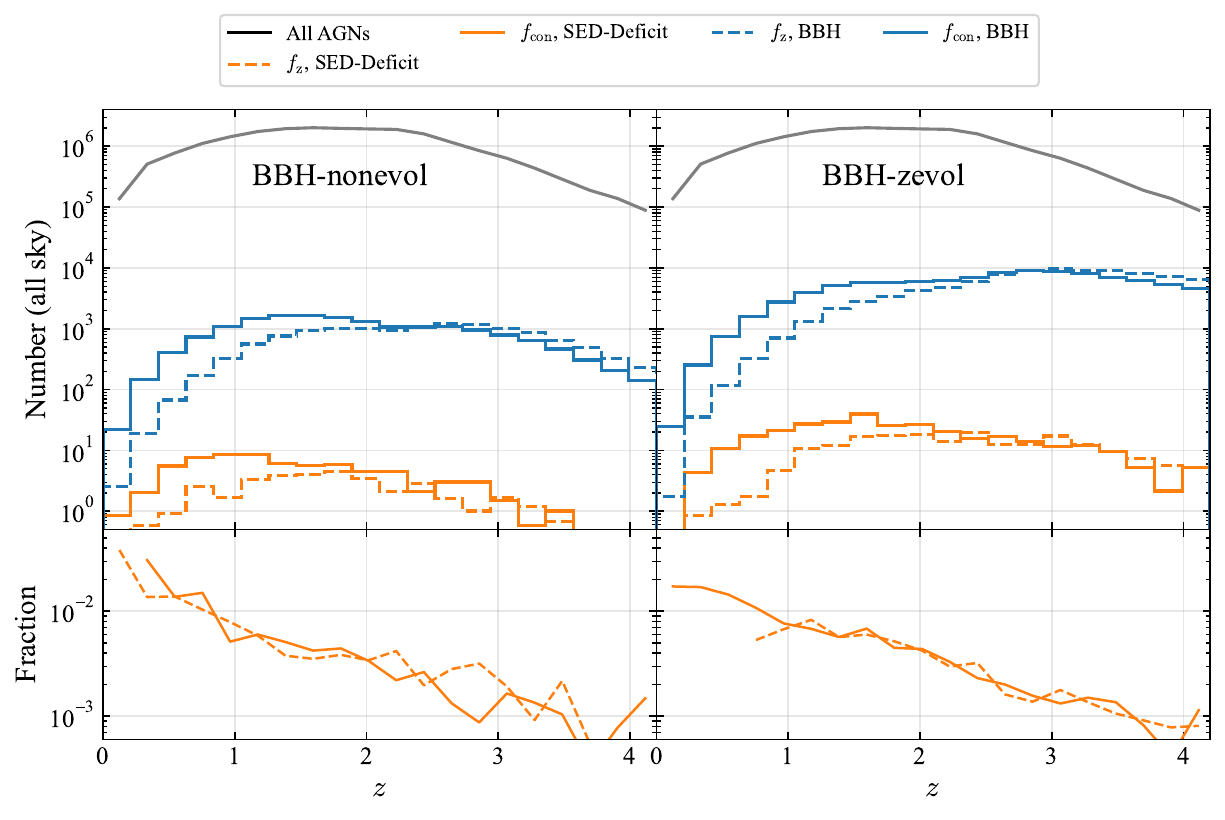}
\caption{
Number distribution of SED-deficit SMBBHs as a function of redshift. The top left and top right panels show the results obtained from the BBH-nonevol and BBH-zevol models, respectively. Solid and dashed colored lines correspond to models assuming either a redshift-independent ($f_{\mathrm{con}}$) or redshift-dependent model ($f_{\mathrm{z}}$) accretion rate model. The fraction of detectable SED-deficit SMBBHs (orange line) within the mock sample of magnitude-limited observable SMBBHs with $a_{\mathrm{BBH}}<10^4R_{\mathrm{g}}$ (blue line) decrease with redshift, with the exact ratios shown in the two bottom panels (see Section~\ref{subsec:detecability} for details). The results shown here are obtained by averaging over ten realizations. The black line at the top of each panel shows the total number of observed AGNs with V-band magnitude $\lesssim 24$ as a function of redshift (following \citealt{2023MNRAS.520.3476Q}).
}
\label{fig:f5}
\end{figure*}

We apply the diagnostic introduced in Section~\ref{subsubsec:diagnostics} to identify and analyze those SMBBHs exhibiting detectable SED deficits among the mock sample. Figure~\ref{fig:f3} presents the distributions of key parameters for SED-deficit SMBBHs, including the semi-major axis and the primary SMBH mass. We find that most sources are located at separations between $100$ and $1000 R_{\mathrm{g}}$, with the primary mass distribution peaking at $10^{8.5}M_{\odot}$ and $10^{8.9} M_{\odot}$ for the BBH-nonevol and BBH-zevol models, respectively. Our diagnostic method favors systems with relatively large SMBH masses, consistent with previous studies of SMBBHs exhibiting spectral features \citep{2019ApJ...879..110K}. In addition, under the assumption of our model, binary systems with smaller mass ratios are more likely to exhibit SED deficits, as in such systems the small secondary SMBH accretes the majority of the infalling material from the outer circumbinary disk, leading to more significant SED-deficit signatures. Consequently, the majority of the selected SED-deficit SMBBHs have $q \lesssim 0.1$. This preference mainly reflects the selection effect of our SED-deficit diagnostic under the adopted photometric band coverage.

However, our diagnostic requires the SED deficit to fall within the observational bands of the instruments, imposing strong constraints on the parameter space of these detectable SMBBHs. As shown in Figure~\ref{fig:f4}, most detectable SED-deficit SMBBHs have orbital periods shorter than $30$ years, corresponding to semimajor axes $\lesssim 0.1$\,pc. These limitations introduce a model-dependent ceiling on the total number of candidates. For example, assuming a redshift-dependent accretion rate $f_{\mathrm{z}}$, we estimate that the total number of detectable SED-deficit SMBBHs under the criteria set in Section~\ref{subsubsec:diagnostics} is about $36$ and $201$ under the BBH-nonevol and BBH-zevol models, respectively (see Table~\ref{tab:tabeltwo}). 

To evaluate the selection effects introduced by our method, we construct a control SMBBH sample with semi-major axes smaller than $10^4 R_{\mathrm{g}}$ and meeting the same detection criteria (i.e., $<24$ mag across all CSST bands). For SMBBHs in this sample, the observed orbital period correlates positively with redshift, primarily due to cosmological time dilation (proportional to ($1+z$)). However, the SED-deficit selected SMBBHs do not exhibit this trend. This is because they are selected based on their SED-deficit features falling within specific observational bands in the observer frame, a criterion that preferentially selects systems with shorter rest-frame periods at higher redshifts (proportional to $1/(1+z)$). Consequently, the distribution of their observed orbital periods shows no significant increase with redshift because the selection effect cancels the time-dilation effect (see Figure~\ref{fig:f4}). These selection effects, beyond the considerations of specific evolutionary stages and radiation capabilities, further limit the detection rate of SED-deficit SMBBHs. 

Figure~\ref{fig:f5} shows the redshift distribution of selected SED-deficit SMBBHs under different models, incorporating both redshift-dependent and redshift-independent accretion rate distributions. While the overall trends in the redshift distributions of SED-deficit SMBBHs resulting from different models are similar, the peaks of these distributions differ: for the BBH-nonevol model with redshift-independent accretion assumptions, the number density peaks at $z_{\mathrm{peak}} \sim 1$, whereas for the BBH-zevol model, the peak shifts to $z_{\mathrm{peak}} \sim 2$. 
The larger predicted number in the BBH-zevol model is not solely a direct consequence of the peak of the scaling relation. The adopted redshift evolution assigns larger BH masses to host galaxies at higher redshifts, allowing a broader range of host galaxies to contribute to the massive-binary population relevant to our selection. This intrinsic population shift is further amplified by selection effects. In particular, for the same range of $a_{\rm BBH}/R_{\rm g}$, more massive binaries have longer residence times, leaving more systems in the evolutionary stage where SED-deficit signatures are identifiable. In addition, more massive systems tend to be more luminous and are therefore more likely to pass the magnitude limit. 
These differences emphasize the influence of assumptions about scaling relations and accretion models, suggesting that the formation, evolution, and emission modeling of SMBBHs play an important role in determining the abundance and detectability of such SED-deficit systems. 

We also note that the fraction of SED-deficit SMBBHs among the magnitude-limited observable close SMBBH population decreases with increasing redshift. As previously noted, the SED-deficit features can only be identified within specific bands monitored by certain survey telescopes in the observer's frame. Comparing with those SED-deficit SMBBHs at low redshifts, those at high redshifts must be more compact (i.e., having relatively smaller semimajor axis), thus their SED deficits at higher frequencies in their rest frames are shifted to the telescope's bands. However, the residence time for SMBBHs with smaller separations is relatively shorter, leading to statistically fewer sources at these evolutionary stages. As a result, the fraction of SED-deficit SMBBHs among magnitude-limited SMBBHs shows a continuous trend of decreasing with redshift across all models, as seen from the bottom panels in Figure~\ref{fig:f5}. 

Finally, we compare the number of detectable SED-deficit SMBBHs with that of the all-sky AGN population. Using results from the miniJPAS mock catalog \citep{2023MNRAS.520.3476Q}, based on the Quasar luminosity function (QLF) and adopting a similar $r$-band sensitivity of $24$\,mag, we show the distribution of all AGNs as the black curve in Figure~\ref{fig:f5} and estimate that the overall detection rate of SED-deficit SMBBHs is approximately $\sim 0.0002\%$-$0.001\%$ among all AGNs. We adopt these results based on normalized mock observations to avoid potential deviations introduced by manual reproduction. This fraction, as also mentioned earlier, corresponds to tens to several hundred candidates (for various models) across the entire sky. 

\subsection{False positive control}
\label{subsec:false alarm}

Our diagnostic serves as a simplified and ideal form of hypothesis testing, and there may be significant associated false alarm rate for detection due to various factors. For our criteria, considering the faintest sources with y-band apparent magnitude $m_{y}=24$~mag (based on the power-law SED relation, single BH accretion disks are fainter at the longer wavelengths), we simulate $10^6$ SEDs of single BH accretion disks with the CSST sensitivity (assuming SNR$\sim$10 for spectroscopy to model the random noise) and find a false alarm rate of less than $0.001\%$. If we relax the restrictions to $\Delta _{\mathrm{min}}=0.5$, the false alarm rate will rise up to $0.37\%$, making the detection ineffective.
Our selection of $\Delta_{\mathrm{min}}$ and $\Delta_{\mathrm{min}}^{\mathrm{all}}$ provides not only a detailed criterion for identifying SED deficits, but also reflects a comprehensive consideration aimed at balancing efficiency and reliability.

Other significant complexities also persist in observations. For instance, if AGN variability appears asynchronous across different bands, or if emission lines contribute substantially to the flux measured in specific bands, additional uncertainties will arise beyond the simple noise model discussed earlier. However, the luminosity enhancements caused by such effects typically do not persist across consecutive bands, thus the false positives can be excluded if we observe a noticeable decline in the spectral index between other bands. Assuming a Gaussian-distributed asynchronous variation with a standard deviation of $\Delta F_\lambda/F_\lambda= 5\%$ in each band, the false alarm rate can reach as high as $13\%$ when adopting a threshold of ($\Delta _{\mathrm{min}}=1.0$, $\Delta _{\mathrm{min}}^{\mathrm{all}} = 1.2$). After introducing the index decline control, e.g., requiring the minimum observed index change to satisfy $\bigl\{\Delta k \bigr\}_{\min} > -0.1$, the false alarm rate falls below $0.05\%$. By adjusting the index decline control parameter, in combination with appropriately choosing the threshold ($\Delta _{\mathrm{min}},\Delta _{\mathrm{min}}^{\mathrm{all}}$), we can select SMBBH candidates despite observational complexities, though still accompanied with a large number of false positives. Although real line contamination may be stronger than our simple perturbation model, this test suggests that our diagnostic can partly suppress false positives caused by localized band-to-band flux perturbations.

Dust extinction might appear as an astrophysical mimic of an SED-deficit-like feature. To quantify its impact on our deficit diagnostic, we apply a standard Calzetti attenuation law \citep[][]{2000ApJ...533..682C} to the single-BH SEDs simulated above and recompute $\Delta k$ with the same procedure. We scan a representative reddening range of $E(B-V)=0-0.3$, and find that dust reddening induces a smooth, monotonic modification of the spectral index across all bands, without producing additional slope changes $\Delta k$ between adjacent bands. Consequently, the resulting $\Delta k$ values remain below our adopted identification thresholds, and the false-positive rate changes by less than 5\% relative to the unreddened case within the tested typical $E(B-V)$ range. 

We further test the robustness of the SED-deficit selection against X-ray reprocessing. Using the irradiated circumbinary-disk temperature profile described in Section~\ref{subsubsec:accretion model}, we recompute the SEDs of the mock SMBBHs and apply the same selection criteria. We find that the number of selected SED-deficit SMBBHs changes by less than 1\% compared with the fiducial calculation. Therefore, standard X-ray reprocessing from AGN-like coronas and shocks does not significantly alter our detectability estimates. This weak impact can be understood from the fact that the irradiation flux is much smaller than the local viscous flux near the inner edge of the circumbinary disk, and cannot efficiently refill the optical/UV deficit produced by the gap. However, our calculations serve as a representative test, as the detailed X-ray emission from SMBBH accretion flows is highly uncertain in realistic systems \citep[e.g.,][]{2020ApJ...900..148S}.

\subsection{Periodic variation SMBBHs}
\label{subsec:variableBBHs}

\begin{figure*}
\centering
\includegraphics[width=0.9\linewidth]{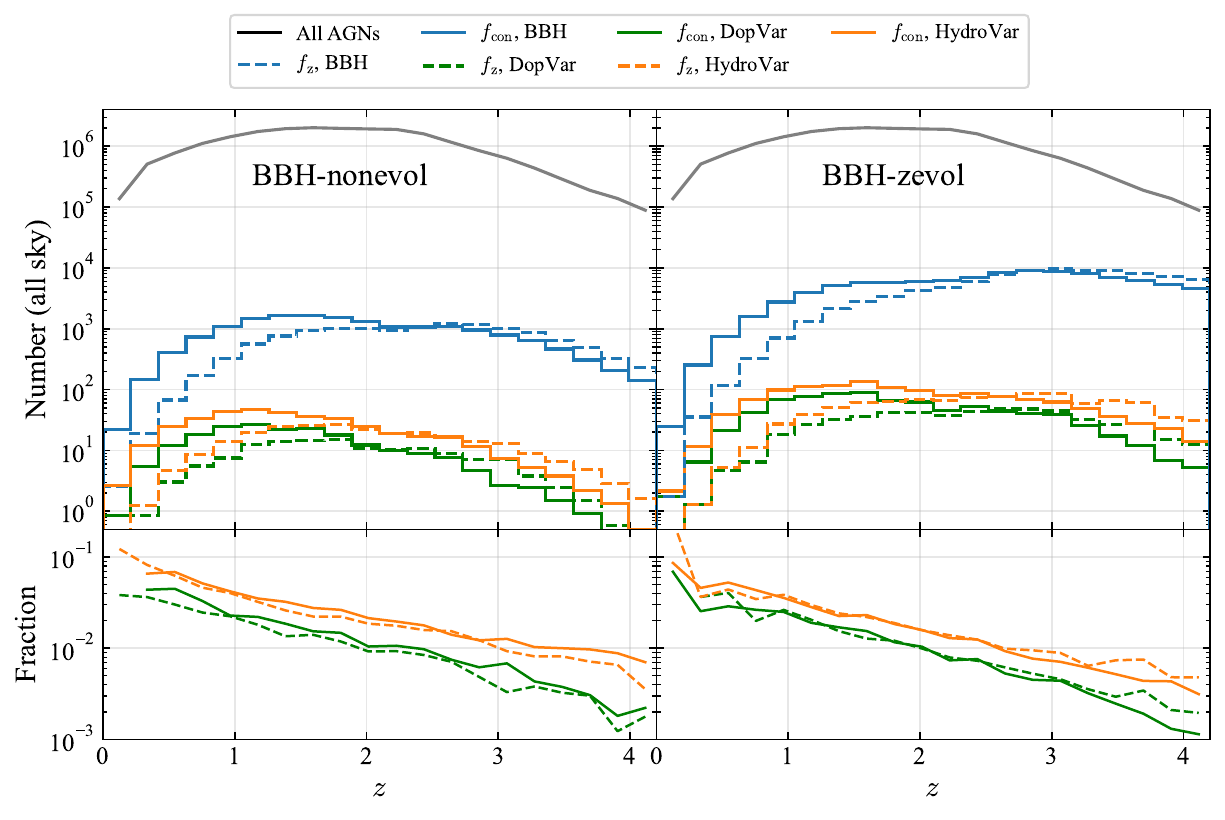}
\caption{
Redshift distributions of detectable variable SMBBHs. The top left and right panels show the results obtained from the BBH-nonevol and BBH-zevol models, respectively. Solid and dashed color lines still correspond to models assuming either redshift-independent ($f_{\mathrm{con}}$) or redshift-dependent ($f_{\mathrm{z}}$) accretion rate model. The fractions of detectable DopVar SMBBHs (green line) and HydroVar SMBBHs (orange line) in the mock sample of magnitude-limited observable SMBBHs with $a_{\mathrm{BBH}}<10^4R_{\mathrm{g}}$ (blue line) decreases with redshift, with the exact ratios shown in the two bottom panels. The black line at the top in each panel shows the total number of observed AGNs like before. The results shown here are obtained by averaging over ten realizations to avoid large fluctuations due to small number statistics. See Section~\ref{subsubsec:evolution} for detailed descriptions of these models.
}
\label{fig:f6}
\end{figure*}

\begin{figure}
\includegraphics[width=0.9\linewidth]{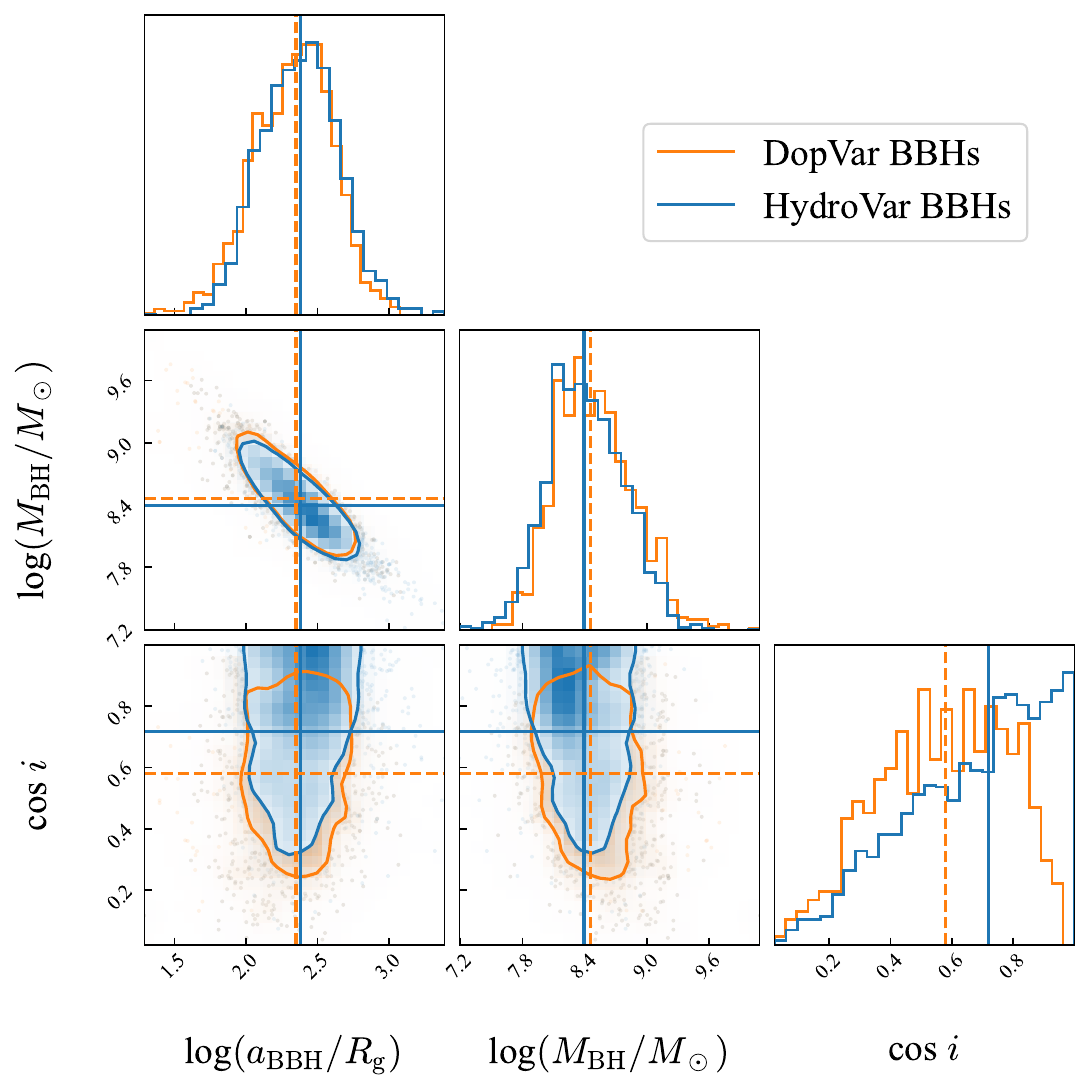}
\caption{
Semi-major axis, mass and inclination angle distributions for those mock SMBBHs selected via the Doppler-induced periodic variations (DopVar SMBBHs) and the Hydrodynamic-induced periodic variations (HydroVar SMBBHs). The distributions are obtained from the combination of $10$ realizations in order to mitigate fluctuations due to small number statistics. In the top, middle-right and bottom-right panels, the orange and blue curves represent the one dimensional distributions of the DopVar and HydroVar SMBBHs, respectively. In the remaining panels, the DopVar SMBBHs and HydroVar SMBBHs are marked by orange and blue dots, with the orange and blue curves showing the $68\%$ probability regions, respectively. The vertical orange dashed lines and solid blue lines indicate the mean values of the corresponding parameters.
}
\label{fig:f7}
\end{figure}

\begin{figure*}
\subfigure[BBH-nonevol model]{
\includegraphics[width=0.45\textwidth]{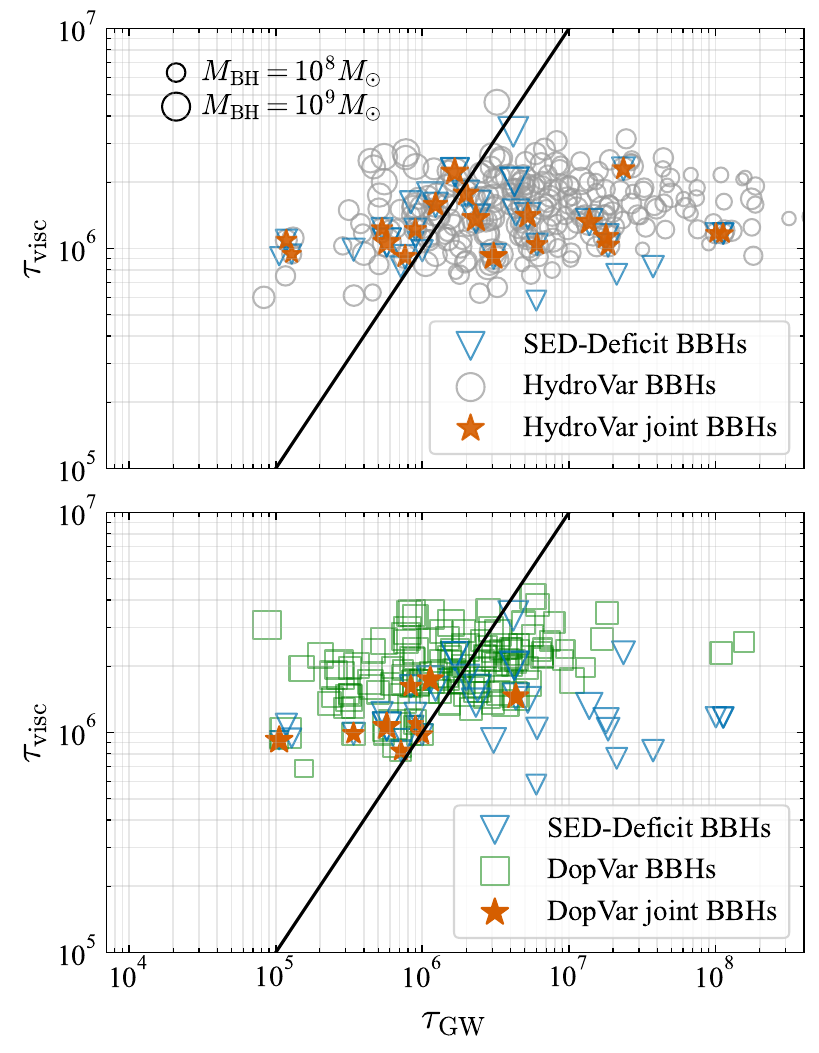} }
\subfigure[BBH-zevol model]{
\includegraphics[width=0.45\textwidth]{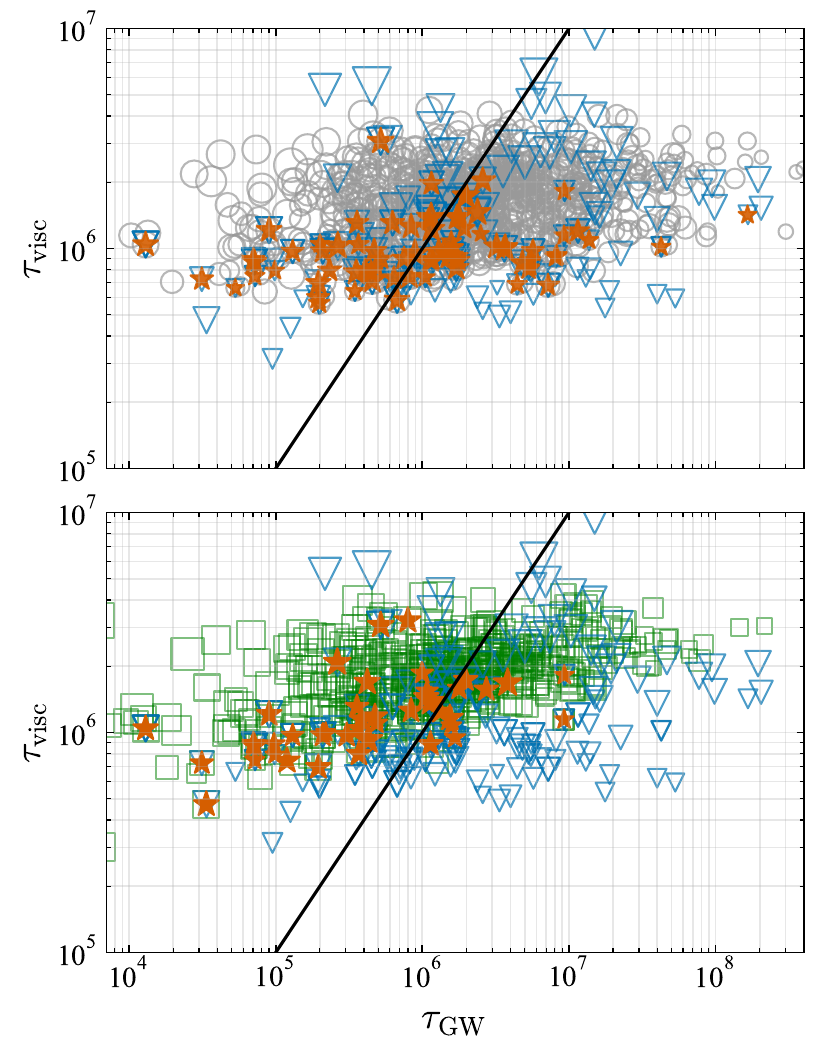} }
\caption{
Orbital decay timescales due to the GW radiation and the viscous disk interactions for those mock SMBBHs. Left and right panels show the results of a single realization under the BBH-nonevol and BBH-zevol models, respectively. In the top panels, gray circles represent SMBBHs selected from the mock sample via Hydrodynamic-induced periodic variations, and in the bottom panels green squares represent the SMBBHs selected from the mock sample via Doppler-induced periodic variations. In all panels, the blue triangles denote the mock SMBBHs selected via SED deficits, and the orange stars indicate those SMBBHs selected via both SED deficits and periodic variations. In each panel, the solid line represents $\tau_{\rm GW}=\tau_{\rm visc}$, separating the region of orbital decay dominated by GW emission (left of the solid line) from that dominated by viscous disk interactions (right of the line). Open circles mark the mock SMBBHs, and the larger the sizes of the symbols, the higher the masses, as indicated by the example markers shown in the upper-left corner of the figure.
}
\label{fig:f8}
\end{figure*}

We apply the periodic variation approach (see Section~\ref{subsubsec:variation}) to the mock SMBBH sample generated from population synthesis model and derive the number distributions for both DopVar and HydroVar SMBBHs. Figure~\ref{fig:f6} shows the number distributions of detectable SMBBHs selected via Doppler-induced periodic variations (DopVar) or Hydrodynamic-induced periodic variations (HydroVar). As seen from this figure, the candidate number drops rapidly beyond $z \gtrsim 2$, consistent with the trend reported in \citet{2019MNRAS.485.1579K} and \citet{2025arXiv250821510C}. This is mainly due to the cosmological redshift effects: SMBBHs at higher redshifts must have intrinsically shorter orbital periods to ensure that in the observer frame their variation periods are below the identification threshold (set to $5$ years in this work and also in \citealt{2019MNRAS.485.1579K}). This preference for more compact SMBBHs at high redshifts, correspondingly faster evolution and shorter residence time, reduces the overall number of detectable ones. Additionally, there exist other model-dependent preferences, the most apparent of which is that DopVar candidates are viewed at inclinations closer to edge-on (compared with HydroVar candidates, see Fig.~\ref{fig:f7}), same as the case in BELs \citep[][but more complicated BLRs \citealt{2024A&A...687A..57G}]{2020MNRAS.491.4023S, 2021RAA....21..219J}. 

The total numbers of detectable SMBBHs via Doppler-induced periodic variations (DopVar) or Hydrodynamics-induced periodic variations (HydroVar) are listed in Table~\ref{tab:tabeltwo}. The estimated numbers are $\sim 135-810$ and $\sim 254-1323$ for the former and the latter ones, respectively, and the exact numbers depend on the choices of the accretion rate model and the relationship between SMBH masses and host galaxy masses. In terms of the overall detectable SMBBH number, the BBH-zevol model predicts approximately 3-5 times more variable SMBBHs than the BBH-nonevol model does, reflecting the significant impact of the different choices of the SMBH–host galaxy scaling relation. If adopting the BBH-nonevol model, the resulting fraction of detectable HydroVar SMBBHs among all AGNs is $1.3\times10^{-5}$, which is similar to the fraction (i.e., $\sim 5\times 10^{-6}$) reported in \citet[][]{2019MNRAS.485.1579K}, while the fraction of detectable DopVar SMBBHs is $6.8\times10^{-6}$, about $8.5$ times larger than that obtained for Doppler variable binaries (i.e., $\sim 8\times 10^{-7}$) in \citet{2019MNRAS.485.1579K}, due to the differences in modelings. When adopting the BBH-zevol model, the fraction of detectable HydroVar SMBBHs increases to $6.6\times10^{-5}$, about $5$ times higher than that obtained for the BBH-nonevol model, while that of detectable DopVar SMBBHs rises to $4.1\times10^{-5}$, about $3$ times larger. 

It has long been recognized that the predicted number of periodically variable SMBBHs \citep[][]{2018ApJ...856...42S, 2019MNRAS.485.1579K, 2025ApJ...987..106C,2025arXiv250821510C} falls well below the observed population of variable AGNs \citep{2015MNRAS.453.1562G,2016MNRAS.463.2145C}. This discrepancy is largely attributed to the high false positive rate associated with periodicity searches \citep[e.g., ][in which a false alarm rate of $\sim0.001$ was obtained by considering processes like the damped random walk (DRW) for variations of normal AGNs]{2016MNRAS.461.3145V}, a conclusion further supported by recent detection results \citep{2025A&A...699A..55F} that have ruled out many previously claimed candidates. These false positives (produced by single SMBH AGNs) mimic periodic variability from active SMBBHs and therefore impede the detection of genuine SMBBHs. The recent discovery of $181$ periodic AGNs among $770,000$ AGNs by {\it Gaia} \citep{2025arXiv250516884H}, with careful consideration about the false positives due to DRW-like variability, gives an upper limit on the fraction of periodically variable SMBBHs, i.e., $\sim 0.02\%$, higher than the fraction we estimate above by a factor of one order of magnitude or more. This might suggest that a large fraction of these $181$ periodicity AGNs are perhaps still false positives rather than true SMBBHs, even though periodicities due to factors like DRW variations were analyzed and compared in \citet{2025arXiv250516884H}. This is further supported by \citet{2025arXiv250910601E}, in which the majority of these variable candidates from {\it Gaia} are considered to be safely excluded. Even if we assume that all the candidates in \citet{2025arXiv250516884H} are false positives, the upper limit for the false positive rate can be estimated as $\sim 0.02\%$.

Figure~\ref{fig:f8} shows the comparison between the orbital evolution timescale driven by disk viscosity and that driven by GW radiation for each SMBBH selected by SED deficits, Doppler- or hydrodynamic-induced periodic variations, and for those jointly selected by SED deficits and periodic variations. As seen from this figure, the evolution timescales due to viscosity distribute in a relatively narrower range ($\sim 5\times 10^5-10^7$\,yr) while those due to GW radiation distribute in a much wider range ($\sim 10^4-10^8$\,yr), mainly because the GW driven evolution is much more sensitive to the semi-major axis than that by viscosity. Most of the selected DopVar and HydroVar SMBBHs are around the transition period from the viscous disk dominant stage to the GW dominant stage, similar to the SED-deficit selected SMBBHs. Those periodic variation selected SMBBHs have relatively smaller evolution timescale compared with SED-deficit SMBBHs, mainly because that the period cutoff $P\le 5$\,yr in the selection criteria limits the SMBBH separation. The residence time for those SED-deficit and periodic variable SMBBHs is not too short (typically $10^6$\,yr) compared with the total lifetime of AGNs ($\sim10^8$\,yr), but the fraction of the selected SMBBHs with respect to all AGNs is small, which is mainly caused by the fact that only a small fraction of AGN host SMBBHs (consistent with \citealt{2019MNRAS.485.1579K}) with semi-major axes $<10^4R_{\rm g}$ (see Figs.~\ref{fig:f5} and \ref{fig:f6}).

To better understand the differences and complementarity between the SED-deficit and periodic variation approaches, we compare the resulting numbers and property distributions of the SMBBHs selected by these two approaches. First, the SED-deficit approach identifies fewer SMBBHs overall compared with the periodic variation approach, and favors SMBBHs with smaller mass ratios, slightly higher masses and closer separations in unit of $R_{\mathrm{g}}$ compared with those selected from the periodic variation approach. Second, periodic variation selected SMBBHs are constrained by the setting on their orbital periods, which is $<5$ years in the observer's frame in this work. In contrast, SED-deficit detection is governed by more complex inter-dependencies among system parameters (as discussed in Section~\ref{subsec:detecability}). Consequently, the SED-deficit approach results in SMBBHs covering a wider range of orbital periods (Figure~\ref{fig:f4}). These findings indicate that the SED-deficit approach can uncover some sources missed by the periodic variation approach, particularly the higher-mass, wider-separation systems. Moreover, the overlap between SED-deficit and periodically variable SMBBHs provides an independent means of verification to SMBBH candidates selected from either of the two approaches. 

\subsection{BBHs with both SED-deficits and periodic variations}
\label{subsec:jointBBHs}

\begin{table*}
\centering
\caption{
Predicted numbers of mock SMBBHs selected via different methods. 
}
\label{tab:tabeltwo}
\begin{tabular}{c c c c c c c c }
\toprule\hline
\multicolumn{2}{c}{} & \multirow{2}{*}{SED-Deficit} & \multirow{2}{*}{DopVar} &\multirow{2}{*}{HydroVar}  & \multicolumn{1}{c}{} & \multicolumn{2}{c}{Joint SMBBHs} \\ \cline{7-8}
\multicolumn{2}{c}{Models}  &   &  & & & DopVar+SED-deficit & HydroVar+SED-deficit \\ \midrule
\multirow{2}{*}{$f_{\mathrm{z}}$} & BBH-nonevol & 36 (53)   & 135  & 254 & & 9 (15)   & 19 (27)  \\
 & BBH-zevol   & 201 (313) & 569 & 1030 & & 42 (68) & 75 (115) \\ \midrule
\multirow{2}{*}{$f_{\mathrm{con}}$} & BBH-nonevol & 70 (134)   & 199 & 381 & & 18 (35)  & 35 (61) \\
 & BBH-zevol   & 318 (489) & 810 & 1323 & & 74 (107) & 123 (185)  \\ \bottomrule\hline
\smallskip
\end{tabular}
\begin{minipage}{\textwidth}
\textbf{Notes.} First column lists the SMBBH models, with both the BBH-nonevol and BBH-zevol models, each adopting either the redshift-independent ($f_{\rm con}$) or redshift-dependent ($f_{\rm z}$) accretion rate distribution. The second to fourth columns list the total number of mock SMBBHs selected via the SED-deficit, Doppler-induced periodic variation (DopVar), and Hydrodynamic-induced periodic variation (HydroVar) approaches, respectively. The fifth and sixth columns list the total number of mock SMBBHs selected simultaneously via the SED-deficit and Doppler-induced periodic variation method, and via the SED-deficit and Hydrodynamic-induced periodic variation method. Values in brackets correspond to results obtained after relaxing the continuous control criterion to $\bigl\{\Delta k \bigr\}_{\min} > -0.3$.
\end{minipage}
\end{table*}

We further assess the feasibility of joint detection by identifying SMBBHs that simultaneously exhibit SED deficits and periodic variation (DopVar or HydroVar). Adopting the limiting magnitudes of CSST and LSST, the BBH-nonevol model with redshift dependent Eddington ratio distribution yields $9$ or $19$ DopVar+SED-deficit or HydroVar+SED-deficit joint SMBBHs across all sky, while the BBH-zevol model with redshift dependent Eddington ratio distribution yields correspondingly a larger number of $42$ or $75$ (see Tab.~\ref{tab:tabeltwo}). Roughly $7\%-9\%$ of periodic variable SMBBHs by both variability mechanisms are indeed accompanied by identifiable SED deficits with the criteria set in this paper. Among SED-deficit SMBBHs, however, about $37\%-53\%$ are hydrodynamic variable BBHs and about $20\%-26\%$ are Doppler variable BBHs.

The combination of these two independent approaches may help mitigate the intrinsic uncertainties associated with each individual method, and may serve as a cross validation to offer robust evidence for sub-parsec SMBBHs. The false positive rate for either the SED-deficit approach ($0.05\%$) or the periodic variation approach (at most $0.02\%$ based on observations by \citet{2025arXiv250516884H}, if assuming that all candidates therein were false positives), could be much higher than the correspondingly expected detection fraction of SMBBHs. Assuming these two approaches are statistically independent from each other (see discussion in Section~\ref{subsec:assumptions}), then the false positive rate for SMBBH candidates jointly selected by both SED-deficits and periodic variations (hereafter joint candidates) is $\sim 0.05\% \times 0.02\% \sim 10^{-7}$, which is roughly an order of magnitude or more lower than the expected fraction of SMBBHs identifiable through the joint detection of SED-deficits and periodic variations. This indicates that the joint approach is effective for selecting genuine SMBBHs, with negligible contamination from false positives.  

The total number of jointly selected SMBBHs depends on the adopted criteria. For instance, if relaxing the false positive threshold for SED-deficit SMBBHs from the fiducial choice $\bigl\{\Delta k \bigr\}_{\min} > -0.1$ to $\bigl\{\Delta k \bigr\}_{\min} > -0.3$, the number of selected SMBBHs increases by approximately $50\%$ (see Table~\ref{tab:tabeltwo}). However, looser criteria naturally lead to higher false positive rate. Based on our tests, this relaxed threshold results in a false positive rate of $\sim 0.4\%$ among single AGNs, corresponding to a joint detection false positive rate of $\sim 0.4\% \times 0.02\% \simeq 8 \times 10^{-7}$, which is still below the fraction of the selected SMBBH population (under the BBH-zevol model). Determining the detection threshold therefore requires a careful balance between completeness and reliability.

\begin{figure*}
\centering
\includegraphics[width=0.8\linewidth]{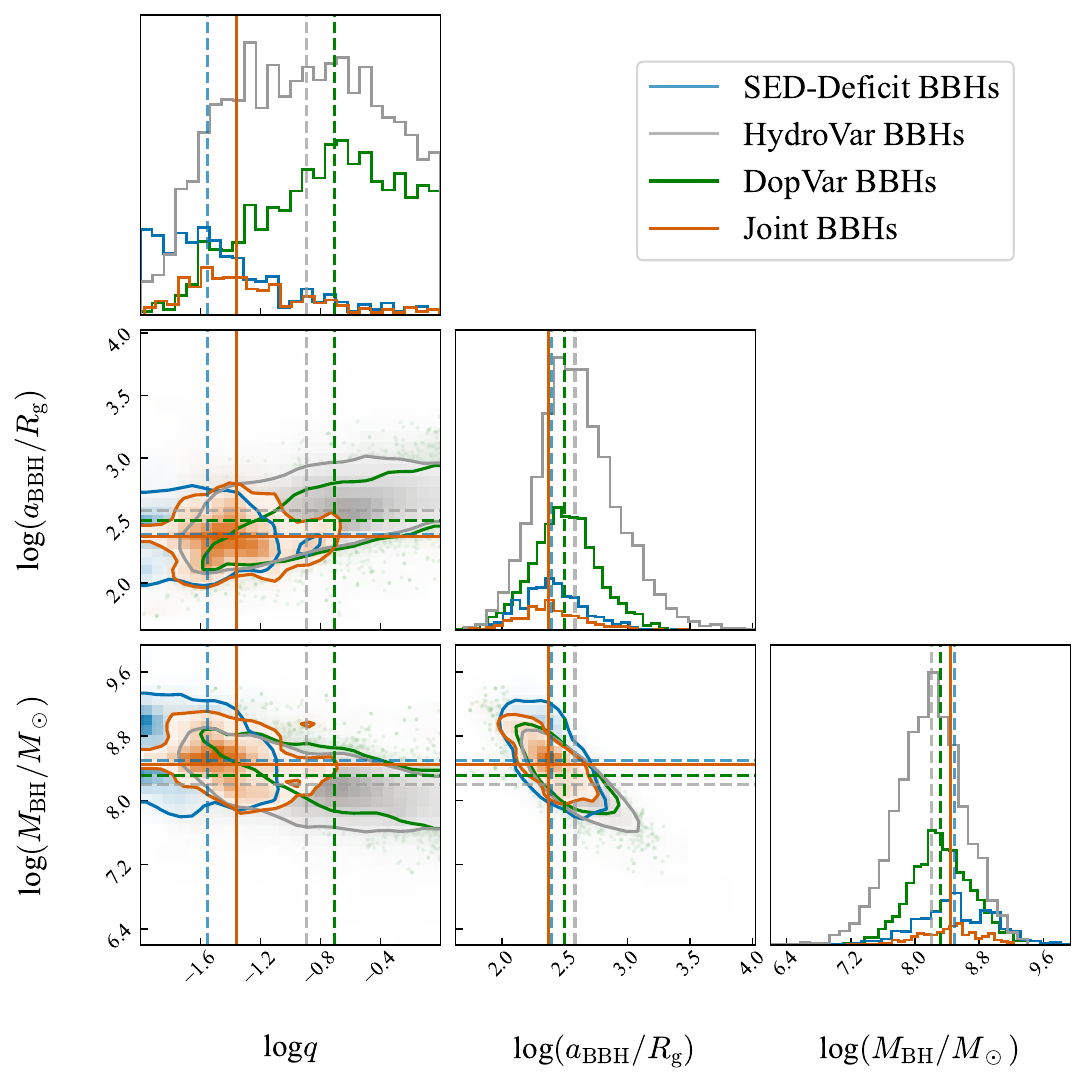}
\caption{
Distributions of the primary mass ($M_{\rm BH}$), mass ratio ($q$), and semimajor axis ($a_{\rm BBH}$ in units of $R_{\rm g}$) of SMBBHs jointly selected via both the periodic variation and SED-deficit (orange points) approaches in the BBH-nonevol model for ten realizations, and their comparison with those for SMBBHs selected via the Hydrodynamic variation (grey points), Doppler variation (green points) or SED deficit (blue points) approach only. 
}
\label{fig:f9}
\end{figure*}

We then examine the parameter distributions of mock SMBBHs that can be jointly selected by both SED-deficits and periodic variations. In general, these systems show moderate shifts toward the typical properties of SED-deficit SMBBHs (compared with the periodic variation systems). Figure~\ref{fig:f9} shows the distributions for the SED-deficit, HydroVar, DopVar, and HydroVar/DopVar+SED-deficit (joint detection) SMBBHs, respectively. As seen from this Figure, the jointly selected population tends to favor systems with tighter orbits than either individual population. This is consistent with the additional constraints imposed by the need for both detectable periodic variation and significant SED deficit. These jointly selected SMBBHs also favor smaller mass ratios, due to the preference of SED deficit detection, when compared with DopVar SMBBHs (see also Fig.~\ref{fig:f9}). This preference in $q$ imposes strong constraints, significantly reducing the number of detectable SMBBHs selected jointly by these two different approaches. Another implication is that adopting a hydrodynamic induced variability period longer than the orbital period for high–mass-ratio SMBBHs, as suggested by simulations and discussed in Section~\ref{subsubsec:variation}, has little impact on our joint results.

There are fewer additional constraints for joint SMBBHs derived from HydroVar SMBBHs compared with those from DopVar SMBBHs. So both the mass and separation distributions of joint SMBBHs derived from HydroVar SMBBHs are more relaxed and nearly identical to those of the SED-deficit SMBBHs. As the HydroVar method initially selects more SMBBHs than the DopVar method does, the expected number of HydroVar+SED-deficit SMBBHs is $\sim2$ times that of the DopVar+SED-deficit systems (see Table~\ref{tab:tabeltwo}).  

\subsection{GW observations of the selected SMBBHs by PTAs}
\label{subsec:GWPTA}

\begin{figure*}
\subfigure[Gas-rich SMBBHs $<10^4R_{\mathrm{G}}$]{
\includegraphics[width=0.49\linewidth]{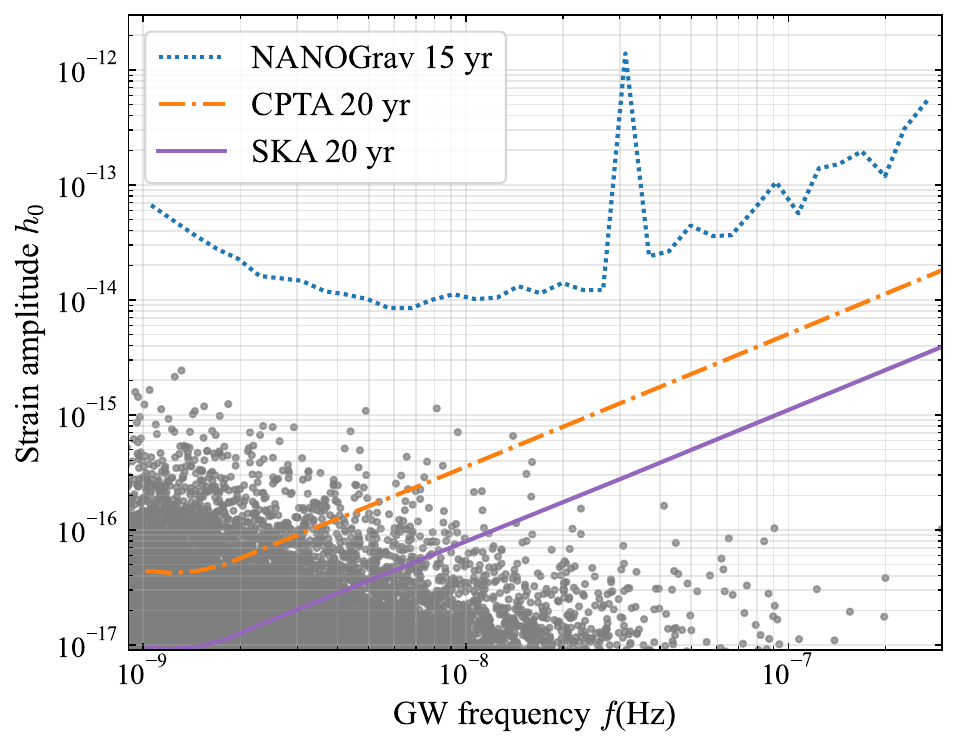}
}
\subfigure[EM detectable SMBBHs]{
\includegraphics[width=0.49\linewidth]{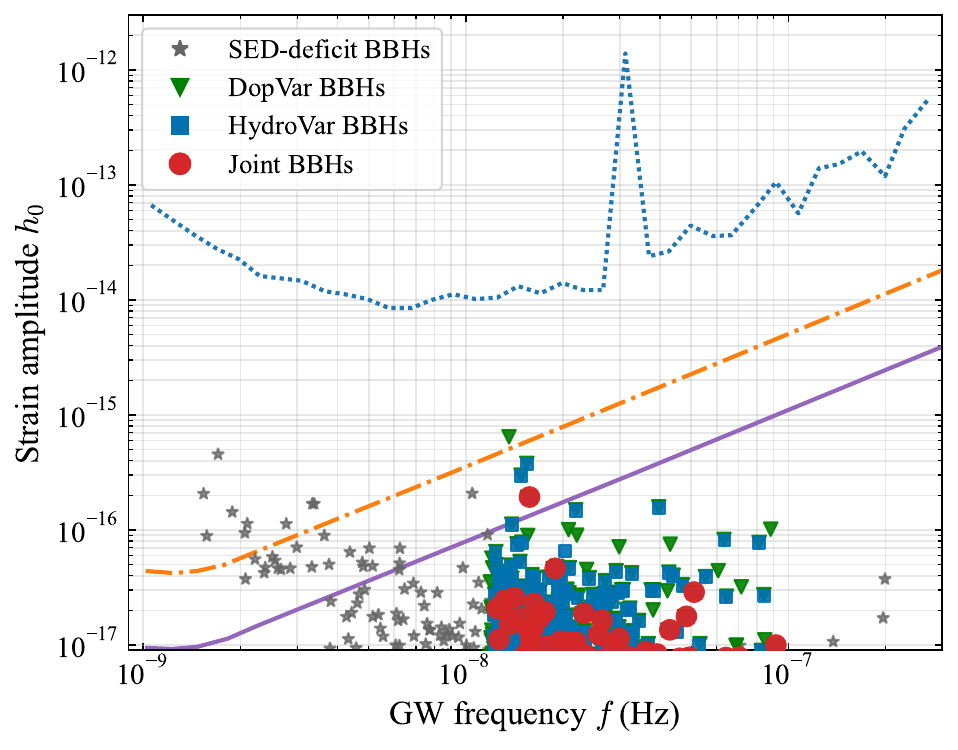}
}
\caption{
GW characteristic strains and frequencies of individual mock SMBBHs from gas-rich galaxy mergers with semimajor axes $<10^4R_{\mathrm{g}}$ in one realization (left panel) and those among them that can be selected by their periodic variation and/or SED-deficit (right panel) features. In the right panel, grey, blue, green, and red symbols specifically mark those SMBBHs selected via SED-deficits, hydrodynamic-induced periodic variations, Doppler-induced periodic variations, and both SED-deficits and periodic variations, respectively. The blue dotted curve shows the sensitivity threshold corresponding to the $95\%$ upper limit on individual SMBBHs, based on the NANOGrav 15-year data set \citep{2023ApJ...952L..37A}. The orange dash-dotted and purple solid lines represent the expected sensitivity curves for individual sources of CPTA and Square Kilometre Array (SKA) PTA over a total observational period of $20$\,years, respectively (see rough estimations in \citealt{2025ApJ...978..104G}). Apparently, some individual SMBBHs selected via SED-deficits can be detected at frequencies $\lesssim 10^{-8}$\,Hz by CPTA and SKA-PTA through $20$-year observations, however, at most a few that selected out either by Doppler-induced or Hydrodynamic-induced periodic variations can be marginally detected at $f>10^{-8}$\,Hz. The results shown in this figure are based on the BBH-zevol model. 
}
\label{fig:f10}
\end{figure*}

In this subsection, we further investigate the GW signals from those selected mock SMBBHs and check whether they could be detected by PTA experiments. Figure~\ref{fig:f10} shows the strain amplitudes of those SMBBHs selected by either periodic variations or SED deficits. As seen from this Figure, all the selected SMBBHs are below the sensitivity curve of NANOGrav 15 year data set \citep{2023ApJ...952L..37A, 2024ApJ...974..261C} and cannot be detected by current NANOGrav observations, which is consistent with that no individual GW signal from SMBBHs have been detected to date. 

Considering CPTA \citep{2016ASPC..502...19L} and SKA-PTA \citep[e.g., ][]{2010CQGra..27h4016S}, the GW signals from many of those selected SMBBHs may be detectable over a long period of observations. For CPTA, we assume that the total number of pulsars monitored is $N_{\rm p}=100$, timing precision $\sigma_t=20$\,ns, and cadence $\Delta t=0.04$\,yr, while for SKA-PTA, $N_{\rm p}=10^3$,  $\sigma_t=10$\,ns, and $\Delta t=0.04$\,yr. With these PTA configurations, we estimate that about $4$ or $15$ SED-deficit SMBBHs can also be detected with 20\,yr observations by CPTA or SKA-PTA with SNR$>3$, given that totally about $306$ or $1717$ SMBBHs can be detected with 20\,yr observations by CPTA or SKA-PTA overall. If we shorten the observation period to $10$\,years, the detection prospects for SED-deficit SMBBHs change to $0$ or $1.3$ (among $2.1$ or $49$ detectable SMBBHs overall) for CPTA or SKA-PTA. However, no SED-deficit SMBBHs can be detected when adopting the BBH-nonevol model. Among these SMBBHs, the total masses are typically concentrated at around $10^{10} M_{\odot}$, with a distribution among $10^{9.5}-10^{10.5} M_{\odot}$. As for redshifts, most of them are located within the redshift range $z \in(0.8,3.2)$ (taking SKA-PTA detectable ones as examples). Those mock SMBBHs selected via periodic variations (either DopVar or HydroVar) with orbital periods $<5$\,yr, radiating GWs at frequencies $\gtrsim 10^{-8}$\,Hz, are not easy to be detected, mainly because these PTA experiments are not sensitive enough at such high frequencies and the total number of the SMBBHs emitting at this frequency range with sufficiently large amplitudes is small. The most sensitive one among PTAs, i.e., SKA-PTA, may be able to detect one or at most a few such periodic variation SMBBHs within $20$-year observations. 

Future PTA observations are expected to reveal a large number of SMBBHs \citep[e.g.,][]{2023ApJ...955..132C}, and a small part of these SMBBHs are expected to be active and accompanied with SED-deficit features to be selected by targeted surveys. Among those candidates detectable by PTAs, quite a large number of active SMBBHs are at relatively low frequencies $\sim 10^{-9}$\,Hz, with about 1\% of them exhibiting SED features that meet our selection criteria. This highlights the importance of EM searches for sub-parsec SMBBHs in multimessenger astronomy. The periods of those candidates would be several tens of years and require long observation time. As a result, it is hard for PTAs to detect those SMBBHs with periodic variations ($P\lesssim 5$\,yr), primarily due to the limitation on the variation periods by EM observations, let alone those that exhibit both periodic variations and SED deficits. If future observations are able to monitor and identify periodic variable SMBBHs with longer periods, e.g., $>10$ years, the prospects for multimessenger observations will be considerably more promising, especially when the joint EM selection approach is employed.

\section{Conclusions and Discussions} 
\label{sec:discussion}

In this paper, we investigate the detectability of active sub-pc SMBBHs with SED-deficit features originating from the unique triple-disk SMBBH structure and/or the periodic variations induced by either the orbital motion modulated Doppler boosting or periodic variations of the accretion rates onto the inner mini-disks, as well as the joint detection of SMBBHs with both SED-deficits and periodic variations. To this end, we construct a population synthesis model to generate mock SMBBHs across the cosmic time, adopting the widely used procedures, i.e., modeling galaxy mergers based on the MRPG, GSMF, and SMBH-host galaxy scaling relations linked to galaxy morphology, and considering SMBBH orbital evolution particularly during the viscosity driven stage. We also calculate the SED of each mock SMBBH system with triple-disk structure and estimate the variability of its light curves due to either Doppler boosting or accretion rate variation. We summarize the main conclusions of this work as follows, and then discuss the limitations of the adopted models and compare our results with those in previous studies.
\begin{enumerate}
\item Using optical SED-deficit signature can help identify active SMBBHs with semimajor axes $\lesssim 10^3R_{\rm g}$, consistent with \citet{2019ApJ...879..110K}. After carefully modeling the SMBBH accretion disk structure and applying diagnostic criteria for the SED deficit signatures, we find that up to $\sim 318$ active SMBBHs across the entire sky may be detected by the multiband observations using filters as those for CSST with a magnitude limit of $24$\,mag for each band in our fiducial model (BBH-zevol). Most of these SMBBHs have a primary BH with mass $\sim 10^{8.5}$–$10^{9}M_{\odot}$ and a small mass ratio of $q \lesssim 0.1$. 
Forthcoming wide-field surveys (e.g., CSST, J-PAS, and LSST) are expected to uncover hundreds of such SMBBHs, but with substantial false positives due to uncertainties introduced by AGN variability, emission lines, extinctions and obscurations, and SED modelling for active SMBBH systems. 
\item We find that periodic variability, as a powerful probe to sub-pc SMBBHs, may yield up to more than a thousand of SMBBHs by future surveys. We predict a fraction of detectable SMBBHs via periodicities (by a factor of several times) smaller than that currently obtained from observations, in line with other theoretical studies \citep[e.g.,][]{2019MNRAS.485.1579K}, suggesting that the false positives are substantial among the periodic variation selected SMBBH candidates.
\item The joint detection of SMBBHs with SED-deficits and periodic variations provides a more reliable way to select SMBBH candidates through cross-validation. We find that such systems with joint detection account for only a small fraction (about $7\%-9\%$, correspondingly a number of up to one hundred) of the overall periodically variable SMBBH population due to the selection effects of the two approaches, which is compatible with the negative result of searches for SED-deficit features in periodic AGNs by \citet{2020MNRAS.492.2910G}. Moreover, this fraction increases to approximately $20\%$–$53\%$ for jointly selected SMBBHs among the SED-deficit selected SMBBHs, which exhibit similar mass distributions but tend to have higher mass ratios compared with those SED-deficit ones.
\item Considering the uncertainties in the SMBH-host galaxy scaling relation (especially its evolution), the expected detection number of SMBBHs could vary by a large factor. When adopting our fiducial redshift-dependent scaling relation constrained by the current GWB observations \citep{2023ApJ...955..132C}, the expected detection number of SMBBHs selected via SED-deficit (or periodic variation) is a factor $\sim 4.5-5.6$ more than that obtained by adopting the local baseline SMBH-host scaling relation without redshift evolution. Correspondingly, the expected number of joint detection of SMBBHs by adopting the redshift-dependent SMBH-host scaling relation is a factor of $\sim 3.5-4.7$ times that by adopting a redshift-independent scaling relation. In all cases, SMBBHs tend to be more massive when the redshift-dependent scaling relation is adopted. 
\item About $4$ and $15$ SED-deficit selected SMBBHs could be detected by CPTA and SKA-PTA over $20$ year observation period with SNR$>3$. If considering those SMBBHs that can be jointly selected by both SED-deficit and periodic variations with period $<5$ years, however, none of them are expected to be detected by either PTAs. Without considering the selection effects, however, the total number of active SMBBHs that can be detected by CPTA and SKA-PTA over $20$ year observation period with SNR$>3$ are about $306$ and $1717$. This may raise an important problem for electromagnetically identifying the hosts of hundreds of PTA detected SMBBHs \citep[e.g.,][]{2017NatAs...1..886M, 2018MNRAS.477..964K, 2024ApJ...974..261C} in the near future.
\end{enumerate}

\subsection{Model assumptions and limitations}
\label{subsec:assumptions}

We identify tens to hundreds of detectable SED-deficit SMBBHs in our mock sample assuming a limiting magnitude of $24$ mag across all bands, and with various settings depending on the SMBBH model. Taking into account the sky coverage, i.e., about 17,500 square degrees for CSST and 18,000 square degrees for LSST \citep{2019ApJ...873..111I}, the candidate number should be scaled by a factor of $\sim 40\%$ for both deficit SMBBHs and joint SMBBHs. Our predictions are inevitably dependent on the adopted models, particularly the galaxy merger rate and the binary evolution timescale during the disk accretion stage, both may significantly impact the predictions if they differ by factors of a few. Conversely, future detections of SED-deficit and/or periodically variable SMBBHs may help to put constraints on the cosmic population of SMBBHs.

When modelling SMBBH populations, we assign morphological types to merging galaxies and exclude gas-poor mergers with two elliptical progenitors. This reduces the total number of mergers by less than $10\%$. Unlike the detection of dual AGNs, where nuclear activity in both progenitors is required, our approach imposes fewer constraints on galaxy types. Under these assumptions, even S0 galaxies are capable of hosting inner disks in the majority of cases \citep[e.g.,][]{2018A&A...617A.113E}, enabling them to supply gas to the central BH or maintain the accreting activity from past galaxy mergers. Furthermore, we also test our outputs by adopting the GSMF fitting relations from other studies, such as \citet{2017MNRAS.471.3098L}, and find only minor differences ($<10\%$) in the resulting source numbers.

Another uncertainty in our population synthesis is the scaling relation for low-mass galaxies. The adopted GSMF formally includes dwarf or low-mass galaxies, which can enter the SMBBH population especially through minor mergers and may affect the properties of the secondary BHs. In this regime, the bulge fraction, BH occupation fraction, and the extrapolation of the $M_{\rm BH}-M_{\rm bulge}$ relation are less well constrained than in massive galaxies \citep[e.g.,][]{2020ARA&A..58..257G}. Some recent studies have extended the BH-bulge mass relation to low-mass galaxies. In particular, \citet{2019ApJ...887..245S} obtained an empirical relation for low-mass galaxies hosting broad-line AGNs, $\log(M_{\rm BH}/M_\odot)=8.80+1.24\log(M_{\rm bulge}/10^{11}M_\odot)$, with an intrinsic scatter of 0.68 dex. This empirical relation is broadly similar to the relation adopted in our BBH-nonevol model, although it has a larger scatter. To quantify the impact of this uncertainty, we repeat the BBH-nonevol calculation by adopting this relation for galaxies with $M_{\rm gal}<10^{10}M_\odot$, while keeping the other parts of the model unchanged. We find that the predicted number changes by only $\sim2.6\%$ for the observable magnitude-limited sample, $\sim7.6\%$ for the periodically variable sample and $\sim 3.3\%$ for the SED-deficit selected sample. Therefore, adopting an empirical low-mass BH-bulge relation has little impact on the baseline non-evolving prediction. 
For the fiducial BBH-zevol model, the selected samples are expected to be even less sensitive to the low-mass regime, because the detectable systems are dominated by relatively massive BHs. In particular, the secondary BHs in the final selected samples are mostly above $\sim10^{6.5}M_\odot$, with very few low-mass secondaries. This indicates that the uncertainties associated with low-mass galaxies mainly affect the low-mass tail of the underlying mock population and have limited impact on the detectable samples considered in this work.

The mass growth of both SMBHs due to accretion before they become a closely bound binary is ignored in our model, which may affect our results. However, recent studies \citep{2020MNRAS.495.4681I,2023MNRAS.519.2083I} suggest that, even if the mass growth is considered during both the pairing and hardening phases, the SMBBHs we concern with relatively high masses remain marginally affected because of the limited time available for substantial mass accumulation (typically $\sim 10^7-10^8$\,yr).

The allocation of accretion inevitably affects our results. In our calculations, we ignore the emission from the primary mini-disk when the accretion rate falls below $10\%$ of the Eddington limit. As shown by \citet{2017MNRAS.467..898Q}, the total emission can be suppressed by a factor $\gtrsim 3$ comparing with that estimated from the standard thin disk. If assuming the emission from all the primary mini-disks with accretion rate smaller than $10\%$ of the Eddington limit is just a factor $1/3$ of that from the standard disk, we find that the number of the SED-deficit selected SMBBHs decreases by at most $40\%$. Alternatively, if we keep the accretion allocation unchanged (see Section~\ref{subsubsec:accretion model}) but impose the Eddington limit on the secondary, thereby limiting the accretion onto the primary the same as the settings adopted in \textsc{L-Galaxies} \citep[e.g.,][]{2023MNRAS.519.2083I}, the number of detectable mock SMBBHs changes to $31$ and $318$ under the BBH-nonevol and BBH-zevol models, respectively. In these cases, the selected SMBBHs tend to have larger mass ratios and higher BH masses, which in turn lead to a lower joint detection fraction of $\sim 14\%$–$30\%$ among SED-deficit SMBBHs, rather than the fraction $\sim 20\%$–$53\%$ in our fiducial model, since periodically variable SMBBHs preferentially occur in systems with relatively large mass ratios. All these tests are performed with the same procedures and false positive controls as in our fiducial model, underscoring the substantial impact of accretion properties on our predictions. Nevertheless, the observational prospects are still promising, and at least a number of SMBBHs can be detected via the joint observations of both the SED-deficits and light curve periodicities.

In addition, fast transient surveys such as LSST are able to provide repeated observations, which can be used to obtain the mean SEDs by averaging multiple observations in each band thus avoid the asynchronous variability across different bands due to AGN intrinsic variations. With this advantage, a looser threshold of ($\Delta _{\mathrm{min}}=0.6$, $\Delta _{\mathrm{min}}^{\mathrm{all}} = 0.8$) may be sufficient to maintain a comparable false positive level, yielding approximately $234$ or $655$ SED-deficit SMBBHs under the BBH-nonevol or BBH-zevol model, which largely exceeds our fiducial estimate. 

We calculate SMBBH evolution at the disk stage based on \citet{2009ApJ...700.1952H}, a widely used model for SMBBH evolution in the accretion disk scale, which also matches the results from hydrodynamic simulations \citep[e.g., ][]{2017MNRAS.469.4258T}. However, several recent studies have challenged these semi-analytical results. \citet{2017MNRAS.466.1170M} reported disk out-drive, a finding supported by subsequent studies \citep{2021ApJ...918L..15B,2025ApJ...984..144T,2024ApJ...970..156D}. Some suggest that parameter settings strongly influence the SMBBH evolution \citep{2017MNRAS.469.4258T, 2020ApJ...900...43T, 2021MNRAS.507.1458F, 2022A&A...660A.101P}, while others emphasize the importance of eccentricity \citep{2021ApJ...914L..21D, 2023MNRAS.522.2707S, 2024MNRAS.534.2609S}. The evolution of SMBBHs embedded in gaseous disks remains an active area of research with many uncertainties, which highlights the importance of observational searching and confirmation of SMBBHs. If SMBBHs become stalled at the disk phase, the expected detection rate of sources with SED features would be more optimistic. For example, an increase in the viscous residence timescale by more than an order of magnitude could enhance the fraction of variable BBHs in our calculations, potentially approaching the fraction of $Gaia$ candidates reported by \citet{2025arXiv250516884H}. However, as the viscous timescale increases, GW begins to dominate at earlier evolutionary stages for these variable SMBBHs, making the increase in numbers less sensitive to further extensions of the viscous decay timescale.

For simplicity, we do not consider the eccentricity or misaligned disks \citep{2023ARA&A..61..517L}, nor do we account for the radiation of accretion streams within the gap \citep{2015MNRAS.446L..36F}, which has been shown to be negligible \citep{2016ApJ...832...22S}. Recent studies suggest the existence of an eccentricity equilibrium during SMBBH accretion \citep{2021ApJ...909L..13Z, 2023MNRAS.522.2707S, 2024MNRAS.534.2609S}, leading to `preferred' regions in the $[e,q]$ (i.e., binary eccentricity and mass ratio) parameter space, which deserves further investigation in future work. To assess the impact of disk truncation on our results, we repeat the calculations with a truncation radius of $R_{\mathrm{H}}$ instead of the fiducial $2R_{\mathrm{H}}$ motivated by hydrodynamic simulations. The change boosts the predicted number of SMBBH candidates by about 15\%-20\%, since some SMBBHs at larger separations (and being thus more abundant) can now exhibit their SED-deficit signatures within the survey wavelength range. 

We adopt the simple accretion disk model for calculating the SED from triple disk-SMBBH systems, which cannot fully reflect the reality. If the accretion disk(s) associated with SMBBH systems differ from the standard thin disk model \citep[e.g., see Section 5.3.1 in ][]{2018ASSL..454.....S}, the resulting SED may be substantially different from what we calculated from our simple model. Nevertheless, we argue that our method may still remain effective in identifying SED-deficit features if the disks have a significant power-law dependence on radius, i.e., cases where $R(\lambda)$ follows an exponent different from the standard $R\propto\lambda^{4/3}$. This is because our diagnostics focus on the SED index change (corresponding to the second-order derivative) caused by the disk structure and emission ``dip'', to recognize the SED-deficits, rather than assuming a fixed spectral index. In practical observations, the diagnostics can be adapted to identify ``outliers'' directly from observed AGN SEDs, providing a method for selecting SMBBH candidates like that in \citet{2020MNRAS.492.2910G}. 

Considering the complexities of AGN emission, our theoretical estimates for SMBBH detection are not yet conclusive for real observations. We have discussed the control of false positives caused by noise and AGN asynchronous variability. However, factors such as AGN obscuration (dominated by torus), color-dependent extinctions, and pseudo-continuum formed by emission lines from broad emission lines (BELs) \citep{2015ApJ...809..117Y, 2016ApJ...825...42V, 2019ApJ...879..110K, 2022ApJ...927....3F} remain to be considered, making it more challenging to detect SED-deficit features. Dust reddening \citep{2016ApJ...829....4L, 2022ApJ...927....3F, 2023arXiv231016896D} appears to be the most significant challenge in identifying such SED-deficit features, similar to the false positive issue in the search for periodically variable SMBBHs \citep{2018ApJ...856...42S, 2019ApJ...879..110K}. While the joint detection offers a means of cross-validation, addressing these challenges remains a critical and evolving area in the study of sub-parsec SMBBHs. Dealing with the emissions case by case can also mitigate some of these obstacles \citep{2015ARA&A..53..365N, 2016MNRAS.463.2064H, 2023NatAs...7.1506C, 2023MNRAS.521L..11J} during actual surveys. Spectroscopic observations can further help remove the impact of emission lines. The joint selection approach is also intended to address such challenges.

Finally, we note that the false positives of the SED-deficit and periodic-variability diagnostics may not be fully independent in all physical scenarios. In ordinary single-BH AGN, variable line-of-sight obscuration by clouds or outflows can alter the observed continuum shape while also inducing flux variability. Such changing-obscuration events are observed in AGN and are generally associated with variations in the line-of-sight column density caused by clouds or outflows eclipsing the central engine \citep[e.g.,][]{2018MNRAS.478.1660G, 2023NatAs...7.1282R}. Radio-loud quasars and blazars provide another possible source of correlated contamination, since jet-dominated emission can modify the broadband SED, while precessing radio jets may generate quasi-periodic variability in candidate periodic quasars \citep[e.g.,][]{2020MNRAS.492.2910G}. Nevertheless, these contaminants are not expected to generically reproduce the full set of binary-consistent properties. In the binary interpretation, the SED deficit and periodic variability should be linked to the same underlying binary parameters, and hence the deficit feature and variability timescale should be mutually consistent within a binary accretion model. False-positive systems may instead yield inconsistent binary parameters when the SED and time-domain signals are modeled jointly. Long-term multi-band monitoring and consistency checks between the SED-inferred and variability-inferred binary properties are therefore essential for rejecting correlated false positives, although such tests remain observationally expensive and model-dependent.

\subsection{Comparison with previous studies}
\label{subsec:comparison}

Our estimate of the detectable number of SMBBHs via SED-deficits is more than two orders of magnitude lower compared with the estimates in a previous study by \citet{2019ApJ...879..110K}, in which it was estimated based on the calculation of `notch', i.e., the characteristic energy of missing photons. They predicted approximately $100$ SMBBHs detectable at a magnitude limit of $m_{\mathrm{V}} \simeq 18$\,mag in the original SDSS Quasar Survey. In their approach, an SMBBH is considered detectable if either the low-energy cutoff or high-energy recovery of the SEDs fell within the observational bands. In contrast, we compute the continuum emission from triple-disk SMBBH systems and quantify their deviation from those of single BH systems \citep[][]{2012ApJ...761...90G, 2015ApJ...809..117Y}, offering a more direct and realistic comparison with observations. Under our criteria, we predict only about $0.05$ and $0.12$ SMBBHs with distinguishable SED-deficits under the same flux limit $m_{\mathrm{V}}=18$\,mag, which is more than two orders of magnitude fewer. This significant reduction in detectable SMBBH numbers mainly comes from our more detailed consideration of disk radiation, which makes the SED-deficit feature harder to detect due to integration of the emission from all the triple disks. In fact, the diagnostic criteria introduced in this paper for changes in the power-law index can mitigate overestimation.

Although our predicted total number of SED-deficit SMBBHs differs from that reported by \citet{2019ApJ...879..110K}, the underlying parameter distributions show close agreement. For example, both our BBH-nonevol model and the study of \citet{2019ApJ...879..110K} indicate that the SED-deficit SMBBHs have high-masses, with mass distributions peaking at $\sim10^{8.5} M_{\odot}$. Note that the SED deficit signatures can also be produced solely by the emission of two mini-disks according to our model. Such SED deficits are generally shallower and more challenging to verify compared to those produced by the gap between the inner mini-disks and the outer circumbinary disk. 

\citet{2019MNRAS.485.1579K} systematically forecasted the population of periodic variable AGNs containing SMBBHs. They classified accretion disks based on Eddington ratios and explored a wide range of model assumptions. Our results are broadly consistent with theirs: under LSST-like sensitivity, they predicted $\sim 30$ Doppler and $\sim200$ hydrodynamic variable SMBBHs over the full sky, while our BBH-nonevol model yields $135$ Doppler and $245$ hydrodynamic variable SMBBHs, respectively. These discrepancies likely arise from different model settings and sampling strategies. When we instead adopt the BBH-zevol model, motivated by the constraints from recent GWB detections, the number of detectable periodicity SMBBHs increases by several times, highlighting the significant influence of SMBBH modeling. Such an increase may help narrow the gap noted by \citet{2019MNRAS.485.1579K}, where the predicted number was much smaller than the number of observed candidates \citep{2015MNRAS.453.1562G, 2016MNRAS.463.2145C, 2025A&A...699A..55F, 2025arXiv250516884H}, though this gap is likely dominated by false positives. \citet{2025arXiv250821510C} recently reported a more optimistic estimate for periodically variable SMBBHs expected to be detected by LSST, after incorporating eccentricity and accretion history based on the \textsc{L-Galaxies} model.

\citet{2020MNRAS.492.2910G} performed a systematic search for joint sources in observations. Starting from the observed $138$ periodic variable AGNs, they collected SED data for each source and tried two approaches. In the first approach, they combined the normalized magnitude together to compare with a control sample of AGNs without periodicity. No direct statistical difference was found between the two samples, which may be not surprising given our analysis. As our results suggest that only a small fraction ($7\%-9\%$) of periodically variable AGNs are expected to exhibit identifiable SED deficits, and SED deviations from single BH AGNs may occur at different wavelengths for different systems since SED deficits are sensitive to SMBBH parameters (see Section~\ref{subsec:detecability}), thus the SED-deficit should be weakened or even washed out in the mixed SEDs. In their second approach, they selected abnormally red quasars with a color-based criterion, but found no inconsistency with the control sample in fraction. Apparently, the dominating factor is that most periodic variation AGNs may not really host SMBBHs \citep{2016MNRAS.461.3145V, 2018ApJ...856...42S, 2020MNRAS.492.2910G,2025arXiv250910601E}, which is also supported by studies based on SMBBH evolution \citep{2019MNRAS.485.1579K} and GWB merger rate reconstruction \citep{2025ApJ...987..106C}. Our analysis suggests that only a small fraction of periodic variation SMBBHs are expected to be accompanied with SED deficits. No such signatures were found by \citet{2020MNRAS.492.2910G} in the observational selected periodic quasar sample also supports that most of the sources in this sample, as SMBBH candidates, may be just false positives.

The negative search results for SED deficits among observed periodical variable AGNs also underscores the difficulty in sub-pc SMBBH detection \citet{2020MNRAS.492.2910G}. The single color-based criterion cannot effectively validate binary systems in the candidates selected via periodical light variations, if there exists any. This is one of the reasons that we identify the deficit feature in the present work by quantifying the change of the spectral power-law index. We note, however, evidence for periodic variability has been found for Mrk 231 (a quasar that exhibits SED-deficit signature) in both ultraviolet (UV) \citep{2018MNRAS.480.5504Y} and optical bands \citep{2025ApJ...987..106C}, indicating that there may be indeed a number of such systems with both SED-deficit and periodic variation signatures as suggested by the population study in the present paper.  

Our results are highly sensitive to the bands and limiting magnitudes of the adopted telescopes. In the above analysis, we adopt the CSST survey to select triple-disk SMBBH systems via SED-deficits, taking into account the simulated false positive rate. Here we further explore an alternative case by using the infrared bands as those of James Webb Space Telescope (JWST) NIRCam (from $0.70\mu$m to $2.77\mu$m)\footnote{\url{https://jwst-docs.stsci.edu/jwst-near-infrared-camera}} to select deficit SMBBHs. We find that $193$ SMBBHs based on the BBH-nonevol model across the entire sky that can be selected out, which is about $5$ times more than that obtained by using the CSST bands and limiting magnitudes. A similar proportional increase is found under the BBH-zevol model, where the number of candidates rises to $969$. Moreover, most of the SMBBHs detectable by JWST lie at relatively high redshifts, peaking around $z_{\mathrm{peak}}\simeq 3$, where their SED deficits are redshifted into JWST’s detection range. In fact, the triple-disk-SMBBH scenario has already been proposed as an explanation to the ``V-shaped" spectra of the so-called ``little red dots" detected by JWST \citep[][]{2025arXiv250505322I}. These results suggest that if future wide-field surveys achieve sensitivities comparable to JWST, thousands of such SMBBH systems could potentially be discovered.

\section*{Acknowledgements}

This work is partly supported by the National SKA Program of China and the National Key Program for Science and Technology Research and Development (grant no. 2020YFC2201400, 2020SKA0120101, 2022YFC2205201), the National Astronomical Observatory of China (grant no. E4TG660101), the Strategic Priority Program of the Chinese Academy of Sciences (grant no. XDB0550300), and the China Manned Space Program with grant no. CMS-CSST-2025-A07. This work was performed in part at the Aspen Center for Physics, which is supported by National Science Foundation grant PHY-2210452.

\section*{DATA AVAILABILITY}

The data underlying this article will be shared on reasonable request to the corresponding author.




\bibliographystyle{mnras}
\bibliography{BBH} 

@ARTICLE{2023MNRAS.522.1895C,
       author = {{Chen}, Nianyi and {Di Matteo}, Tiziana and {Ni}, Yueying and {Tremmel}, Michael and {DeGraf}, Colin and {Shen}, Yue and {Holgado}, A. Miguel and {Bird}, Simeon and {Croft}, Rupert and {Feng}, Yu},
        title = "{Properties and evolution of dual and offset AGN in the ASTRID simulation at z   2}",
      journal = {\mnras},
         year = 2023,
        month = jun,
       volume = {522},
       number = {2},
        pages = {1895-1913},
          doi = {10.1093/mnras/stad834},
archivePrefix = {arXiv},
       eprint = {2208.04970},
 primaryClass = {astro-ph.GA},
       adsurl = {https://ui.adsabs.harvard.edu/abs/2023MNRAS.522.1895C}
}

@ARTICLE{1989ApJ...344..115C,
       author = {{Chen}, Kaiyou and {Halpern}, Jules P.},
        title = "{Structure of Line-emitting Accretion Disks in Active Galactic Nuclei: ARP 102B}",
      journal = {\apj},
         year = 1989,
        month = sep,
       volume = {344},
        pages = {115},
          doi = {10.1086/167782},
       adsurl = {https://ui.adsabs.harvard.edu/abs/1989ApJ...344..115C}
}

@ARTICLE{2025MNRAS.536.1489M,
       author = {{Miles}, Matthew T. and {Shannon}, Ryan M. and {Reardon}, Daniel J. and {Bailes}, Matthew and {Champion}, David J. and {Geyer}, Marisa and {Gitika}, Pratyasha and {Grunthal}, Kathrin and {Keith}, Michael J. and {Kramer}, Michael and {Kulkarni}, Atharva D. and {Nathan}, Rowina S. and {Parthasarathy}, Aditya and {Singha}, Jaikhomba and {Theureau}, Gilles and {Thrane}, Eric and {Abbate}, Federico and {Buchner}, Sarah and {Cameron}, Andrew D. and {Camilo}, Fernando and {Moreschi}, Beatrice E. and {Shaifullah}, Golam and {Shamohammadi}, Mohsen and {Possenti}, Andrea and {Krishnan}, Vivek Venkatraman},
        title = "{The MeerKAT Pulsar Timing Array: the first search for gravitational waves with the MeerKAT radio telescope}",
      journal = {\mnras},
         year = 2025,
        month = jan,
       volume = {536},
       number = {2},
        pages = {1489-1500},
          doi = {10.1093/mnras/stae2571},
archivePrefix = {arXiv},
       eprint = {2412.01153},
 primaryClass = {astro-ph.HE},
       adsurl = {https://ui.adsabs.harvard.edu/abs/2025MNRAS.536.1489M}
}

@ARTICLE{1964PhRv..136.1224P,
       author = {{Peters}, P.~C.},
        title = "{Gravitational Radiation and the Motion of Two Point Masses}",
      journal = {Physical Review},
         year = 1964,
        month = nov,
       volume = {136},
       number = {4B},
        pages = {1224-1232},
          doi = {10.1103/PhysRev.136.B1224},
       adsurl = {https://ui.adsabs.harvard.edu/abs/1964PhRv..136.1224P}
}

@ARTICLE{2023ApJ...955..132C,
       author = {{Chen}, Yunfeng and {Yu}, Qingjuan and {Lu}, Youjun},
        title = "{Pulsar Timing Array Detections of Supermassive Binary Black Holes: Implications from the Detected Common Process Signal and Beyond}",
      journal = {\apj},
         year = 2023,
        month = oct,
       volume = {955},
       number = {2},
          eid = {132},
        pages = {132},
          doi = {10.3847/1538-4357/ace59f},
archivePrefix = {arXiv},
       eprint = {2306.10997},
 primaryClass = {astro-ph.HE},
       adsurl = {https://ui.adsabs.harvard.edu/abs/2023ApJ...955..132C}
}

@ARTICLE{2019MNRAS.485.1579K,
       author = {{Kelley}, Luke Zoltan and {Haiman}, Zolt{\'a}n and {Sesana}, Alberto and {Hernquist}, Lars},
        title = "{Massive BH binaries as periodically variable AGN}",
      journal = {\mnras},
         year = 2019,
        month = may,
       volume = {485},
       number = {2},
        pages = {1579-1594},
          doi = {10.1093/mnras/stz150},
archivePrefix = {arXiv},
       eprint = {1809.02138},
 primaryClass = {astro-ph.HE},
       adsurl = {https://ui.adsabs.harvard.edu/abs/2019MNRAS.485.1579K}
}

@ARTICLE{2023arXiv231016896D,
       author = {{D'Orazio}, Daniel J. and {Charisi}, Maria},
        title = "{Observational Signatures of Supermassive Black Hole Binaries}",
      journal = {arXiv e-prints},
         year = 2023,
        month = oct,
          eid = {arXiv:2310.16896},
        pages = {arXiv:2310.16896},
          doi = {10.48550/arXiv.2310.16896},
archivePrefix = {arXiv},
       eprint = {2310.16896},
 primaryClass = {astro-ph.HE},
       adsurl = {https://ui.adsabs.harvard.edu/abs/2023arXiv231016896D}
}

@ARTICLE{2023ApJ...952L..37A,
       author = {{Agazie}, Gabriella and {Anumarlapudi}, Akash and {Archibald}, Anne M. and {Baker}, Paul T. and {B{\'e}csy}, Bence and {Blecha}, Laura and {Bonilla}, Alexander and {Brazier}, Adam and {Brook}, Paul R. and {Burke-Spolaor}, Sarah and {Burnette}, Rand and {Case}, Robin and {Casey-Clyde}, J. Andrew and {Charisi}, Maria and {Chatterjee}, Shami and {Chatziioannou}, Katerina and {Cheeseboro}, Belinda D. and {Chen}, Siyuan and {Cohen}, Tyler and {Cordes}, James M. and {Cornish}, Neil J. and {Crawford}, Fronefield and {Cromartie}, H. Thankful and {Crowter}, Kathryn and {Cutler}, Curt J. and {D'Orazio}, Daniel J. and {Decesar}, Megan E. and {Degan}, Dallas and {Demorest}, Paul B. and {Deng}, Heling and {Dolch}, Timothy and {Drachler}, Brendan and {Ferrara}, Elizabeth C. and {Fiore}, William and {Fonseca}, Emmanuel and {Freedman}, Gabriel E. and {Gardiner}, Emiko and {Garver-Daniels}, Nate and {Gentile}, Peter A. and {Gersbach}, Kyle A. and {Glaser}, Joseph and {Good}, Deborah C. and {G{\"u}ltekin}, Kayhan and {Hazboun}, Jeffrey S. and {Hourihane}, Sophie and {Islo}, Kristina and {Jennings}, Ross J. and {Johnson}, Aaron and {Jones}, Megan L. and {Kaiser}, Andrew R. and {Kaplan}, David L. and {Kelley}, Luke Zoltan and {Kerr}, Matthew and {Key}, Joey S. and {Laal}, Nima and {Lam}, Michael T. and {Lamb}, William G. and {Lazio}, T. Joseph W. and {Lewandowska}, Natalia and {Littenberg}, Tyson B. and {Liu}, Tingting and {Luo}, Jing and {Lynch}, Ryan S. and {Ma}, Chung-Pei and {Madison}, Dustin R. and {McEwen}, Alexander and {McKee}, James W. and {McLaughlin}, Maura A. and {McMann}, Natasha and {Meyers}, Bradley W. and {Meyers}, Patrick M. and {Mingarelli}, Chiara M.~F. and {Mitridate}, Andrea and {Natarajan}, Priyamvada and {Ng}, Cherry and {Nice}, David J. and {Ocker}, Stella Koch and {Olum}, Ken D. and {Pennucci}, Timothy T. and {Perera}, Benetge B.~P. and {Petrov}, Polina and {Pol}, Nihan S. and {Radovan}, Henri A. and {Ransom}, Scott M. and {Ray}, Paul S. and {Romano}, Joseph D. and {Runnoe}, Jessie C. and {Sardesai}, Shashwat C. and {Schmiedekamp}, Ann and {Schmiedekamp}, Carl and {Schmitz}, Kai and {Schult}, Levi and {Shapiro-Albert}, Brent J. and {Siemens}, Xavier and {Simon}, Joseph and {Siwek}, Magdalena S. and {Stairs}, Ingrid H. and {Stinebring}, Daniel R. and {Stovall}, Kevin and {Sun}, Jerry P. and {Susobhanan}, Abhimanyu and {Swiggum}, Joseph K. and {Taylor}, Jacob and {Taylor}, Stephen R. and {Turner}, Jacob E. and {Unal}, Caner and {Vallisneri}, Michele and {Vigeland}, Sarah J. and {Wachter}, Jeremy M. and {Wahl}, Haley M. and {Wang}, Qiaohong and {Witt}, Caitlin A. and {Wright}, David and {Young}, Olivia and {Nanograv Collaboration}},
        title = "{The NANOGrav 15 yr Data Set: Constraints on Supermassive Black Hole Binaries from the Gravitational-wave Background}",
      journal = {\apjl},
         year = 2023,
        month = aug,
       volume = {952},
       number = {2},
          eid = {L37},
        pages = {L37},
          doi = {10.3847/2041-8213/ace18b},
archivePrefix = {arXiv},
       eprint = {2306.16220},
 primaryClass = {astro-ph.HE},
       adsurl = {https://ui.adsabs.harvard.edu/abs/2023ApJ...952L..37A}
}

@ARTICLE{2023ApJ...951L...8A,
       author = {{Agazie}, Gabriella and {Anumarlapudi}, Akash and {Archibald}, Anne M. and {Arzoumanian}, Zaven and {Baker}, Paul T. and {B{\'e}csy}, Bence and {Blecha}, Laura and {Brazier}, Adam and {Brook}, Paul R. and {Burke-Spolaor}, Sarah and {Burnette}, Rand and {Case}, Robin and {Charisi}, Maria and {Chatterjee}, Shami and {Chatziioannou}, Katerina and {Cheeseboro}, Belinda D. and {Chen}, Siyuan and {Cohen}, Tyler and {Cordes}, James M. and {Cornish}, Neil J. and {Crawford}, Fronefield and {Cromartie}, H. Thankful and {Crowter}, Kathryn and {Cutler}, Curt J. and {Decesar}, Megan E. and {Degan}, Dallas and {Demorest}, Paul B. and {Deng}, Heling and {Dolch}, Timothy and {Drachler}, Brendan and {Ellis}, Justin A. and {Ferrara}, Elizabeth C. and {Fiore}, William and {Fonseca}, Emmanuel and {Freedman}, Gabriel E. and {Garver-Daniels}, Nate and {Gentile}, Peter A. and {Gersbach}, Kyle A. and {Glaser}, Joseph and {Good}, Deborah C. and {G{\"u}ltekin}, Kayhan and {Hazboun}, Jeffrey S. and {Hourihane}, Sophie and {Islo}, Kristina and {Jennings}, Ross J. and {Johnson}, Aaron D. and {Jones}, Megan L. and {Kaiser}, Andrew R. and {Kaplan}, David L. and {Kelley}, Luke Zoltan and {Kerr}, Matthew and {Key}, Joey S. and {Klein}, Tonia C. and {Laal}, Nima and {Lam}, Michael T. and {Lamb}, William G. and {Lazio}, T. Joseph W. and {Lewandowska}, Natalia and {Littenberg}, Tyson B. and {Liu}, Tingting and {Lommen}, Andrea and {Lorimer}, Duncan R. and {Luo}, Jing and {Lynch}, Ryan S. and {Ma}, Chung-Pei and {Madison}, Dustin R. and {Mattson}, Margaret A. and {McEwen}, Alexander and {McKee}, James W. and {McLaughlin}, Maura A. and {McMann}, Natasha and {Meyers}, Bradley W. and {Meyers}, Patrick M. and {Mingarelli}, Chiara M.~F. and {Mitridate}, Andrea and {Natarajan}, Priyamvada and {Ng}, Cherry and {Nice}, David J. and {Ocker}, Stella Koch and {Olum}, Ken D. and {Pennucci}, Timothy T. and {Perera}, Benetge B.~P. and {Petrov}, Polina and {Pol}, Nihan S. and {Radovan}, Henri A. and {Ransom}, Scott M. and {Ray}, Paul S. and {Romano}, Joseph D. and {Sardesai}, Shashwat C. and {Schmiedekamp}, Ann and {Schmiedekamp}, Carl and {Schmitz}, Kai and {Schult}, Levi and {Shapiro-Albert}, Brent J. and {Siemens}, Xavier and {Simon}, Joseph and {Siwek}, Magdalena S. and {Stairs}, Ingrid H. and {Stinebring}, Daniel R. and {Stovall}, Kevin and {Sun}, Jerry P. and {Susobhanan}, Abhimanyu and {Swiggum}, Joseph K. and {Taylor}, Jacob and {Taylor}, Stephen R. and {Turner}, Jacob E. and {Unal}, Caner and {Vallisneri}, Michele and {van Haasteren}, Rutger and {Vigeland}, Sarah J. and {Wahl}, Haley M. and {Wang}, Qiaohong and {Witt}, Caitlin A. and {Young}, Olivia and {Nanograv Collaboration}},
        title = "{The NANOGrav 15 yr Data Set: Evidence for a Gravitational-wave Background}",
      journal = {\apjl},
         year = 2023,
        month = jul,
       volume = {951},
       number = {1},
          eid = {L8},
        pages = {L8},
          doi = {10.3847/2041-8213/acdac6},
archivePrefix = {arXiv},
       eprint = {2306.16213},
 primaryClass = {astro-ph.HE},
       adsurl = {https://ui.adsabs.harvard.edu/abs/2023ApJ...951L...8A}
}

@ARTICLE{2017MNRAS.464.3131K,
       author = {{Kelley}, Luke Zoltan and {Blecha}, Laura and {Hernquist}, Lars},
        title = "{Massive black hole binary mergers in dynamical galactic environments}",
      journal = {\mnras},
         year = 2017,
        month = jan,
       volume = {464},
       number = {3},
        pages = {3131-3157},
          doi = {10.1093/mnras/stw2452},
archivePrefix = {arXiv},
       eprint = {1606.01900},
 primaryClass = {astro-ph.HE},
       adsurl = {https://ui.adsabs.harvard.edu/abs/2017MNRAS.464.3131K}
}

@ARTICLE{2025ApJ...991...71C,
       author = {{Chan}, Chi-Ho and {Tiwari}, Vishal and {Bogdanovi{\'c}}, Tamara and {Jiang}, Yan-Fei and {Davis}, Shane W.},
        title = "{Radiative Magnetohydrodynamics Simulation of Minidisks in Equal-mass Massive Black Hole Binaries}",
      journal = {\apj},
         year = 2025,
        month = sep,
       volume = {991},
       number = {1},
          eid = {71},
        pages = {71},
          doi = {10.3847/1538-4357/adf4c9},
archivePrefix = {arXiv},
       eprint = {2505.02919},
 primaryClass = {astro-ph.HE},
       adsurl = {https://ui.adsabs.harvard.edu/abs/2025ApJ...991...71C}
}

@ARTICLE{2025A&A...699A..55F,
       author = {{Foustoul}, V. and {Webb}, N.~A. and {Mignon-Risse}, R. and {Kammoun}, E. and {Volonteri}, M. and {Dong-P{\'a}ez}, C.~A.},
        title = "{A catalogue of candidate milliparsec-separation massive black hole binaries from long-term optical photometric monitoring}",
      journal = {\aap},
         year = 2025,
        month = jul,
       volume = {699},
          eid = {A55},
        pages = {A55},
          doi = {10.1051/0004-6361/202452818},
archivePrefix = {arXiv},
       eprint = {2505.06656},
 primaryClass = {astro-ph.GA},
       adsurl = {https://ui.adsabs.harvard.edu/abs/2025A&A...699A..55F}
}

@ARTICLE{1980Natur.287..307B,
       author = {{Begelman}, M.~C. and {Blandford}, R.~D. and {Rees}, M.~J.},
        title = "{Massive black hole binaries in active galactic nuclei}",
      journal = {\nat},
         year = 1980,
        month = sep,
       volume = {287},
       number = {5780},
        pages = {307-309},
          doi = {10.1038/287307a0},
       adsurl = {https://ui.adsabs.harvard.edu/abs/1980Natur.287..307B}
}

@ARTICLE{2020ApJ...897...86C,
       author = {{Chen}, Yunfeng and {Yu}, Qingjuan and {Lu}, Youjun},
        title = "{Dynamical Evolution of Cosmic Supermassive Binary Black Holes and Their Gravitational-wave Radiation}",
      journal = {\apj},
         year = 2020,
        month = jul,
       volume = {897},
       number = {1},
          eid = {86},
        pages = {86},
          doi = {10.3847/1538-4357/ab9594},
archivePrefix = {arXiv},
       eprint = {2005.10818},
 primaryClass = {astro-ph.HE},
       adsurl = {https://ui.adsabs.harvard.edu/abs/2020ApJ...897...86C}
}

@ARTICLE{2020ApJ...889...79Y,
       author = {{Yan}, Changshuo and {Zhao}, Wen and {Lu}, Youjun},
        title = "{On Using Inspiraling Supermassive Binary Black Holes in the PTA Frequency Band as Standard Sirens to Constrain Dark Energy}",
      journal = {\apj},
         year = 2020,
        month = feb,
       volume = {889},
       number = {2},
          eid = {79},
        pages = {79},
          doi = {10.3847/1538-4357/ab60a6},
archivePrefix = {arXiv},
       eprint = {1912.04103},
 primaryClass = {astro-ph.GA},
       adsurl = {https://ui.adsabs.harvard.edu/abs/2020ApJ...889...79Y}
}

@ARTICLE{2013ARA&A..51..511K,
       author = {{Kormendy}, John and {Ho}, Luis C.},
        title = "{Coevolution (Or Not) of Supermassive Black Holes and Host Galaxies}",
      journal = {\araa},
         year = 2013,
        month = aug,
       volume = {51},
       number = {1},
        pages = {511-653},
          doi = {10.1146/annurev-astro-082708-101811},
archivePrefix = {arXiv},
       eprint = {1304.7762},
 primaryClass = {astro-ph.CO},
       adsurl = {https://ui.adsabs.harvard.edu/abs/2013ARA&A..51..511K}
}

@ARTICLE{2017MNRAS.471.3098L,
       author = {{Lopes}, Amanda R. and {Gruppioni}, C. and {Ribeiro}, M.~B. and {Pozzetti}, L. and {February}, S. and {Ilbert}, O. and {Pozzi}, F.},
        title = "{Effect of different cosmologies on the galaxy stellar mass function}",
      journal = {\mnras},
         year = 2017,
        month = nov,
       volume = {471},
       number = {3},
        pages = {3098-3111},
          doi = {10.1093/mnras/stx1799},
       adsurl = {https://ui.adsabs.harvard.edu/abs/2017MNRAS.471.3098L}
}

@ARTICLE{2015ApJ...809..117Y,
       author = {{Yan}, Chang-Shuo and {Lu}, Youjun and {Dai}, Xinyu and {Yu}, Qingjuan},
        title = "{A Probable Milli-parsec Supermassive Binary Black Hole in the Nearest Quasar Mrk 231}",
      journal = {\apj},
         year = 2015,
        month = aug,
       volume = {809},
       number = {2},
          eid = {117},
        pages = {117},
          doi = {10.1088/0004-637X/809/2/117},
archivePrefix = {arXiv},
       eprint = {1508.06292},
 primaryClass = {astro-ph.HE},
       adsurl = {https://ui.adsabs.harvard.edu/abs/2015ApJ...809..117Y}
}

@ARTICLE{2002MNRAS.331..935Y,
       author = {{Yu}, Qingjuan},
        title = "{Evolution of massive binary black holes}",
      journal = {\mnras},
         year = 2002,
        month = apr,
       volume = {331},
       number = {4},
        pages = {935-958},
          doi = {10.1046/j.1365-8711.2002.05242.x},
archivePrefix = {arXiv},
       eprint = {astro-ph/0109530},
 primaryClass = {astro-ph},
       adsurl = {https://ui.adsabs.harvard.edu/abs/2002MNRAS.331..935Y}
}

@ARTICLE{2008ApJ...689..732Y,
       author = {{Yu}, Qingjuan and {Lu}, Youjun},
        title = "{Toward Precise Constraints on the Growth of Massive Black Holes}",
      journal = {\apj},
         year = 2008,
        month = dec,
       volume = {689},
       number = {2},
        pages = {732-754},
          doi = {10.1086/592770},
archivePrefix = {arXiv},
       eprint = {0808.3777},
 primaryClass = {astro-ph},
       adsurl = {https://ui.adsabs.harvard.edu/abs/2008ApJ...689..732Y}
}

@ARTICLE{2013MNRAS.436.2997D,
       author = {{D'Orazio}, Daniel J. and {Haiman}, Zolt{\'a}n and {MacFadyen}, Andrew},
        title = "{Accretion into the central cavity of a circumbinary disc}",
      journal = {\mnras},
         year = 2013,
        month = dec,
       volume = {436},
       number = {4},
        pages = {2997-3020},
          doi = {10.1093/mnras/stt1787},
archivePrefix = {arXiv},
       eprint = {1210.0536},
 primaryClass = {astro-ph.GA},
       adsurl = {https://ui.adsabs.harvard.edu/abs/2013MNRAS.436.2997D}
}

@ARTICLE{1983ApJ...268..368E,
       author = {{Eggleton}, P.~P.},
        title = "{Aproximations to the radii of Roche lobes.}",
      journal = {\apj},
         year = 1983,
        month = may,
       volume = {268},
        pages = {368-369},
          doi = {10.1086/160960},
       adsurl = {https://ui.adsabs.harvard.edu/abs/1983ApJ...268..368E}
}

@ARTICLE{2014ApJ...783..134F,
       author = {{Farris}, Brian D. and {Duffell}, Paul and {MacFadyen}, Andrew I. and {Haiman}, Zoltan},
        title = "{Binary Black Hole Accretion from a Circumbinary Disk: Gas Dynamics inside the Central Cavity}",
      journal = {\apj},
         year = 2014,
        month = mar,
       volume = {783},
       number = {2},
          eid = {134},
        pages = {134},
          doi = {10.1088/0004-637X/783/2/134},
archivePrefix = {arXiv},
       eprint = {1310.0492},
 primaryClass = {astro-ph.HE},
       adsurl = {https://ui.adsabs.harvard.edu/abs/2014ApJ...783..134F}
}

@ARTICLE{2015MNRAS.446L..36F,
       author = {{Farris}, B.~D. and {Duffell}, P. and {MacFadyen}, A.~I. and {Haiman}, Z.},
        title = "{Characteristic signatures in the thermal emission from accreting binary black holes.}",
      journal = {\mnras},
         year = 2015,
        month = jan,
       volume = {446},
        pages = {L36-L40},
          doi = {10.1093/mnrasl/slu160},
archivePrefix = {arXiv},
       eprint = {1406.0007},
 primaryClass = {astro-ph.HE},
       adsurl = {https://ui.adsabs.harvard.edu/abs/2015MNRAS.446L..36F}
}

@ARTICLE{2012ApJ...761...90G,
       author = {{G{\"u}ltekin}, Kayhan and {Miller}, Jon M.},
        title = "{Observable Consequences of Merger-driven Gaps and Holes in Black Hole Accretion Disks}",
      journal = {\apj},
         year = 2012,
        month = dec,
       volume = {761},
       number = {2},
          eid = {90},
        pages = {90},
          doi = {10.1088/0004-637X/761/2/90},
archivePrefix = {arXiv},
       eprint = {1207.0296},
 primaryClass = {astro-ph.HE},
       adsurl = {https://ui.adsabs.harvard.edu/abs/2012ApJ...761...90G}
}

@ARTICLE{1996ApJ...467L..77A,
       author = {{Artymowicz}, Pawel and {Lubow}, Stephen H.},
        title = "{Mass Flow through Gaps in Circumbinary Disks}",
      journal = {\apjl},
         year = 1996,
        month = aug,
       volume = {467},
        pages = {L77},
          doi = {10.1086/310200},
       adsurl = {https://ui.adsabs.harvard.edu/abs/1996ApJ...467L..77A}
}

@ARTICLE{2002ApJ...567L...9A,
       author = {{Armitage}, Philip J. and {Natarajan}, Priyamvada},
        title = "{Accretion during the Merger of Supermassive Black Holes}",
      journal = {\apjl},
         year = 2002,
        month = mar,
       volume = {567},
       number = {1},
        pages = {L9-L12},
          doi = {10.1086/339770},
archivePrefix = {arXiv},
       eprint = {astro-ph/0201318},
 primaryClass = {astro-ph},
       adsurl = {https://ui.adsabs.harvard.edu/abs/2002ApJ...567L...9A}
}

@ARTICLE{2008ApJ...672...83M,
       author = {{MacFadyen}, Andrew I. and {Milosavljevi{\'c}}, Milo{\v{s}}},
        title = "{An Eccentric Circumbinary Accretion Disk and the Detection of Binary Massive Black Holes}",
      journal = {\apj},
         year = 2008,
        month = jan,
       volume = {672},
       number = {1},
        pages = {83-93},
          doi = {10.1086/523869},
archivePrefix = {arXiv},
       eprint = {astro-ph/0607467},
 primaryClass = {astro-ph},
       adsurl = {https://ui.adsabs.harvard.edu/abs/2008ApJ...672...83M}
}

@ARTICLE{2023ARA&A..61..517L,
       author = {{Lai}, Dong and {Mu{\~n}oz}, Diego J.},
        title = "{Circumbinary Accretion: From Binary Stars to Massive Binary Black Holes}",
      journal = {\araa},
         year = 2023,
        month = aug,
       volume = {61},
        pages = {517-560},
          doi = {10.1146/annurev-astro-052622-022933},
archivePrefix = {arXiv},
       eprint = {2211.00028},
 primaryClass = {astro-ph.HE},
       adsurl = {https://ui.adsabs.harvard.edu/abs/2023ARA&A..61..517L}
}

@ARTICLE{2009ApJ...700.1952H,
       author = {{Haiman}, Zolt{\'a}n and {Kocsis}, Bence and {Menou}, Kristen},
        title = "{The Population of Viscosity- and Gravitational Wave-driven Supermassive Black Hole Binaries Among Luminous Active Galactic Nuclei}",
      journal = {\apj},
         year = 2009,
        month = aug,
       volume = {700},
       number = {2},
        pages = {1952-1969},
          doi = {10.1088/0004-637X/700/2/1952},
archivePrefix = {arXiv},
       eprint = {0904.1383},
 primaryClass = {astro-ph.CO},
       adsurl = {https://ui.adsabs.harvard.edu/abs/2009ApJ...700.1952H}
}

@ARTICLE{1973A&A....24..337S,
       author = {{Shakura}, N.~I. and {Sunyaev}, R.~A.},
        title = "{Black holes in binary systems. Observational appearance.}",
      journal = {\aap},
         year = 1973,
        month = jan,
       volume = {24},
        pages = {337-355},
       adsurl = {https://ui.adsabs.harvard.edu/abs/1973A&A....24..337S}
}

@ARTICLE{2020ApJ...901...25D,
       author = {{Duffell}, Paul C. and {D'Orazio}, Daniel and {Derdzinski}, Andrea and {Haiman}, Zoltan and {MacFadyen}, Andrew and {Rosen}, Anna L. and {Zrake}, Jonathan},
        title = "{Circumbinary Disks: Accretion and Torque as a Function of Mass Ratio and Disk Viscosity}",
      journal = {\apj},
         year = 2020,
        month = sep,
       volume = {901},
       number = {1},
          eid = {25},
        pages = {25},
          doi = {10.3847/1538-4357/abab95},
archivePrefix = {arXiv},
       eprint = {1911.05506},
 primaryClass = {astro-ph.SR},
       adsurl = {https://ui.adsabs.harvard.edu/abs/2020ApJ...901...25D}
}

@ARTICLE{2023MNRAS.519.2083I,
       author = {{Izquierdo-Villalba}, David and {Sesana}, Alberto and {Colpi}, Monica},
        title = "{Unveiling the hosts of parsec-scale massive black hole binaries: morphology and electromagnetic signatures}",
      journal = {\mnras},
         year = 2023,
        month = feb,
       volume = {519},
       number = {2},
        pages = {2083-2100},
          doi = {10.1093/mnras/stac3677},
archivePrefix = {arXiv},
       eprint = {2207.04064},
 primaryClass = {astro-ph.GA},
       adsurl = {https://ui.adsabs.harvard.edu/abs/2023MNRAS.519.2083I}
}

@ARTICLE{2021ApJ...918L..15B,
       author = {{Bortolas}, Elisa and {Franchini}, Alessia and {Bonetti}, Matteo and {Sesana}, Alberto},
        title = "{The Competing Effect of Gas and Stars in the Evolution of Massive Black Hole Binaries}",
      journal = {\apjl},
         year = 2021,
        month = sep,
       volume = {918},
       number = {1},
          eid = {L15},
        pages = {L15},
          doi = {10.3847/2041-8213/ac1c0c},
archivePrefix = {arXiv},
       eprint = {2108.13436},
 primaryClass = {astro-ph.HE},
       adsurl = {https://ui.adsabs.harvard.edu/abs/2021ApJ...918L..15B}
}

@ARTICLE{2015MNRAS.447.2772R,
       author = {{Ravi}, V. and {Wyithe}, J.~S.~B. and {Shannon}, R.~M. and {Hobbs}, G.},
        title = "{Prospects for gravitational-wave detection and supermassive black hole astrophysics with pulsar timing arrays}",
      journal = {\mnras},
         year = 2015,
        month = mar,
       volume = {447},
       number = {3},
        pages = {2772-2783},
          doi = {10.1093/mnras/stu2659},
archivePrefix = {arXiv},
       eprint = {1406.5297},
 primaryClass = {astro-ph.CO},
       adsurl = {https://ui.adsabs.harvard.edu/abs/2015MNRAS.447.2772R}
}

@ARTICLE{2015MNRAS.449...49R,
       author = {{Rodriguez-Gomez}, Vicente and {Genel}, Shy and {Vogelsberger}, Mark and {Sijacki}, Debora and {Pillepich}, Annalisa and {Sales}, Laura V. and {Torrey}, Paul and {Snyder}, Greg and {Nelson}, Dylan and {Springel}, Volker and {Ma}, Chung-Pei and {Hernquist}, Lars},
        title = "{The merger rate of galaxies in the Illustris simulation: a comparison with observations and semi-empirical models}",
      journal = {\mnras},
         year = 2015,
        month = may,
       volume = {449},
       number = {1},
        pages = {49-64},
          doi = {10.1093/mnras/stv264},
archivePrefix = {arXiv},
       eprint = {1502.01339},
 primaryClass = {astro-ph.GA},
       adsurl = {https://ui.adsabs.harvard.edu/abs/2015MNRAS.449...49R}
}

@ARTICLE{2022LRR....25....3B,
       author = {{Bogdanovi{\'c}}, Tamara and {Miller}, M. Coleman and {Blecha}, Laura},
        title = "{Electromagnetic counterparts to massive black-hole mergers}",
      journal = {Living Reviews in Relativity},
         year = 2022,
        month = dec,
       volume = {25},
       number = {1},
          eid = {3},
        pages = {3},
          doi = {10.1007/s41114-022-00037-8},
archivePrefix = {arXiv},
       eprint = {2109.03262},
 primaryClass = {astro-ph.HE},
       adsurl = {https://ui.adsabs.harvard.edu/abs/2022LRR....25....3B}
}

@ARTICLE{2012ApJ...745...83A,
       author = {{Antonini}, Fabio and {Merritt}, David},
        title = "{Dynamical Friction around Supermassive Black Holes}",
      journal = {\apj},
         year = 2012,
        month = jan,
       volume = {745},
       number = {1},
          eid = {83},
        pages = {83},
          doi = {10.1088/0004-637X/745/1/83},
archivePrefix = {arXiv},
       eprint = {1108.1163},
 primaryClass = {astro-ph.GA},
       adsurl = {https://ui.adsabs.harvard.edu/abs/2012ApJ...745...83A}
}

@ARTICLE{1996NewA....1...35Q,
       author = {{Quinlan}, Gerald D.},
        title = "{The dynamical evolution of massive black hole binaries I. Hardening in a fixed stellar background}",
      journal = {\na},
         year = 1996,
        month = jul,
       volume = {1},
       number = {1},
        pages = {35-56},
          doi = {10.1016/S1384-1076(96)00003-6},
archivePrefix = {arXiv},
       eprint = {astro-ph/9601092},
 primaryClass = {astro-ph},
       adsurl = {https://ui.adsabs.harvard.edu/abs/1996NewA....1...35Q}
}

@ARTICLE{2011ApJ...738...92Y,
       author = {{Yu}, Qingjuan and {Lu}, Youjun and {Mohayaee}, Roya and {Colin}, Jacques},
        title = "{The Low Frequency of Dual Active Galactic Nuclei versus the High Merger Rate of Galaxies: A Phenomenological Model}",
      journal = {\apj},
         year = 2011,
        month = sep,
       volume = {738},
       number = {1},
          eid = {92},
        pages = {92},
          doi = {10.1088/0004-637X/738/1/92},
archivePrefix = {arXiv},
       eprint = {1105.1963},
 primaryClass = {astro-ph.CO},
       adsurl = {https://ui.adsabs.harvard.edu/abs/2011ApJ...738...92Y}
}

@ARTICLE{2015MNRAS.452..575S,
       author = {{Sijacki}, Debora and {Vogelsberger}, Mark and {Genel}, Shy and {Springel}, Volker and {Torrey}, Paul and {Snyder}, Gregory F. and {Nelson}, Dylan and {Hernquist}, Lars},
        title = "{The Illustris simulation: the evolving population of black holes across cosmic time}",
      journal = {\mnras},
         year = 2015,
        month = sep,
       volume = {452},
       number = {1},
        pages = {575-596},
          doi = {10.1093/mnras/stv1340},
archivePrefix = {arXiv},
       eprint = {1408.6842},
 primaryClass = {astro-ph.GA},
       adsurl = {https://ui.adsabs.harvard.edu/abs/2015MNRAS.452..575S}
}

@ARTICLE{2023MNRAS.520.3476Q,
       author = {{Queiroz}, Carolina and {Abramo}, L. Raul and {Rodrigues}, Nat{\'a}lia V.~N. and {P{\'e}rez-R{\`a}fols}, Ignasi and {Mart{\'\i}nez-Solaeche}, Gin{\'e}s and {Hern{\'a}n-Caballero}, Antonio and {Hern{\'a}ndez-Monteagudo}, Carlos and {Lumbreras-Calle}, Alejandro and {Pieri}, Matthew M. and {Morrison}, Sean S. and {Bonoli}, Silvia and {Chaves-Montero}, Jon{\'a}s and {Chies-Santos}, Ana L. and {D{\'\i}az-Garc{\'\i}a}, L.~A. and {Fernandez-Soto}, Alberto and {Gonz{\'a}lez Delgado}, Rosa M. and {Alcaniz}, Jailson and {Ben{\'\i}tez}, Narciso and {Cenarro}, A. Javier and {Civera}, Tamara and {Dupke}, Renato A. and {Ederoclite}, Alessandro and {L{\'o}pez-Sanjuan}, Carlos and {Mar{\'\i}n-Franch}, Antonio and {Mendes de Oliveira}, Claudia and {Moles}, Mariano and {Muniesa}, David and {Sodr{\'e}}, Laerte and {Taylor}, Keith and {Varela}, Jes{\'u}s and {V{\'a}zquez Rami{\'o}}, H{\'e}ctor},
        title = "{The miniJPAS survey quasar selection - I. Mock catalogues for classification}",
      journal = {\mnras},
         year = 2023,
        month = apr,
       volume = {520},
       number = {3},
        pages = {3476-3493},
          doi = {10.1093/mnras/stac2962},
archivePrefix = {arXiv},
       eprint = {2202.00103},
 primaryClass = {astro-ph.GA},
       adsurl = {https://ui.adsabs.harvard.edu/abs/2023MNRAS.520.3476Q}
}

@ARTICLE{2023NatAs...7.1376Z,
       author = {{Zhuang}, Ming-Yang and {Ho}, Luis C.},
        title = "{Evolutionary paths of active galactic nuclei and their host galaxies}",
      journal = {Nature Astronomy},
         year = 2023,
        month = nov,
       volume = {7},
        pages = {1376-1389},
          doi = {10.1038/s41550-023-02051-4},
archivePrefix = {arXiv},
       eprint = {2308.08603},
 primaryClass = {astro-ph.GA},
       adsurl = {https://ui.adsabs.harvard.edu/abs/2023NatAs...7.1376Z}
}

@ARTICLE{2024arXiv240514843G,
       author = {{Guti{\'e}rrez}, Eduardo M. and {Combi}, Luciano and {Ryan}, Geoffrey},
        title = "{Accretion onto supermassive black hole binaries}",
      journal = {arXiv e-prints},
         year = 2024,
        month = may,
          eid = {arXiv:2405.14843},
        pages = {arXiv:2405.14843},
          doi = {10.48550/arXiv.2405.14843},
archivePrefix = {arXiv},
       eprint = {2405.14843},
 primaryClass = {astro-ph.HE},
       adsurl = {https://ui.adsabs.harvard.edu/abs/2024arXiv240514843G}
}

@ARTICLE{2024MNRAS.534.2609S,
       author = {{Siwek}, Magdalena and {Kelley}, Luke Zoltan and {Hernquist}, Lars},
        title = "{Signatures of circumbinary disc dynamics in multimessenger population studies of massive black hole binaries}",
      journal = {\mnras},
         year = 2024,
        month = nov,
       volume = {534},
       number = {3},
        pages = {2609-2620},
          doi = {10.1093/mnras/stae2251},
archivePrefix = {arXiv},
       eprint = {2403.08871},
 primaryClass = {astro-ph.HE},
       adsurl = {https://ui.adsabs.harvard.edu/abs/2024MNRAS.534.2609S}
}

@ARTICLE{2025ApJ...984..144T,
       author = {{Tiede}, Christopher and {Zrake}, Jonathan and {MacFadyen}, Andrew and {Haiman}, Zolt{\'a}n},
        title = "{Suppressed Accretion onto Massive Black Hole Binaries Surrounded by Thin Disks}",
      journal = {\apj},
         year = 2025,
        month = may,
       volume = {984},
       number = {2},
          eid = {144},
        pages = {144},
          doi = {10.3847/1538-4357/adc727},
archivePrefix = {arXiv},
       eprint = {2410.03830},
 primaryClass = {astro-ph.GA},
       adsurl = {https://ui.adsabs.harvard.edu/abs/2025ApJ...984..144T}
}

@ARTICLE{2020NatAs...4..282S,
       author = {{Shankar}, Francesco and {Allevato}, Viola and {Bernardi}, Mariangela and {Marsden}, Christopher and {Lapi}, Andrea and {Menci}, Nicola and {Grylls}, Philip J. and {Krumpe}, Mirko and {Zanisi}, Lorenzo and {Ricci}, Federica and {La Franca}, Fabio and {Baldi}, Ranieri D. and {Moreno}, Jorge and {Sheth}, Ravi K.},
        title = "{Constraining black hole-galaxy scaling relations and radiative efficiency from galaxy clustering}",
      journal = {Nature Astronomy},
         year = 2020,
        month = jan,
       volume = {4},
        pages = {282-291},
          doi = {10.1038/s41550-019-0949-y},
archivePrefix = {arXiv},
       eprint = {1910.10175},
 primaryClass = {astro-ph.GA},
       adsurl = {https://ui.adsabs.harvard.edu/abs/2020NatAs...4..282S}
}

@ARTICLE{2022ApJ...928..137G,
       author = {{Guti{\'e}rrez}, Eduardo M. and {Combi}, Luciano and {Noble}, Scott C. and {Campanelli}, Manuela and {Krolik}, Julian H. and {L{\'o}pez Armengol}, Federico and {Garc{\'\i}a}, Federico},
        title = "{Electromagnetic Signatures from Supermassive Binary Black Holes Approaching Merger}",
      journal = {\apj},
         year = 2022,
        month = apr,
       volume = {928},
       number = {2},
          eid = {137},
        pages = {137},
          doi = {10.3847/1538-4357/ac56de},
archivePrefix = {arXiv},
       eprint = {2112.09773},
 primaryClass = {astro-ph.HE},
       adsurl = {https://ui.adsabs.harvard.edu/abs/2022ApJ...928..137G}
}

@ARTICLE{2021MNRAS.506.2408X,
       author = {{Xin}, Chengcheng and {Haiman}, Zolt{\'a}n},
        title = "{Ultra-short-period massive black hole binary candidates in LSST as LISA 'verification binaries'}",
      journal = {\mnras},
         year = 2021,
        month = sep,
       volume = {506},
       number = {2},
        pages = {2408-2417},
          doi = {10.1093/mnras/stab1856},
archivePrefix = {arXiv},
       eprint = {2105.00005},
 primaryClass = {astro-ph.HE},
       adsurl = {https://ui.adsabs.harvard.edu/abs/2021MNRAS.506.2408X}
}

@ARTICLE{2024MNRAS.527.4690L,
       author = {{Li}, Junyao and {Silverman}, John D. and {Merloni}, Andrea and {Salvato}, Mara and {Buchner}, Johannes and {Goulding}, Andy and {Liu}, Teng and {Arcodia}, Riccardo and {Comparat}, Johan and {Ding}, Xuheng and {Ichikawa}, Kohei and {Imanishi}, Masatoshi and {Kawaguchi}, Toshihiro and {Kawinwanichakij}, Lalitwadee and {Toba}, Yoshiki},
        title = "{The eROSITA final equatorial-depth survey (eFEDS): host-galaxy demographics of X-ray AGNs with Subaru Hyper Suprime-Cam}",
      journal = {\mnras},
         year = 2024,
        month = jan,
       volume = {527},
       number = {3},
        pages = {4690-4704},
          doi = {10.1093/mnras/stad3438},
archivePrefix = {arXiv},
       eprint = {2302.12438},
 primaryClass = {astro-ph.GA},
       adsurl = {https://ui.adsabs.harvard.edu/abs/2024MNRAS.527.4690L}
}

@ARTICLE{2019MNRAS.488.3143B,
       author = {{Behroozi}, Peter and {Wechsler}, Risa H. and {Hearin}, Andrew P. and {Conroy}, Charlie},
        title = "{UNIVERSEMACHINE: The correlation between galaxy growth and dark matter halo assembly from z = 0-10}",
      journal = {\mnras},
         year = 2019,
        month = sep,
       volume = {488},
       number = {3},
        pages = {3143-3194},
          doi = {10.1093/mnras/stz1182},
archivePrefix = {arXiv},
       eprint = {1806.07893},
 primaryClass = {astro-ph.GA},
       adsurl = {https://ui.adsabs.harvard.edu/abs/2019MNRAS.488.3143B}
}

@ARTICLE{2025ApJ...978...86L,
       author = {{Luo}, Di and {Jiang}, Ning and {Liu}, Xin},
        title = "{A Systematic Search for Candidate Supermassive Black Hole Binaries Using Periodic Mid-infrared Light Curves of Active Galactic Nuclei}",
      journal = {\apj},
         year = 2025,
        month = jan,
       volume = {978},
       number = {1},
          eid = {86},
        pages = {86},
          doi = {10.3847/1538-4357/ad9245},
archivePrefix = {arXiv},
       eprint = {2411.06902},
 primaryClass = {astro-ph.HE},
       adsurl = {https://ui.adsabs.harvard.edu/abs/2025ApJ...978...86L}
}

@ARTICLE{2023MNRAS.519.5031S,
       author = {{Steinle}, Nathan and {Gerosa}, Davide},
        title = "{The Bardeen-Petterson effect, disc breaking, and the spin orientations of supermassive black hole binaries}",
      journal = {\mnras},
         year = 2023,
        month = mar,
       volume = {519},
       number = {4},
        pages = {5031-5042},
          doi = {10.1093/mnras/stac3821},
archivePrefix = {arXiv},
       eprint = {2211.00044},
 primaryClass = {astro-ph.HE},
       adsurl = {https://ui.adsabs.harvard.edu/abs/2023MNRAS.519.5031S}
}

@ARTICLE{2019MNRAS.487.4985B,
       author = {{Biava}, Nadia and {Colpi}, Monica and {Capelo}, Pedro R. and {Bonetti}, Matteo and {Volonteri}, Marta and {Tamfal}, Tomas and {Mayer}, Lucio and {Sesana}, Alberto},
        title = "{The lifetime of binary black holes in S{\'e}rsic galaxy models}",
      journal = {\mnras},
         year = 2019,
        month = aug,
       volume = {487},
       number = {4},
        pages = {4985-4994},
          doi = {10.1093/mnras/stz1614},
archivePrefix = {arXiv},
       eprint = {1903.05682},
 primaryClass = {astro-ph.GA},
       adsurl = {https://ui.adsabs.harvard.edu/abs/2019MNRAS.487.4985B}
}

@ARTICLE{2010A&A...523A..13P,
       author = {{Pozzetti}, L. and {Bolzonella}, M. and {Zucca}, E. and {Zamorani}, G. and {Lilly}, S. and {Renzini}, A. and {Moresco}, M. and {Mignoli}, M. and {Cassata}, P. and {Tasca}, L. and {Lamareille}, F. and {Maier}, C. and {Meneux}, B. and {Halliday}, C. and {Oesch}, P. and {Vergani}, D. and {Caputi}, K. and {Kova{\v{c}}}, K. and {Cimatti}, A. and {Cucciati}, O. and {Iovino}, A. and {Peng}, Y. and {Carollo}, M. and {Contini}, T. and {Kneib}, J. -P. and {Le F{\'e}vre}, O. and {Mainieri}, V. and {Scodeggio}, M. and {Bardelli}, S. and {Bongiorno}, A. and {Coppa}, G. and {de la Torre}, S. and {de Ravel}, L. and {Franzetti}, P. and {Garilli}, B. and {Kampczyk}, P. and {Knobel}, C. and {Le Borgne}, J. -F. and {Le Brun}, V. and {Pell{\`o}}, R. and {Perez Montero}, E. and {Ricciardelli}, E. and {Silverman}, J.~D. and {Tanaka}, M. and {Tresse}, L. and {Abbas}, U. and {Bottini}, D. and {Cappi}, A. and {Guzzo}, L. and {Koekemoer}, A.~M. and {Leauthaud}, A. and {Maccagni}, D. and {Marinoni}, C. and {McCracken}, H.~J. and {Memeo}, P. and {Porciani}, C. and {Scaramella}, R. and {Scarlata}, C. and {Scoville}, N.},
        title = "{zCOSMOS - 10k-bright spectroscopic sample. The bimodality in the galaxy stellar mass function: exploring its evolution with redshift}",
      journal = {\aap},
         year = 2010,
        month = nov,
       volume = {523},
          eid = {A13},
        pages = {A13},
          doi = {10.1051/0004-6361/200913020},
archivePrefix = {arXiv},
       eprint = {0907.5416},
 primaryClass = {astro-ph.CO},
       adsurl = {https://ui.adsabs.harvard.edu/abs/2010A&A...523A..13P}
}

@ARTICLE{2020MNRAS.495.4681I,
       author = {{Izquierdo-Villalba}, David and {Bonoli}, Silvia and {Dotti}, Massimo and {Sesana}, Alberto and {Rosas-Guevara}, Yetli and {Spinoso}, Daniele},
        title = "{From galactic nuclei to the halo outskirts: tracing supermassive black holes across cosmic history and environments}",
      journal = {\mnras},
         year = 2020,
        month = jul,
       volume = {495},
       number = {4},
        pages = {4681-4706},
          doi = {10.1093/mnras/staa1399},
archivePrefix = {arXiv},
       eprint = {2001.10548},
 primaryClass = {astro-ph.GA},
       adsurl = {https://ui.adsabs.harvard.edu/abs/2020MNRAS.495.4681I}
}

@ARTICLE{2015ARA&A..53..365N,
       author = {{Netzer}, Hagai},
        title = "{Revisiting the Unified Model of Active Galactic Nuclei}",
      journal = {\araa},
         year = 2015,
        month = aug,
       volume = {53},
        pages = {365-408},
          doi = {10.1146/annurev-astro-082214-122302},
archivePrefix = {arXiv},
       eprint = {1505.00811},
 primaryClass = {astro-ph.GA},
       adsurl = {https://ui.adsabs.harvard.edu/abs/2015ARA&A..53..365N}
}

@ARTICLE{2014ARA&A..52..529Y,
       author = {{Yuan}, Feng and {Narayan}, Ramesh},
        title = "{Hot Accretion Flows Around Black Holes}",
      journal = {\araa},
         year = 2014,
        month = aug,
       volume = {52},
        pages = {529-588},
          doi = {10.1146/annurev-astro-082812-141003},
archivePrefix = {arXiv},
       eprint = {1401.0586},
 primaryClass = {astro-ph.HE},
       adsurl = {https://ui.adsabs.harvard.edu/abs/2014ARA&A..52..529Y}
}

@ARTICLE{2023MNRAS.522.2707S,
       author = {{Siwek}, Magdalena and {Weinberger}, Rainer and {Hernquist}, Lars},
        title = "{Orbital evolution of binaries in circumbinary discs}",
      journal = {\mnras},
         year = 2023,
        month = jun,
       volume = {522},
       number = {2},
        pages = {2707-2717},
          doi = {10.1093/mnras/stad1131},
archivePrefix = {arXiv},
       eprint = {2302.01785},
 primaryClass = {astro-ph.HE},
       adsurl = {https://ui.adsabs.harvard.edu/abs/2023MNRAS.522.2707S}
}

@ARTICLE{2016MNRAS.455.1989G,
       author = {{Goicovic}, F.~G. and {Cuadra}, J. and {Sesana}, A. and {Stasyszyn}, F. and {Amaro-Seoane}, P. and {Tanaka}, T.~L.},
        title = "{Infalling clouds on to supermassive black hole binaries - I. Formation of discs, accretion and gas dynamics}",
      journal = {\mnras},
         year = 2016,
        month = jan,
       volume = {455},
       number = {2},
        pages = {1989-2003},
          doi = {10.1093/mnras/stv2470},
archivePrefix = {arXiv},
       eprint = {1507.05596},
 primaryClass = {astro-ph.HE},
       adsurl = {https://ui.adsabs.harvard.edu/abs/2016MNRAS.455.1989G}
}

@ARTICLE{2024APh...15402892C,
       author = {{Cattorini}, Federico and {Giacomazzo}, Bruno},
        title = "{GRMHD study of accreting massive black hole binaries in astrophysical environment: A review}",
      journal = {Astroparticle Physics},
         year = 2024,
        month = jan,
       volume = {154},
          eid = {102892},
        pages = {102892},
          doi = {10.1016/j.astropartphys.2023.102892},
archivePrefix = {arXiv},
       eprint = {2401.02521},
 primaryClass = {astro-ph.HE},
       adsurl = {https://ui.adsabs.harvard.edu/abs/2024APh...15402892C}
}

@ARTICLE{2017MNRAS.466.1170M,
       author = {{Miranda}, Ryan and {Mu{\~n}oz}, Diego J. and {Lai}, Dong},
        title = "{Viscous hydrodynamics simulations of circumbinary accretion discs: variability, quasi-steady state and angular momentum transfer}",
      journal = {\mnras},
         year = 2017,
        month = apr,
       volume = {466},
       number = {1},
        pages = {1170-1191},
          doi = {10.1093/mnras/stw3189},
archivePrefix = {arXiv},
       eprint = {1610.07263},
 primaryClass = {astro-ph.SR},
       adsurl = {https://ui.adsabs.harvard.edu/abs/2017MNRAS.466.1170M}
}

@ARTICLE{2016MNRAS.463.2064H,
       author = {{Hern{\'a}n-Caballero}, Antonio and {Hatziminaoglou}, Evanthia and {Alonso-Herrero}, Almudena and {Mateos}, Silvia},
        title = "{The near-to-mid infrared spectrum of quasars}",
      journal = {\mnras},
         year = 2016,
        month = dec,
       volume = {463},
       number = {2},
        pages = {2064-2078},
          doi = {10.1093/mnras/stw2107},
archivePrefix = {arXiv},
       eprint = {1605.04867},
 primaryClass = {astro-ph.GA},
       adsurl = {https://ui.adsabs.harvard.edu/abs/2016MNRAS.463.2064H}
}

@ARTICLE{2025ApJ...987..106C,
       author = {{Casey-Clyde}, J. Andrew and {Mingarelli}, Chiara M.~F. and {Greene}, Jenny E. and {Goulding}, Andy D. and {Chen}, Siyuan and {Trump}, Jonathan R.},
        title = "{Quasars Can Signpost Supermassive Black Hole Binaries}",
      journal = {\apj},
         year = 2025,
        month = jul,
       volume = {987},
       number = {2},
          eid = {106},
        pages = {106},
          doi = {10.3847/1538-4357/adce05},
archivePrefix = {arXiv},
       eprint = {2405.19406},
 primaryClass = {astro-ph.HE},
       adsurl = {https://ui.adsabs.harvard.edu/abs/2025ApJ...987..106C}
}

@ARTICLE{2023NatAs...7.1506C,
       author = {{Cai}, Zhen-Yi and {Wang}, Jun-Xian},
        title = "{A universal average spectral energy distribution for quasars from the optical to the extreme ultraviolet}",
      journal = {Nature Astronomy},
         year = 2023,
        month = dec,
       volume = {7},
        pages = {1506-1516},
          doi = {10.1038/s41550-023-02088-5},
archivePrefix = {arXiv},
       eprint = {2309.01541},
 primaryClass = {astro-ph.GA},
       adsurl = {https://ui.adsabs.harvard.edu/abs/2023NatAs...7.1506C}
}

@ARTICLE{2023MNRAS.521L..11J,
       author = {{Jalan}, Priyanka and {Rakshit}, Suvendu and {Woo}, Jong-Jak and {Kotilainen}, Jari and {Stalin}, C.~S.},
        title = "{An empirical relation to estimate host galaxy stellar light from AGN spectra}",
      journal = {\mnras},
         year = 2023,
        month = may,
       volume = {521},
       number = {1},
        pages = {L11-L16},
          doi = {10.1093/mnrasl/slad014},
archivePrefix = {arXiv},
       eprint = {2302.01948},
 primaryClass = {astro-ph.GA},
       adsurl = {https://ui.adsabs.harvard.edu/abs/2023MNRAS.521L..11J}
}

@ARTICLE{2023Natur.616...45C,
       author = {{Chen}, Yu-Ching and {Liu}, Xin and {Foord}, Adi and {Shen}, Yue and {Oguri}, Masamune and {Chen}, Nianyi and {Di Matteo}, Tiziana and {Holgado}, Miguel and {Hwang}, Hsiang-Chih and {Zakamska}, Nadia},
        title = "{A close quasar pair in a disk-disk galaxy merger at z = 2.17}",
      journal = {\nat},
         year = 2023,
        month = apr,
       volume = {616},
       number = {7955},
        pages = {45-49},
          doi = {10.1038/s41586-023-05766-6},
archivePrefix = {arXiv},
       eprint = {2209.11249},
 primaryClass = {astro-ph.CO},
       adsurl = {https://ui.adsabs.harvard.edu/abs/2023Natur.616...45C}
}

@ARTICLE{2020MNRAS.492.2910G,
       author = {{Guo}, Hengxiao and {Liu}, Xin and {Zafar}, Tayyaba and {Liao}, Wei-Ting},
        title = "{Spectral energy distributions of candidate periodically variable quasars: testing the binary black hole hypothesis}",
      journal = {\mnras},
         year = 2020,
        month = feb,
       volume = {492},
       number = {2},
        pages = {2910-2923},
          doi = {10.1093/mnras/stz3566},
archivePrefix = {arXiv},
       eprint = {1907.06676},
 primaryClass = {astro-ph.GA},
       adsurl = {https://ui.adsabs.harvard.edu/abs/2020MNRAS.492.2910G}
}

@ARTICLE{2019NewAR..8601525D,
       author = {{De Rosa}, Alessandra and {Vignali}, Cristian and {Bogdanovi{\'c}}, Tamara and {Capelo}, Pedro R. and {Charisi}, Maria and {Dotti}, Massimo and {Husemann}, Bernd and {Lusso}, Elisabeta and {Mayer}, Lucio and {Paragi}, Zsolt and {Runnoe}, Jessie and {Sesana}, Alberto and {Steinborn}, Lisa and {Bianchi}, Stefano and {Colpi}, Monica and {del Valle}, Luciano and {Frey}, S{\'a}ndor and {Gab{\'a}nyi}, Krisztina {\'E}. and {Giustini}, Margherita and {Guainazzi}, Matteo and {Haiman}, Zoltan and {Herrera Ruiz}, Noelia and {Herrero-Illana}, Rub{\'e}n and {Iwasawa}, Kazushi and {Komossa}, S. and {Lena}, Davide and {Loiseau}, Nora and {Perez-Torres}, Miguel and {Piconcelli}, Enrico and {Volonteri}, Marta},
        title = "{The quest for dual and binary supermassive black holes: A multi-messenger view}",
      journal = {\nar},
         year = 2019,
        month = dec,
       volume = {86},
          eid = {101525},
        pages = {101525},
          doi = {10.1016/j.newar.2020.101525},
archivePrefix = {arXiv},
       eprint = {2001.06293},
 primaryClass = {astro-ph.GA},
       adsurl = {https://ui.adsabs.harvard.edu/abs/2019NewAR..8601525D}
}

@ARTICLE{2019ApJ...879..110K,
       author = {{Krolik}, Julian H. and {Volonteri}, Marta and {Dubois}, Yohan and {Devriendt}, Julien},
        title = "{Population Estimates for Electromagnetically Distinguishable Supermassive Binary Black Holes}",
      journal = {\apj},
         year = 2019,
        month = jul,
       volume = {879},
       number = {2},
          eid = {110},
        pages = {110},
          doi = {10.3847/1538-4357/ab24c9},
archivePrefix = {arXiv},
       eprint = {1905.10450},
 primaryClass = {astro-ph.GA},
       adsurl = {https://ui.adsabs.harvard.edu/abs/2019ApJ...879..110K}
}

@ARTICLE{2012ApJ...746..169S,
       author = {{Shen}, Yue and {Kelly}, Brandon C.},
        title = "{The Demographics of Broad-line Quasars in the Mass-Luminosity Plane. I. Testing FWHM-based Virial Black Hole Masses}",
      journal = {\apj},
         year = 2012,
        month = feb,
       volume = {746},
       number = {2},
          eid = {169},
        pages = {169},
          doi = {10.1088/0004-637X/746/2/169},
archivePrefix = {arXiv},
       eprint = {1107.4372},
 primaryClass = {astro-ph.CO},
       adsurl = {https://ui.adsabs.harvard.edu/abs/2012ApJ...746..169S}
}

@ARTICLE{2012MNRAS.420..860S,
       author = {{Sesana}, A. and {Roedig}, C. and {Reynolds}, M.~T. and {Dotti}, M.},
        title = "{Multimessenger astronomy with pulsar timing and X-ray observations of massive black hole binaries}",
      journal = {\mnras},
         year = 2012,
        month = feb,
       volume = {420},
       number = {1},
        pages = {860-877},
          doi = {10.1111/j.1365-2966.2011.20097.x},
archivePrefix = {arXiv},
       eprint = {1107.2927},
 primaryClass = {astro-ph.CO},
       adsurl = {https://ui.adsabs.harvard.edu/abs/2012MNRAS.420..860S}
}

@ARTICLE{2014ApJ...785..115R,
       author = {{Roedig}, Constanze and {Krolik}, Julian H. and {Miller}, M. Coleman},
        title = "{Observational Signatures of Binary Supermassive Black Holes}",
      journal = {\apj},
         year = 2014,
        month = apr,
       volume = {785},
       number = {2},
          eid = {115},
        pages = {115},
          doi = {10.1088/0004-637X/785/2/115},
archivePrefix = {arXiv},
       eprint = {1402.7098},
 primaryClass = {astro-ph.HE},
       adsurl = {https://ui.adsabs.harvard.edu/abs/2014ApJ...785..115R}
}

@ARTICLE{2015MNRAS.453.1562G,
       author = {{Graham}, Matthew J. and {Djorgovski}, S.~G. and {Stern}, Daniel and {Drake}, Andrew J. and {Mahabal}, Ashish A. and {Donalek}, Ciro and {Glikman}, Eilat and {Larson}, Steve and {Christensen}, Eric},
        title = "{A systematic search for close supermassive black hole binaries in the Catalina Real-time Transient Survey}",
      journal = {\mnras},
         year = 2015,
        month = oct,
       volume = {453},
       number = {2},
        pages = {1562-1576},
          doi = {10.1093/mnras/stv1726},
archivePrefix = {arXiv},
       eprint = {1507.07603},
 primaryClass = {astro-ph.GA},
       adsurl = {https://ui.adsabs.harvard.edu/abs/2015MNRAS.453.1562G}
}

@ARTICLE{2007MNRAS.379..956D,
       author = {{Dotti}, M. and {Colpi}, M. and {Haardt}, F. and {Mayer}, L.},
        title = "{Supermassive black hole binaries in gaseous and stellar circumnuclear discs: orbital dynamics and gas accretion}",
      journal = {\mnras},
         year = 2007,
        month = aug,
       volume = {379},
       number = {3},
        pages = {956-962},
          doi = {10.1111/j.1365-2966.2007.12010.x},
archivePrefix = {arXiv},
       eprint = {astro-ph/0612505},
 primaryClass = {astro-ph},
       adsurl = {https://ui.adsabs.harvard.edu/abs/2007MNRAS.379..956D}
}

@ARTICLE{2012MNRAS.420..705T,
       author = {{Tanaka}, Takamitsu and {Menou}, Kristen and {Haiman}, Zolt{\'a}n.},
        title = "{Electromagnetic counterparts of supermassive black hole binaries resolved by pulsar timing arrays}",
      journal = {\mnras},
         year = 2012,
        month = feb,
       volume = {420},
       number = {1},
        pages = {705-719},
          doi = {10.1111/j.1365-2966.2011.20083.x},
archivePrefix = {arXiv},
       eprint = {1107.2937},
 primaryClass = {astro-ph.CO},
       adsurl = {https://ui.adsabs.harvard.edu/abs/2012MNRAS.420..705T}
}

@ARTICLE{2012MNRAS.427.2660K,
       author = {{Kocsis}, Bence and {Haiman}, Zolt{\'a}n. and {Loeb}, Abraham},
        title = "{Gas pile-up, gap overflow and Type 1.5 migration in circumbinary discs: general theory}",
      journal = {\mnras},
         year = 2012,
        month = dec,
       volume = {427},
       number = {3},
        pages = {2660-2679},
          doi = {10.1111/j.1365-2966.2012.22129.x},
archivePrefix = {arXiv},
       eprint = {1205.4714},
 primaryClass = {astro-ph.EP},
       adsurl = {https://ui.adsabs.harvard.edu/abs/2012MNRAS.427.2660K}
}

@ARTICLE{2002MNRAS.335..965Y,
       author = {{Yu}, Qingjuan and {Tremaine}, Scott},
        title = "{Observational constraints on growth of massive black holes}",
      journal = {\mnras},
         year = 2002,
        month = oct,
       volume = {335},
       number = {4},
        pages = {965-976},
          doi = {10.1046/j.1365-8711.2002.05532.x},
archivePrefix = {arXiv},
       eprint = {astro-ph/0203082},
 primaryClass = {astro-ph},
       adsurl = {https://ui.adsabs.harvard.edu/abs/2002MNRAS.335..965Y}
}

@ARTICLE{2024ApJ...970..156D,
       author = {{Duffell}, Paul C. and {Dittmann}, Alexander J. and {D'Orazio}, Daniel J. and {Franchini}, Alessia and {Kratter}, Kaitlin M. and {Penzlin}, Anna B.~T. and {Ragusa}, Enrico and {Siwek}, Magdalena and {Tiede}, Christopher and {Wang}, Haiyang and {Zrake}, Jonathan and {Dempsey}, Adam M. and {Haiman}, Zoltan and {Lupi}, Alessandro and {Pirog}, Michal and {Ryan}, Geoffrey},
        title = "{The Santa Barbara Binary‑disk Code Comparison}",
      journal = {\apj},
         year = 2024,
        month = aug,
       volume = {970},
       number = {2},
          eid = {156},
        pages = {156},
          doi = {10.3847/1538-4357/ad5a7e},
archivePrefix = {arXiv},
       eprint = {2402.13039},
 primaryClass = {astro-ph.SR},
       adsurl = {https://ui.adsabs.harvard.edu/abs/2024ApJ...970..156D}
}

@ARTICLE{2021MNRAS.507.1458F,
       author = {{Franchini}, Alessia and {Sesana}, Alberto and {Dotti}, Massimo},
        title = "{Circumbinary disc self-gravity governing supermassive black hole binary mergers}",
      journal = {\mnras},
         year = 2021,
        month = oct,
       volume = {507},
       number = {1},
        pages = {1458-1467},
          doi = {10.1093/mnras/stab2234},
archivePrefix = {arXiv},
       eprint = {2106.13253},
 primaryClass = {astro-ph.HE},
       adsurl = {https://ui.adsabs.harvard.edu/abs/2021MNRAS.507.1458F}
}

@ARTICLE{2020ApJ...900...43T,
       author = {{Tiede}, Christopher and {Zrake}, Jonathan and {MacFadyen}, Andrew and {Haiman}, Zoltan},
        title = "{Gas-driven Inspiral of Binaries in Thin Accretion Disks}",
      journal = {\apj},
         year = 2020,
        month = sep,
       volume = {900},
       number = {1},
          eid = {43},
        pages = {43},
          doi = {10.3847/1538-4357/aba432},
archivePrefix = {arXiv},
       eprint = {2005.09555},
 primaryClass = {astro-ph.GA},
       adsurl = {https://ui.adsabs.harvard.edu/abs/2020ApJ...900...43T}
}

@ARTICLE{2017MNRAS.469.4258T,
       author = {{Tang}, Yike and {MacFadyen}, Andrew and {Haiman}, Zolt{\'a}n},
        title = "{On the orbital evolution of supermassive black hole binaries with circumbinary accretion discs}",
      journal = {\mnras},
         year = 2017,
        month = aug,
       volume = {469},
       number = {4},
        pages = {4258-4267},
          doi = {10.1093/mnras/stx1130},
archivePrefix = {arXiv},
       eprint = {1703.03913},
 primaryClass = {astro-ph.HE},
       adsurl = {https://ui.adsabs.harvard.edu/abs/2017MNRAS.469.4258T}
}

@ARTICLE{2022A&A...660A.101P,
       author = {{Penzlin}, Anna B.~T. and {Kley}, Wilhelm and {Audiffren}, Hugo and {Sch{\"a}fer}, Christoph M.},
        title = "{Binary orbital evolution driven by a circumbinary disc}",
      journal = {\aap},
         year = 2022,
        month = apr,
       volume = {660},
          eid = {A101},
        pages = {A101},
          doi = {10.1051/0004-6361/202141399},
archivePrefix = {arXiv},
       eprint = {2202.06681},
 primaryClass = {astro-ph.SR},
       adsurl = {https://ui.adsabs.harvard.edu/abs/2022A&A...660A.101P}
}

@ARTICLE{2016ApJ...832...22S,
       author = {{Shi}, Ji-Ming and {Krolik}, Julian H.},
        title = "{How Bright are the Gaps in Circumbinary Disk Systems?}",
      journal = {\apj},
         year = 2016,
        month = nov,
       volume = {832},
       number = {1},
          eid = {22},
        pages = {22},
          doi = {10.3847/0004-637X/832/1/22},
archivePrefix = {arXiv},
       eprint = {1609.07110},
 primaryClass = {astro-ph.HE},
       adsurl = {https://ui.adsabs.harvard.edu/abs/2016ApJ...832...22S}
}

@article{Zhan2021,
  author = {Hu Zhan},
  title = {The wide-field multiband imaging and slitless spectroscopy survey to be carried out by the Survey Space Telescope of China Manned Space Program},
  journal = {Chinese Science Bulletin},
  year = {2021},
  volume = {66},
  number = {11},
  pages = {1290-1298},
  url = {http://www.sciengine.com/publisher/Science China Press/journal/Chinese Science Bulletin/66/11/10.1360/TB-2021-0016},
  doi = {10.1360/TB-2021-0016}
}

@ARTICLE{2010MNRAS.404.2087B,
       author = {{Bernardi}, M. and {Shankar}, F. and {Hyde}, J.~B. and {Mei}, S. and {Marulli}, F. and {Sheth}, R.~K.},
        title = "{Galaxy luminosities, stellar masses, sizes, velocity dispersions as a function of morphological type}",
      journal = {\mnras},
         year = 2010,
        month = jun,
       volume = {404},
       number = {4},
        pages = {2087-2122},
          doi = {10.1111/j.1365-2966.2010.16425.x},
archivePrefix = {arXiv},
       eprint = {0910.1093},
 primaryClass = {astro-ph.CO},
       adsurl = {https://ui.adsabs.harvard.edu/abs/2010MNRAS.404.2087B}
}

@ARTICLE{2005AJ....130.1516D,
       author = {{De Propris}, Roberto and {Liske}, Jochen and {Driver}, Simon P. and {Allen}, Paul D. and {Cross}, Nicholas J.~G.},
        title = "{The Millennium Galaxy Catalogue: Dynamically Close Pairs of Galaxies and the Global Merger Rate}",
      journal = {\aj},
         year = 2005,
        month = oct,
       volume = {130},
       number = {4},
        pages = {1516-1523},
          doi = {10.1086/433169},
archivePrefix = {arXiv},
       eprint = {astro-ph/0506635},
 primaryClass = {astro-ph},
       adsurl = {https://ui.adsabs.harvard.edu/abs/2005AJ....130.1516D}
}

@ARTICLE{2009ApJ...696..411W,
       author = {{Weinzirl}, Tim and {Jogee}, Shardha and {Khochfar}, Sadegh and {Burkert}, Andreas and {Kormendy}, John},
        title = "{Bulge n and B/T in High-Mass Galaxies: Constraints on the Origin of Bulges in Hierarchical Models}",
      journal = {\apj},
         year = 2009,
        month = may,
       volume = {696},
       number = {1},
        pages = {411-447},
          doi = {10.1088/0004-637X/696/1/411},
archivePrefix = {arXiv},
       eprint = {0807.0040},
 primaryClass = {astro-ph},
       adsurl = {https://ui.adsabs.harvard.edu/abs/2009ApJ...696..411W}
}

@ARTICLE{2015Natur.525..351D,
       author = {{D'Orazio}, Daniel J. and {Haiman}, Zolt{\'a}n and {Schiminovich}, David},
        title = "{Relativistic boost as the cause of periodicity in a massive black-hole binary candidate}",
      journal = {\nat},
         year = 2015,
        month = sep,
       volume = {525},
       number = {7569},
        pages = {351-353},
          doi = {10.1038/nature15262},
archivePrefix = {arXiv},
       eprint = {1509.04301},
 primaryClass = {astro-ph.HE},
       adsurl = {https://ui.adsabs.harvard.edu/abs/2015Natur.525..351D}
}

@ARTICLE{2018A&A...617A.113E,
       author = {{Eliche-Moral}, M.~C. and {Rodr{\'\i}guez-P{\'e}rez}, C. and {Borlaff}, A. and {Querejeta}, M. and {Tapia}, T.},
        title = "{Formation of S0 galaxies through mergers. Morphological properties: tidal relics, lenses, ovals, and other inner components}",
      journal = {\aap},
         year = 2018,
        month = oct,
       volume = {617},
          eid = {A113},
        pages = {A113},
          doi = {10.1051/0004-6361/201832911},
archivePrefix = {arXiv},
       eprint = {1806.06070},
 primaryClass = {astro-ph.GA},
       adsurl = {https://ui.adsabs.harvard.edu/abs/2018A&A...617A.113E}
}

@ARTICLE{2024A&A...687A..57G,
       author = {{Ge}, Junqiang and {Lu}, Youjun and {Yan}, Changshuo and {Liu}, Jifeng},
        title = "{Variations of light curves and broad emission lines for periodic QSOs from co-rotating supermassive binary black holes in elliptical orbits}",
      journal = {\aap},
         year = 2024,
        month = jul,
       volume = {687},
          eid = {A57},
        pages = {A57},
          doi = {10.1051/0004-6361/202348303},
archivePrefix = {arXiv},
       eprint = {2404.05975},
 primaryClass = {astro-ph.GA},
       adsurl = {https://ui.adsabs.harvard.edu/abs/2024A&A...687A..57G}
}

@ARTICLE{2020MNRAS.494.4069K,
       author = {{Kova{\v{c}}evi{\'c}}, Andjelka B. and {Yi}, Tignfeng and {Dai}, Xinyu and {Yang}, Xing and {{\v{C}}vorovi{\'c}-Hajdinjak}, Iva and {Popovi{\'c}}, Luka {\v{C}}.},
        title = "{Confirmed short periodic variability of subparsec supermassive binary black hole candidate Mrk 231}",
      journal = {\mnras},
         year = 2020,
        month = may,
       volume = {494},
       number = {3},
        pages = {4069-4076},
          doi = {10.1093/mnras/staa737},
archivePrefix = {arXiv},
       eprint = {2003.06359},
 primaryClass = {astro-ph.GA},
       adsurl = {https://ui.adsabs.harvard.edu/abs/2020MNRAS.494.4069K}
}

@ARTICLE{2021ApJ...910..101J,
       author = {{Ji}, Xiang and {Lu}, Youjun and {Ge}, Junqiang and {Yan}, Changshuo and {Song}, Zihao},
        title = "{Variation of Broad Emission Lines from QSOs with Optical/UV Periodicity to Test the Interpretation of Supermassive Binary Black Holes}",
      journal = {\apj},
         year = 2021,
        month = apr,
       volume = {910},
       number = {2},
          eid = {101},
        pages = {101},
          doi = {10.3847/1538-4357/abe386},
archivePrefix = {arXiv},
       eprint = {2103.16448},
 primaryClass = {astro-ph.GA},
       adsurl = {https://ui.adsabs.harvard.edu/abs/2021ApJ...910..101J}
}

@ARTICLE{2018MNRAS.480.5504Y,
       author = {{Yang}, Lilan and {Dai}, Xinyu and {Lu}, Youjun and {Zhu}, Zong-Hong and {Shankar}, Francesco},
        title = "{Swift monitoring observations of Mrk 231: detection of ultraviolet variability}",
      journal = {\mnras},
         year = 2018,
        month = nov,
       volume = {480},
       number = {4},
        pages = {5504-5510},
          doi = {10.1093/mnras/sty2253},
archivePrefix = {arXiv},
       eprint = {1808.05235},
 primaryClass = {astro-ph.GA},
       adsurl = {https://ui.adsabs.harvard.edu/abs/2018MNRAS.480.5504Y}
}

@ARTICLE{2021ApJ...914L..21D,
       author = {{D'Orazio}, Daniel J. and {Duffell}, Paul C.},
        title = "{Orbital Evolution of Equal-mass Eccentric Binaries due to a Gas Disk: Eccentric Inspirals and Circular Outspirals}",
      journal = {\apjl},
         year = 2021,
        month = jun,
       volume = {914},
       number = {1},
          eid = {L21},
        pages = {L21},
          doi = {10.3847/2041-8213/ac0621},
archivePrefix = {arXiv},
       eprint = {2103.09251},
 primaryClass = {astro-ph.HE},
       adsurl = {https://ui.adsabs.harvard.edu/abs/2021ApJ...914L..21D}
}

@ARTICLE{2021ApJ...909L..13Z,
       author = {{Zrake}, Jonathan and {Tiede}, Christopher and {MacFadyen}, Andrew and {Haiman}, Zolt{\'a}n},
        title = "{Equilibrium Eccentricity of Accreting Binaries}",
      journal = {\apjl},
         year = 2021,
        month = mar,
       volume = {909},
       number = {1},
          eid = {L13},
        pages = {L13},
          doi = {10.3847/2041-8213/abdd1c},
archivePrefix = {arXiv},
       eprint = {2010.09707},
 primaryClass = {astro-ph.HE},
       adsurl = {https://ui.adsabs.harvard.edu/abs/2021ApJ...909L..13Z}
}

@ARTICLE{2016MNRAS.463.2145C,
       author = {{Charisi}, M. and {Bartos}, I. and {Haiman}, Z. and {Price-Whelan}, A.~M. and {Graham}, M.~J. and {Bellm}, E.~C. and {Laher}, R.~R. and {M{\'a}rka}, S.},
        title = "{A population of short-period variable quasars from PTF as supermassive black hole binary candidates}",
      journal = {\mnras},
         year = 2016,
        month = dec,
       volume = {463},
       number = {2},
        pages = {2145-2171},
          doi = {10.1093/mnras/stw1838},
archivePrefix = {arXiv},
       eprint = {1604.01020},
 primaryClass = {astro-ph.GA},
       adsurl = {https://ui.adsabs.harvard.edu/abs/2016MNRAS.463.2145C}
}

@ARTICLE{2020MNRAS.491.4023S,
       author = {{Song}, Zihao and {Ge}, Junqiang and {Lu}, Youjun and {Ji}, Xiang},
        title = "{Testing the relativistic Doppler boost hypothesis for supermassive binary black holes candidates via broad emission line profiles}",
      journal = {\mnras},
         year = 2020,
        month = jan,
       volume = {491},
       number = {3},
        pages = {4023-4030},
          doi = {10.1093/mnras/stz3354},
archivePrefix = {arXiv},
       eprint = {1912.00560},
 primaryClass = {astro-ph.GA},
       adsurl = {https://ui.adsabs.harvard.edu/abs/2020MNRAS.491.4023S}
}

@ARTICLE{2021A&A...645A..15S,
       author = {{Song}, Zihao and {Ge}, Junqiang and {Lu}, Youjun and {Yan}, Changshuo and {Ji}, Xiang},
        title = "{Broad-line region configuration of the supermassive binary black hole candidate PG1302-102 in the relativistic Doppler boosting scenario}",
      journal = {\aap},
         year = 2021,
        month = jan,
       volume = {645},
          eid = {A15},
        pages = {A15},
          doi = {10.1051/0004-6361/202039300},
archivePrefix = {arXiv},
       eprint = {2010.07512},
 primaryClass = {astro-ph.GA},
       adsurl = {https://ui.adsabs.harvard.edu/abs/2021A&A...645A..15S}
}

@ARTICLE{2020MNRAS.496.1683X,
       author = {{Xin}, Chengcheng and {Charisi}, Maria and {Haiman}, Zolt{\'a}n and {Schiminovich}, David and {Graham}, Matthew J. and {Stern}, Daniel and {D'Orazio}, Daniel J.},
        title = "{Testing the relativistic Doppler boost hypothesis for the binary candidate quasar PG1302-102 with multiband Swift data}",
      journal = {\mnras},
         year = 2020,
        month = aug,
       volume = {496},
       number = {2},
        pages = {1683-1696},
          doi = {10.1093/mnras/staa1643},
archivePrefix = {arXiv},
       eprint = {1907.11246},
 primaryClass = {astro-ph.GA},
       adsurl = {https://ui.adsabs.harvard.edu/abs/2020MNRAS.496.1683X}
}

@ARTICLE{2017NatAs...1..886M,
       author = {{Mingarelli}, Chiara M.~F. and {Lazio}, T. Joseph W. and {Sesana}, Alberto and {Greene}, Jenny E. and {Ellis}, Justin A. and {Ma}, Chung-Pei and {Croft}, Steve and {Burke-Spolaor}, Sarah and {Taylor}, Stephen R.},
        title = "{The local nanohertz gravitational-wave landscape from supermassive black hole binaries}",
      journal = {Nature Astronomy},
         year = 2017,
        month = nov,
       volume = {1},
        pages = {886-892},
          doi = {10.1038/s41550-017-0299-6},
archivePrefix = {arXiv},
       eprint = {1708.03491},
 primaryClass = {astro-ph.GA},
       adsurl = {https://ui.adsabs.harvard.edu/abs/2017NatAs...1..886M}
}

@ARTICLE{2009ApJ...702L..82C,
       author = {{Comerford}, Julia M. and {Griffith}, Roger L. and {Gerke}, Brian F. and {Cooper}, Michael C. and {Newman}, Jeffrey A. and {Davis}, Marc and {Stern}, Daniel},
        title = "{1.75 h $^{-1}$ kpc Separation Dual Active Galactic Nuclei at z = 0.36 in the Cosmos Field}",
      journal = {\apjl},
         year = 2009,
        month = sep,
       volume = {702},
       number = {1},
        pages = {L82-L86},
          doi = {10.1088/0004-637X/702/1/L82},
archivePrefix = {arXiv},
       eprint = {0906.3517},
 primaryClass = {astro-ph.CO},
       adsurl = {https://ui.adsabs.harvard.edu/abs/2009ApJ...702L..82C}
}

@ARTICLE{2012ApJS..201...31G,
       author = {{Ge}, Jun-Qiang and {Hu}, Chen and {Wang}, Jian-Min and {Bai}, Jin-Ming and {Zhang}, Shu},
        title = "{Double-peaked Narrow Emission-line Galaxies from the Sloan Digital Sky Survey. I. Sample and Basic Properties}",
      journal = {\apjs},
         year = 2012,
        month = aug,
       volume = {201},
       number = {2},
          eid = {31},
        pages = {31},
          doi = {10.1088/0067-0049/201/2/31},
archivePrefix = {arXiv},
       eprint = {1208.2485},
 primaryClass = {astro-ph.CO},
       adsurl = {https://ui.adsabs.harvard.edu/abs/2012ApJS..201...31G}
}

@ARTICLE{2021A&A...646A.153S,
       author = {{Severgnini}, P. and {Braito}, V. and {Cicone}, C. and {Saracco}, P. and {Vignali}, C. and {Serafinelli}, R. and {Della Ceca}, R. and {Dotti}, M. and {Cusano}, F. and {Paris}, D. and {Pruto}, G. and {Zaino}, A. and {Ballo}, L. and {Landoni}, M.},
        title = "{A possible sub-kiloparsec dual AGN buried behind the galaxy curtain}",
      journal = {\aap},
         year = 2021,
        month = feb,
       volume = {646},
          eid = {A153},
        pages = {A153},
          doi = {10.1051/0004-6361/202039576},
archivePrefix = {arXiv},
       eprint = {2012.09184},
 primaryClass = {astro-ph.GA},
       adsurl = {https://ui.adsabs.harvard.edu/abs/2021A&A...646A.153S}
}

@ARTICLE{2011ApJ...735...48S,
       author = {{Shen}, Yue and {Liu}, Xin and {Greene}, Jenny E. and {Strauss}, Michael A.},
        title = "{Type 2 Active Galactic Nuclei with Double-peaked [O III] Lines. II. Single AGNs with Complex Narrow-line Region Kinematics are More Common than Binary AGNs}",
      journal = {\apj},
         year = 2011,
        month = jul,
       volume = {735},
       number = {1},
          eid = {48},
        pages = {48},
          doi = {10.1088/0004-637X/735/1/48},
archivePrefix = {arXiv},
       eprint = {1011.5246},
 primaryClass = {astro-ph.CO},
       adsurl = {https://ui.adsabs.harvard.edu/abs/2011ApJ...735...48S}
}

@ARTICLE{2021ApJ...916..110L,
       author = {{Li}, Kunyang and {Ballantyne}, David R. and {Bogdanovi{\'c}}, Tamara},
        title = "{The Detectability of Kiloparsec-scale Dual Active Galactic Nuclei: The Impact of Galactic Structure and Black Hole Orbital Properties}",
      journal = {\apj},
         year = 2021,
        month = aug,
       volume = {916},
       number = {2},
          eid = {110},
        pages = {110},
          doi = {10.3847/1538-4357/ac06a0},
archivePrefix = {arXiv},
       eprint = {2103.02862},
 primaryClass = {astro-ph.GA},
       adsurl = {https://ui.adsabs.harvard.edu/abs/2021ApJ...916..110L}
}

@ARTICLE{2022NatAs...6.1185M,
       author = {{Mannucci}, F. and {Pancino}, E. and {Belfiore}, F. and {Cicone}, C. and {Ciurlo}, A. and {Cresci}, G. and {Lusso}, E. and {Marasco}, A. and {Marconi}, A. and {Nardini}, E. and {Pinna}, E. and {Severgnini}, P. and {Saracco}, P. and {Tozzi}, G. and {Yeh}, S.},
        title = "{Unveiling the population of dual and lensed active galactic nuclei at sub-arcsec separations}",
      journal = {Nature Astronomy},
         year = 2022,
        month = aug,
       volume = {6},
        pages = {1185-1192},
          doi = {10.1038/s41550-022-01761-5},
archivePrefix = {arXiv},
       eprint = {2203.11234},
 primaryClass = {astro-ph.GA},
       adsurl = {https://ui.adsabs.harvard.edu/abs/2022NatAs...6.1185M}
}

@ARTICLE{2023ApJ...951L...6R,
       author = {{Reardon}, Daniel J. and {Zic}, Andrew and {Shannon}, Ryan M. and {Hobbs}, George B. and {Bailes}, Matthew and {Di Marco}, Valentina and {Kapur}, Agastya and {Rogers}, Axl F. and {Thrane}, Eric and {Askew}, Jacob and {Bhat}, N.~D. Ramesh and {Cameron}, Andrew and {Cury{\l}o}, Ma{\l}gorzata and {Coles}, William A. and {Dai}, Shi and {Goncharov}, Boris and {Kerr}, Matthew and {Kulkarni}, Atharva and {Levin}, Yuri and {Lower}, Marcus E. and {Manchester}, Richard N. and {Mandow}, Rami and {Miles}, Matthew T. and {Nathan}, Rowina S. and {Os{\l}owski}, Stefan and {Russell}, Christopher J. and {Spiewak}, Ren{\'e}e and {Zhang}, Songbo and {Zhu}, Xing-Jiang},
        title = "{Search for an Isotropic Gravitational-wave Background with the Parkes Pulsar Timing Array}",
      journal = {\apjl},
         year = 2023,
        month = jul,
       volume = {951},
       number = {1},
          eid = {L6},
        pages = {L6},
          doi = {10.3847/2041-8213/acdd02},
archivePrefix = {arXiv},
       eprint = {2306.16215},
 primaryClass = {astro-ph.HE},
       adsurl = {https://ui.adsabs.harvard.edu/abs/2023ApJ...951L...6R}
}

@ARTICLE{2023A&A...678A..50E,
       author = {{EPTA Collaboration} and {InPTA Collaboration} and {Antoniadis}, J. and {Arumugam}, P. and {Arumugam}, S. and {Babak}, S. and {Bagchi}, M. and {Bak Nielsen}, A. -S. and {Bassa}, C.~G. and {Bathula}, A. and {Berthereau}, A. and {Bonetti}, M. and {Bortolas}, E. and {Brook}, P.~R. and {Burgay}, M. and {Caballero}, R.~N. and {Chalumeau}, A. and {Champion}, D.~J. and {Chanlaridis}, S. and {Chen}, S. and {Cognard}, I. and {Dandapat}, S. and {Deb}, D. and {Desai}, S. and {Desvignes}, G. and {Dhanda-Batra}, N. and {Dwivedi}, C. and {Falxa}, M. and {Ferdman}, R.~D. and {Franchini}, A. and {Gair}, J.~R. and {Goncharov}, B. and {Gopakumar}, A. and {Graikou}, E. and {Grie{\ss}meier}, J. -M. and {Guillemot}, L. and {Guo}, Y.~J. and {Gupta}, Y. and {Hisano}, S. and {Hu}, H. and {Iraci}, F. and {Izquierdo-Villalba}, D. and {Jang}, J. and {Jawor}, J. and {Janssen}, G.~H. and {Jessner}, A. and {Joshi}, B.~C. and {Kareem}, F. and {Karuppusamy}, R. and {Keane}, E.~F. and {Keith}, M.~J. and {Kharbanda}, D. and {Kikunaga}, T. and {Kolhe}, N. and {Kramer}, M. and {Krishnakumar}, M.~A. and {Lackeos}, K. and {Lee}, K.~J. and {Liu}, K. and {Liu}, Y. and {Lyne}, A.~G. and {McKee}, J.~W. and {Maan}, Y. and {Main}, R.~A. and {Mickaliger}, M.~B. and {Ni{\c{t}}u}, I.~C. and {Nobleson}, K. and {Paladi}, A.~K. and {Parthasarathy}, A. and {Perera}, B.~B.~P. and {Perrodin}, D. and {Petiteau}, A. and {Porayko}, N.~K. and {Possenti}, A. and {Prabu}, T. and {Quelquejay Leclere}, H. and {Rana}, P. and {Samajdar}, A. and {Sanidas}, S.~A. and {Sesana}, A. and {Shaifullah}, G. and {Singha}, J. and {Speri}, L. and {Spiewak}, R. and {Srivastava}, A. and {Stappers}, B.~W. and {Surnis}, M. and {Susarla}, S.~C. and {Susobhanan}, A. and {Takahashi}, K. and {Tarafdar}, P. and {Theureau}, G. and {Tiburzi}, C. and {van der Wateren}, E. and {Vecchio}, A. and {Venkatraman Krishnan}, V. and {Verbiest}, J.~P.~W. and {Wang}, J. and {Wang}, L. and {Wu}, Z.},
        title = "{The second data release from the European Pulsar Timing Array. III. Search for gravitational wave signals}",
      journal = {\aap},
         year = 2023,
        month = oct,
       volume = {678},
          eid = {A50},
        pages = {A50},
          doi = {10.1051/0004-6361/202346844},
archivePrefix = {arXiv},
       eprint = {2306.16214},
 primaryClass = {astro-ph.HE},
       adsurl = {https://ui.adsabs.harvard.edu/abs/2023A&A...678A..50E}
}

@ARTICLE{2023RAA....23g5024X,
       author = {{Xu}, Heng and {Chen}, Siyuan and {Guo}, Yanjun and {Jiang}, Jinchen and {Wang}, Bojun and {Xu}, Jiangwei and {Xue}, Zihan and {Caballero}, R. Nicolas and {Yuan}, Jianping and {Xu}, Yonghua and {Wang}, Jingbo and {Hao}, Longfei and {Luo}, Jingtao and {Lee}, Kejia and {Han}, Jinlin and {Jiang}, Peng and {Shen}, Zhiqiang and {Wang}, Min and {Wang}, Na and {Xu}, Renxin and {Wu}, Xiangping and {Manchester}, Richard and {Qian}, Lei and {Guan}, Xin and {Huang}, Menglin and {Sun}, Chun and {Zhu}, Yan},
        title = "{Searching for the Nano-Hertz Stochastic Gravitational Wave Background with the Chinese Pulsar Timing Array Data Release I}",
      journal = {Research in Astronomy and Astrophysics},
         year = 2023,
        month = jul,
       volume = {23},
       number = {7},
          eid = {075024},
        pages = {075024},
          doi = {10.1088/1674-4527/acdfa5},
archivePrefix = {arXiv},
       eprint = {2306.16216},
 primaryClass = {astro-ph.HE},
       adsurl = {https://ui.adsabs.harvard.edu/abs/2023RAA....23g5024X}
}

@ARTICLE{2010ApJ...708..427L,
       author = {{Liu}, Xin and {Shen}, Yue and {Strauss}, Michael A. and {Greene}, Jenny E.},
        title = "{Type 2 Active Galactic Nuclei with Double-Peaked [O III] Lines: Narrow-Line Region Kinematics or Merging Supermassive Black Hole Pairs?}",
      journal = {\apj},
         year = 2010,
        month = jan,
       volume = {708},
       number = {1},
        pages = {427-434},
          doi = {10.1088/0004-637X/708/1/427},
archivePrefix = {arXiv},
       eprint = {0908.2426},
 primaryClass = {astro-ph.CO},
       adsurl = {https://ui.adsabs.harvard.edu/abs/2010ApJ...708..427L}
}

@ARTICLE{2025CQGra..42k3001L,
       author = {{Lu}, Youjun},
        title = "{Supermassive binary black holes and current status of their multimessenger observations}",
      journal = {Classical and Quantum Gravity},
         year = 2025,
        month = jun,
       volume = {42},
       number = {11},
          eid = {113001},
        pages = {113001},
          doi = {10.1088/1361-6382/adc4b0},
       adsurl = {https://ui.adsabs.harvard.edu/abs/2025CQGra..42k3001L}
}

@ARTICLE{2024ApJ...965..164G,
       author = {{Gardiner}, Emiko C. and {Kelley}, Luke Zoltan and {Lemke}, Anna-Malin and {Mitridate}, Andrea},
        title = "{Beyond the Background: Gravitational-wave Anisotropy and Continuous Waves from Supermassive Black Hole Binaries}",
      journal = {\apj},
         year = 2024,
        month = apr,
       volume = {965},
       number = {2},
          eid = {164},
        pages = {164},
          doi = {10.3847/1538-4357/ad2be8},
archivePrefix = {arXiv},
       eprint = {2309.07227},
 primaryClass = {astro-ph.HE},
       adsurl = {https://ui.adsabs.harvard.edu/abs/2024ApJ...965..164G}
}

@ARTICLE{2024PhRvD.109b1302E,
       author = {{Ellis}, John and {Fairbairn}, Malcolm and {H{\"u}tsi}, Gert and {Raidal}, Juhan and {Urrutia}, Juan and {Vaskonen}, Ville and {Veerm{\"a}e}, Hardi},
        title = "{Gravitational waves from supermassive black hole binaries in light of the NANOGrav 15-year data}",
      journal = {\prd},
         year = 2024,
        month = jan,
       volume = {109},
       number = {2},
          eid = {L021302},
        pages = {L021302},
          doi = {10.1103/PhysRevD.109.L021302},
archivePrefix = {arXiv},
       eprint = {2306.17021},
 primaryClass = {astro-ph.CO},
       adsurl = {https://ui.adsabs.harvard.edu/abs/2024PhRvD.109b1302E}
}

@ARTICLE{2025PhRvD.111b3043S,
       author = {{Sato-Polito}, Gabriela and {Zaldarriaga}, Matias},
        title = "{Distribution of the gravitational-wave background from supermassive black holes}",
      journal = {\prd},
         year = 2025,
        month = jan,
       volume = {111},
       number = {2},
          eid = {023043},
        pages = {023043},
          doi = {10.1103/PhysRevD.111.023043},
archivePrefix = {arXiv},
       eprint = {2406.17010},
 primaryClass = {astro-ph.CO},
       adsurl = {https://ui.adsabs.harvard.edu/abs/2025PhRvD.111b3043S}
}

@ARTICLE{2024A&A...691A.270E,
       author = {{Ellis}, John and {Fairbairn}, Malcolm and {H{\"u}tsi}, Gert and {Urrutia}, Juan and {Vaskonen}, Ville and {Veerm{\"a}e}, Hardi},
        title = "{Consistency of JWST black hole observations with NANOGrav gravitational wave measurements}",
      journal = {\aap},
         year = 2024,
        month = nov,
       volume = {691},
          eid = {A270},
        pages = {A270},
          doi = {10.1051/0004-6361/202450846},
archivePrefix = {arXiv},
       eprint = {2403.19650},
 primaryClass = {astro-ph.CO},
       adsurl = {https://ui.adsabs.harvard.edu/abs/2024A&A...691A.270E}
}

@ARTICLE{2025arXiv250109786S,
       author = {{Sato-Polito}, Gabriela and {Zaldarriaga}, Matias and {Quataert}, Eliot},
        title = "{Evolution of SMBHs in light of PTA measurements: implications for growth by mergers and accretion}",
      journal = {arXiv e-prints},
         year = 2025,
        month = jan,
          eid = {arXiv:2501.09786},
        pages = {arXiv:2501.09786},
          doi = {10.48550/arXiv.2501.09786},
archivePrefix = {arXiv},
       eprint = {2501.09786},
 primaryClass = {astro-ph.CO},
       adsurl = {https://ui.adsabs.harvard.edu/abs/2025arXiv250109786S}
}

@ARTICLE{2012ApJS..201...23E,
       author = {{Eracleous}, Michael and {Boroson}, Todd A. and {Halpern}, Jules P. and {Liu}, Jia},
        title = "{A Large Systematic Search for Close Supermassive Binary and Rapidly Recoiling Black Holes}",
      journal = {\apjs},
         year = 2012,
        month = aug,
       volume = {201},
       number = {2},
          eid = {23},
        pages = {23},
          doi = {10.1088/0067-0049/201/2/23},
archivePrefix = {arXiv},
       eprint = {1106.2952},
 primaryClass = {astro-ph.CO},
       adsurl = {https://ui.adsabs.harvard.edu/abs/2012ApJS..201...23E}
}

@ARTICLE{2013ApJ...775...49S,
       author = {{Shen}, Yue and {Liu}, Xin and {Loeb}, Abraham and {Tremaine}, Scott},
        title = "{Constraining Sub-parsec Binary Supermassive Black Holes in Quasars with Multi-epoch Spectroscopy. I. The General Quasar Population}",
      journal = {\apj},
         year = 2013,
        month = sep,
       volume = {775},
       number = {1},
          eid = {49},
        pages = {49},
          doi = {10.1088/0004-637X/775/1/49},
archivePrefix = {arXiv},
       eprint = {1306.4330},
 primaryClass = {astro-ph.CO},
       adsurl = {https://ui.adsabs.harvard.edu/abs/2013ApJ...775...49S}
}

@ARTICLE{2018ApJ...859L..12L,
       author = {{Liu}, Tingting and {Gezari}, Suvi and {Miller}, M. Coleman},
        title = "{Did ASAS-SN Kill the Supermassive Black Hole Binary Candidate PG1302-102?}",
      journal = {\apjl},
         year = 2018,
        month = may,
       volume = {859},
       number = {1},
          eid = {L12},
        pages = {L12},
          doi = {10.3847/2041-8213/aac2ed},
archivePrefix = {arXiv},
       eprint = {1803.05448},
 primaryClass = {astro-ph.HE},
       adsurl = {https://ui.adsabs.harvard.edu/abs/2018ApJ...859L..12L}
}

@ARTICLE{2024ApJ...974..261C,
       author = {{Chen}, Yunfeng and {Yu}, Qingjuan and {Lu}, Youjun},
        title = "{Constraining the Origin of the Nanohertz Gravitational-wave Background by Pulsar Timing Array Observations of Both the Background and Individual Supermassive Binary Black Holes}",
      journal = {\apj},
         year = 2024,
        month = oct,
       volume = {974},
       number = {2},
          eid = {261},
        pages = {261},
          doi = {10.3847/1538-4357/ad7582},
archivePrefix = {arXiv},
       eprint = {2409.18029},
 primaryClass = {astro-ph.HE},
       adsurl = {https://ui.adsabs.harvard.edu/abs/2024ApJ...974..261C}
}

@ARTICLE{2018ApJ...856...42S,
       author = {{Sesana}, Alberto and {Haiman}, Zolt{\'a}n and {Kocsis}, Bence and {Kelley}, Luke Zoltan},
        title = "{Testing the Binary Hypothesis: Pulsar Timing Constraints on Supermassive Black Hole Binary Candidates}",
      journal = {\apj},
         year = 2018,
        month = mar,
       volume = {856},
       number = {1},
          eid = {42},
        pages = {42},
          doi = {10.3847/1538-4357/aaad0f},
archivePrefix = {arXiv},
       eprint = {1703.10611},
 primaryClass = {astro-ph.HE},
       adsurl = {https://ui.adsabs.harvard.edu/abs/2018ApJ...856...42S}
}

@ARTICLE{2018MNRAS.477..964K,
       author = {{Kelley}, Luke Zoltan and {Blecha}, Laura and {Hernquist}, Lars and {Sesana}, Alberto and {Taylor}, Stephen R.},
        title = "{Single sources in the low-frequency gravitational wave sky: properties and time to detection by pulsar timing arrays}",
      journal = {\mnras},
         year = 2018,
        month = jun,
       volume = {477},
       number = {1},
        pages = {964-976},
          doi = {10.1093/mnras/sty689},
archivePrefix = {arXiv},
       eprint = {1711.00075},
 primaryClass = {astro-ph.HE},
       adsurl = {https://ui.adsabs.harvard.edu/abs/2018MNRAS.477..964K}
}

@ARTICLE{2016MNRAS.461.3145V,
       author = {{Vaughan}, S. and {Uttley}, P. and {Markowitz}, A.~G. and {Huppenkothen}, D. and {Middleton}, M.~J. and {Alston}, W.~N. and {Scargle}, J.~D. and {Farr}, W.~M.},
        title = "{False periodicities in quasar time-domain surveys}",
      journal = {\mnras},
         year = 2016,
        month = sep,
       volume = {461},
       number = {3},
        pages = {3145-3152},
          doi = {10.1093/mnras/stw1412},
archivePrefix = {arXiv},
       eprint = {1606.02620},
 primaryClass = {astro-ph.IM},
       adsurl = {https://ui.adsabs.harvard.edu/abs/2016MNRAS.461.3145V}
}

@ARTICLE{2021RAA....21..219J,
       author = {{Ji}, Xiang and {Ge}, Jun-Qiang and {Lu}, You-Jun and {Yan}, Chang-Shuo},
        title = "{Variations of broad emission lines from periodicity QSOs under the interpretation of supermassive binary black holes with misaligned circumbinary broad line regions}",
      journal = {Research in Astronomy and Astrophysics},
         year = 2021,
        month = nov,
       volume = {21},
       number = {9},
          eid = {219},
        pages = {219},
          doi = {10.1088/1674-4527/21/9/219},
archivePrefix = {arXiv},
       eprint = {2110.00834},
 primaryClass = {astro-ph.GA},
       adsurl = {https://ui.adsabs.harvard.edu/abs/2021RAA....21..219J}
}

@ARTICLE{2016A&A...594A..13P,
       author = {{Planck Collaboration} and {Ade}, P.~A.~R. and {Aghanim}, N. and {Arnaud}, M. and {Ashdown}, M. and {Aumont}, J. and {Baccigalupi}, C. and {Banday}, A.~J. and {Barreiro}, R.~B. and {Bartlett}, J.~G. and {Bartolo}, N. and {Battaner}, E. and {Battye}, R. and {Benabed}, K. and {Beno{\^\i}t}, A. and {Benoit-L{\'e}vy}, A. and {Bernard}, J. -P. and {Bersanelli}, M. and {Bielewicz}, P. and {Bock}, J.~J. and {Bonaldi}, A. and {Bonavera}, L. and {Bond}, J.~R. and {Borrill}, J. and {Bouchet}, F.~R. and {Boulanger}, F. and {Bucher}, M. and {Burigana}, C. and {Butler}, R.~C. and {Calabrese}, E. and {Cardoso}, J. -F. and {Catalano}, A. and {Challinor}, A. and {Chamballu}, A. and {Chary}, R. -R. and {Chiang}, H.~C. and {Chluba}, J. and {Christensen}, P.~R. and {Church}, S. and {Clements}, D.~L. and {Colombi}, S. and {Colombo}, L.~P.~L. and {Combet}, C. and {Coulais}, A. and {Crill}, B.~P. and {Curto}, A. and {Cuttaia}, F. and {Danese}, L. and {Davies}, R.~D. and {Davis}, R.~J. and {de Bernardis}, P. and {de Rosa}, A. and {de Zotti}, G. and {Delabrouille}, J. and {D{\'e}sert}, F. -X. and {Di Valentino}, E. and {Dickinson}, C. and {Diego}, J.~M. and {Dolag}, K. and {Dole}, H. and {Donzelli}, S. and {Dor{\'e}}, O. and {Douspis}, M. and {Ducout}, A. and {Dunkley}, J. and {Dupac}, X. and {Efstathiou}, G. and {Elsner}, F. and {En{\ss}lin}, T.~A. and {Eriksen}, H.~K. and {Farhang}, M. and {Fergusson}, J. and {Finelli}, F. and {Forni}, O. and {Frailis}, M. and {Fraisse}, A.~A. and {Franceschi}, E. and {Frejsel}, A. and {Galeotta}, S. and {Galli}, S. and {Ganga}, K. and {Gauthier}, C. and {Gerbino}, M. and {Ghosh}, T. and {Giard}, M. and {Giraud-H{\'e}raud}, Y. and {Giusarma}, E. and {Gjerl{\o}w}, E. and {Gonz{\'a}lez-Nuevo}, J. and {G{\'o}rski}, K.~M. and {Gratton}, S. and {Gregorio}, A. and {Gruppuso}, A. and {Gudmundsson}, J.~E. and {Hamann}, J. and {Hansen}, F.~K. and {Hanson}, D. and {Harrison}, D.~L. and {Helou}, G. and {Henrot-Versill{\'e}}, S. and {Hern{\'a}ndez-Monteagudo}, C. and {Herranz}, D. and {Hildebrandt}, S.~R. and {Hivon}, E. and {Hobson}, M. and {Holmes}, W.~A. and {Hornstrup}, A. and {Hovest}, W. and {Huang}, Z. and {Huffenberger}, K.~M. and {Hurier}, G. and {Jaffe}, A.~H. and {Jaffe}, T.~R. and {Jones}, W.~C. and {Juvela}, M. and {Keih{\"a}nen}, E. and {Keskitalo}, R. and {Kisner}, T.~S. and {Kneissl}, R. and {Knoche}, J. and {Knox}, L. and {Kunz}, M. and {Kurki-Suonio}, H. and {Lagache}, G. and {L{\"a}hteenm{\"a}ki}, A. and {Lamarre}, J. -M. and {Lasenby}, A. and {Lattanzi}, M. and {Lawrence}, C.~R. and {Leahy}, J.~P. and {Leonardi}, R. and {Lesgourgues}, J. and {Levrier}, F. and {Lewis}, A. and {Liguori}, M. and {Lilje}, P.~B. and {Linden-V{\o}rnle}, M. and {L{\'o}pez-Caniego}, M. and {Lubin}, P.~M. and {Mac{\'\i}as-P{\'e}rez}, J.~F. and {Maggio}, G. and {Maino}, D. and {Mandolesi}, N. and {Mangilli}, A. and {Marchini}, A. and {Maris}, M. and {Martin}, P.~G. and {Martinelli}, M. and {Mart{\'\i}nez-Gonz{\'a}lez}, E. and {Masi}, S. and {Matarrese}, S. and {McGehee}, P. and {Meinhold}, P.~R. and {Melchiorri}, A. and {Melin}, J. -B. and {Mendes}, L. and {Mennella}, A. and {Migliaccio}, M. and {Millea}, M. and {Mitra}, S. and {Miville-Desch{\^e}nes}, M. -A. and {Moneti}, A. and {Montier}, L. and {Morgante}, G. and {Mortlock}, D. and {Moss}, A. and {Munshi}, D. and {Murphy}, J.~A. and {Naselsky}, P. and {Nati}, F. and {Natoli}, P. and {Netterfield}, C.~B. and {N{\o}rgaard-Nielsen}, H.~U. and {Noviello}, F. and {Novikov}, D. and {Novikov}, I. and {Oxborrow}, C.~A. and {Paci}, F. and {Pagano}, L. and {Pajot}, F. and {Paladini}, R. and {Paoletti}, D. and {Partridge}, B. and {Pasian}, F. and {Patanchon}, G. and {Pearson}, T.~J. and {Perdereau}, O. and {Perotto}, L. and {Perrotta}, F. and {Pettorino}, V. and {Piacentini}, F. and {Piat}, M. and {Pierpaoli}, E. and {Pietrobon}, D. and {Plaszczynski}, S. and {Pointecouteau}, E. and {Polenta}, G. and {Popa}, L. and {Pratt}, G.~W. and {Pr{\'e}zeau}, G.},
        title = "{Planck 2015 results. XIII. Cosmological parameters}",
      journal = {\aap},
         year = 2016,
        month = sep,
       volume = {594},
          eid = {A13},
        pages = {A13},
          doi = {10.1051/0004-6361/201525830},
archivePrefix = {arXiv},
       eprint = {1502.01589},
 primaryClass = {astro-ph.CO},
       adsurl = {https://ui.adsabs.harvard.edu/abs/2016A&A...594A..13P}
}

@ARTICLE{2014MNRAS.444.2200D,
       author = {{De Propris}, Roberto and {Baldry}, Ivan K. and {Bland-Hawthorn}, Joss and {Brough}, Sarah and {Driver}, Simon P. and {Hopkins}, Andrew M. and {Kelvin}, Lee and {Loveday}, Jon and {Phillipps}, Steve and {Robotham}, Aaron S.~G.},
        title = "{Galaxy and Mass Assembly (GAMA): merging galaxies and their properties}",
      journal = {\mnras},
         year = 2014,
        month = nov,
       volume = {444},
       number = {3},
        pages = {2200-2211},
          doi = {10.1093/mnras/stu1452},
archivePrefix = {arXiv},
       eprint = {1407.4996},
 primaryClass = {astro-ph.GA},
       adsurl = {https://ui.adsabs.harvard.edu/abs/2014MNRAS.444.2200D}
}

@ARTICLE{2022ApJ...927....3F,
       author = {{Foord}, Adi and {Liu}, Xin and {G{\"u}ltekin}, Kayhan and {Whitley}, Kevin and {Shi}, Fangzheng and {Chen}, Yu-Ching},
        title = "{Investigating the Accretion Nature of Binary Supermassive Black Hole Candidate SDSS J025214.67-002813.7}",
      journal = {\apj},
         year = 2022,
        month = mar,
       volume = {927},
       number = {1},
          eid = {3},
        pages = {3},
          doi = {10.3847/1538-4357/ac4af1},
archivePrefix = {arXiv},
       eprint = {2110.02982},
 primaryClass = {astro-ph.HE},
       adsurl = {https://ui.adsabs.harvard.edu/abs/2022ApJ...927....3F}
}

@PROCEEDINGS{2018ASSL..454.....S,
        title = "{Accretion Flows in Astrophysics}",
    booktitle = {Astrophysics and Space Science Library},
         year = 2018,
       editor = {{Shakura}, Nikolay},
       series = {Astrophysics and Space Science Library},
       volume = {454},
        month = jan,
          doi = {10.1007/978-3-319-93009-1},
       adsurl = {https://ui.adsabs.harvard.edu/abs/2018ASSL..454.....S}
}

@ARTICLE{2019ApJ...873..111I,
       author = {{Ivezi{\'c}}, {\v{Z}}eljko and {Kahn}, Steven M. and {Tyson}, J. Anthony and {Abel}, Bob and {Acosta}, Emily and {Allsman}, Robyn and {Alonso}, David and {AlSayyad}, Yusra and {Anderson}, Scott F. and {Andrew}, John and {Angel}, James Roger P. and {Angeli}, George Z. and {Ansari}, Reza and {Antilogus}, Pierre and {Araujo}, Constanza and {Armstrong}, Robert and {Arndt}, Kirk T. and {Astier}, Pierre and {Aubourg}, {\'E}ric and {Auza}, Nicole and {Axelrod}, Tim S. and {Bard}, Deborah J. and {Barr}, Jeff D. and {Barrau}, Aurelian and {Bartlett}, James G. and {Bauer}, Amanda E. and {Bauman}, Brian J. and {Baumont}, Sylvain and {Bechtol}, Ellen and {Bechtol}, Keith and {Becker}, Andrew C. and {Becla}, Jacek and {Beldica}, Cristina and {Bellavia}, Steve and {Bianco}, Federica B. and {Biswas}, Rahul and {Blanc}, Guillaume and {Blazek}, Jonathan and {Blandford}, Roger D. and {Bloom}, Josh S. and {Bogart}, Joanne and {Bond}, Tim W. and {Booth}, Michael T. and {Borgland}, Anders W. and {Borne}, Kirk and {Bosch}, James F. and {Boutigny}, Dominique and {Brackett}, Craig A. and {Bradshaw}, Andrew and {Brandt}, William Nielsen and {Brown}, Michael E. and {Bullock}, James S. and {Burchat}, Patricia and {Burke}, David L. and {Cagnoli}, Gianpietro and {Calabrese}, Daniel and {Callahan}, Shawn and {Callen}, Alice L. and {Carlin}, Jeffrey L. and {Carlson}, Erin L. and {Chandrasekharan}, Srinivasan and {Charles-Emerson}, Glenaver and {Chesley}, Steve and {Cheu}, Elliott C. and {Chiang}, Hsin-Fang and {Chiang}, James and {Chirino}, Carol and {Chow}, Derek and {Ciardi}, David R. and {Claver}, Charles F. and {Cohen-Tanugi}, Johann and {Cockrum}, Joseph J. and {Coles}, Rebecca and {Connolly}, Andrew J. and {Cook}, Kem H. and {Cooray}, Asantha and {Covey}, Kevin R. and {Cribbs}, Chris and {Cui}, Wei and {Cutri}, Roc and {Daly}, Philip N. and {Daniel}, Scott F. and {Daruich}, Felipe and {Daubard}, Guillaume and {Daues}, Greg and {Dawson}, William and {Delgado}, Francisco and {Dellapenna}, Alfred and {de Peyster}, Robert and {de Val-Borro}, Miguel and {Digel}, Seth W. and {Doherty}, Peter and {Dubois}, Richard and {Dubois-Felsmann}, Gregory P. and {Durech}, Josef and {Economou}, Frossie and {Eifler}, Tim and {Eracleous}, Michael and {Emmons}, Benjamin L. and {Fausti Neto}, Angelo and {Ferguson}, Henry and {Figueroa}, Enrique and {Fisher-Levine}, Merlin and {Focke}, Warren and {Foss}, Michael D. and {Frank}, James and {Freemon}, Michael D. and {Gangler}, Emmanuel and {Gawiser}, Eric and {Geary}, John C. and {Gee}, Perry and {Geha}, Marla and {Gessner}, Charles J.~B. and {Gibson}, Robert R. and {Gilmore}, D. Kirk and {Glanzman}, Thomas and {Glick}, William and {Goldina}, Tatiana and {Goldstein}, Daniel A. and {Goodenow}, Iain and {Graham}, Melissa L. and {Gressler}, William J. and {Gris}, Philippe and {Guy}, Leanne P. and {Guyonnet}, Augustin and {Haller}, Gunther and {Harris}, Ron and {Hascall}, Patrick A. and {Haupt}, Justine and {Hernandez}, Fabio and {Herrmann}, Sven and {Hileman}, Edward and {Hoblitt}, Joshua and {Hodgson}, John A. and {Hogan}, Craig and {Howard}, James D. and {Huang}, Dajun and {Huffer}, Michael E. and {Ingraham}, Patrick and {Innes}, Walter R. and {Jacoby}, Suzanne H. and {Jain}, Bhuvnesh and {Jammes}, Fabrice and {Jee}, M. James and {Jenness}, Tim and {Jernigan}, Garrett and {Jevremovi{\'c}}, Darko and {Johns}, Kenneth and {Johnson}, Anthony S. and {Johnson}, Margaret W.~G. and {Jones}, R. Lynne and {Juramy-Gilles}, Claire and {Juri{\'c}}, Mario and {Kalirai}, Jason S. and {Kallivayalil}, Nitya J. and {Kalmbach}, Bryce and {Kantor}, Jeffrey P. and {Karst}, Pierre and {Kasliwal}, Mansi M. and {Kelly}, Heather and {Kessler}, Richard and {Kinnison}, Veronica and {Kirkby}, David and {Knox}, Lloyd and {Kotov}, Ivan V. and {Krabbendam}, Victor L. and {Krughoff}, K. Simon and {Kub{\'a}nek}, Petr and {Kuczewski}, John and {Kulkarni}, Shri and {Ku}, John and {Kurita}, Nadine R. and {Lage}, Craig S. and {Lambert}, Ron and {Lange}, Travis and {Langton}, J. Brian and {Le Guillou}, Laurent and {Levine}, Deborah and {Liang}, Ming and {Lim}, Kian-Tat and {Lintott}, Chris J. and {Long}, Kevin E. and {Lopez}, Margaux and {Lotz}, Paul J. and {Lupton}, Robert H. and {Lust}, Nate B. and {MacArthur}, Lauren A. and {Mahabal}, Ashish and {Mandelbaum}, Rachel and {Markiewicz}, Thomas W. and {Marsh}, Darren S. and {Marshall}, Philip J. and {Marshall}, Stuart and {May}, Morgan and {McKercher}, Robert and {McQueen}, Michelle and {Meyers}, Joshua and {Migliore}, Myriam and {Miller}, Michelle and {Mills}, David J.},
        title = "{LSST: From Science Drivers to Reference Design and Anticipated Data Products}",
      journal = {\apj},
         year = 2019,
        month = mar,
       volume = {873},
       number = {2},
          eid = {111},
        pages = {111},
          doi = {10.3847/1538-4357/ab042c},
archivePrefix = {arXiv},
       eprint = {0805.2366},
 primaryClass = {astro-ph},
       adsurl = {https://ui.adsabs.harvard.edu/abs/2019ApJ...873..111I}
}

@ARTICLE{2016ApJ...829....4L,
       author = {{Leighly}, Karen M. and {Terndrup}, Donald M. and {Gallagher}, Sarah C. and {Lucy}, Adrian B.},
        title = "{The Binary Black Hole Model for Mrk 231 Bites the Dust}",
      journal = {\apj},
         year = 2016,
        month = sep,
       volume = {829},
       number = {1},
          eid = {4},
        pages = {4},
          doi = {10.3847/0004-637X/829/1/4},
archivePrefix = {arXiv},
       eprint = {1604.03456},
 primaryClass = {astro-ph.GA},
       adsurl = {https://ui.adsabs.harvard.edu/abs/2016ApJ...829....4L}
}

@ARTICLE{1995ApJ...452..710N,
       author = {{Narayan}, Ramesh and {Yi}, Insu},
        title = "{Advection-dominated Accretion: Underfed Black Holes and Neutron Stars}",
      journal = {\apj},
         year = 1995,
        month = oct,
       volume = {452},
        pages = {710},
          doi = {10.1086/176343},
archivePrefix = {arXiv},
       eprint = {astro-ph/9411059},
 primaryClass = {astro-ph},
       adsurl = {https://ui.adsabs.harvard.edu/abs/1995ApJ...452..710N}
}

@ARTICLE{2025arXiv250516884H,
       author = {{Huijse}, Pablo and {Davelaar}, Jordy and {De Ridder}, Joris and {Jannsen}, Nicholas and {Aerts}, Conny},
        title = "{Periodic Variability in Space Photometry of 181 New Supermassive Black Hole Binary Candidates}",
      journal = {arXiv e-prints},
         year = 2025,
        month = may,
          eid = {arXiv:2505.16884},
        pages = {arXiv:2505.16884},
          doi = {10.48550/arXiv.2505.16884},
archivePrefix = {arXiv},
       eprint = {2505.16884},
 primaryClass = {astro-ph.HE},
       adsurl = {https://ui.adsabs.harvard.edu/abs/2025arXiv250516884H}
}

@ARTICLE{2024Natur.633..318C,
       author = {{Carniani}, Stefano and {Hainline}, Kevin and {D'Eugenio}, Francesco and {Eisenstein}, Daniel J. and {Jakobsen}, Peter and {Witstok}, Joris and {Johnson}, Benjamin D. and {Chevallard}, Jacopo and {Maiolino}, Roberto and {Helton}, Jakob M. and {Willott}, Chris and {Robertson}, Brant and {Alberts}, Stacey and {Arribas}, Santiago and {Baker}, William M. and {Bhatawdekar}, Rachana and {Boyett}, Kristan and {Bunker}, Andrew J. and {Cameron}, Alex J. and {Cargile}, Phillip A. and {Charlot}, St{\'e}phane and {Curti}, Mirko and {Curtis-Lake}, Emma and {Egami}, Eiichi and {Giardino}, Giovanna and {Isaak}, Kate and {Ji}, Zhiyuan and {Jones}, Gareth C. and {Kumari}, Nimisha and {Maseda}, Michael V. and {Parlanti}, Eleonora and {P{\'e}rez-Gonz{\'a}lez}, Pablo G. and {Rawle}, Tim and {Rieke}, George and {Rieke}, Marcia and {Del Pino}, Bruno Rodr{\'\i}guez and {Saxena}, Aayush and {Scholtz}, Jan and {Smit}, Renske and {Sun}, Fengwu and {Tacchella}, Sandro and {{\"U}bler}, Hannah and {Venturi}, Giacomo and {Williams}, Christina C. and {Willmer}, Christopher N.~A.},
        title = "{Spectroscopic confirmation of two luminous galaxies at a redshift of 14}",
      journal = {\nat},
         year = 2024,
        month = sep,
       volume = {633},
       number = {8029},
        pages = {318-322},
          doi = {10.1038/s41586-024-07860-9},
archivePrefix = {arXiv},
       eprint = {2405.18485},
 primaryClass = {astro-ph.GA},
       adsurl = {https://ui.adsabs.harvard.edu/abs/2024Natur.633..318C}
}

@ARTICLE{2024ApJ...964...39G,
       author = {{Greene}, Jenny E. and {Labbe}, Ivo and {Goulding}, Andy D. and {Furtak}, Lukas J. and {Chemerynska}, Iryna and {Kokorev}, Vasily and {Dayal}, Pratika and {Volonteri}, Marta and {Williams}, Christina C. and {Wang}, Bingjie and {Setton}, David J. and {Burgasser}, Adam J. and {Bezanson}, Rachel and {Atek}, Hakim and {Brammer}, Gabriel and {Cutler}, Sam E. and {Feldmann}, Robert and {Fujimoto}, Seiji and {Glazebrook}, Karl and {de Graaff}, Anna and {Khullar}, Gourav and {Leja}, Joel and {Marchesini}, Danilo and {Maseda}, Michael V. and {Matthee}, Jorryt and {Miller}, Tim B. and {Naidu}, Rohan P. and {Nanayakkara}, Themiya and {Oesch}, Pascal A. and {Pan}, Richard and {Papovich}, Casey and {Price}, Sedona H. and {van Dokkum}, Pieter and {Weaver}, John R. and {Whitaker}, Katherine E. and {Zitrin}, Adi},
        title = "{UNCOVER Spectroscopy Confirms the Surprising Ubiquity of Active Galactic Nuclei in Red Sources at z > 5}",
      journal = {\apj},
         year = 2024,
        month = mar,
       volume = {964},
       number = {1},
          eid = {39},
        pages = {39},
          doi = {10.3847/1538-4357/ad1e5f},
archivePrefix = {arXiv},
       eprint = {2309.05714},
 primaryClass = {astro-ph.GA},
       adsurl = {https://ui.adsabs.harvard.edu/abs/2024ApJ...964...39G}
}

@ARTICLE{2024A&A...691A.145M,
       author = {{Maiolino}, Roberto and {Scholtz}, Jan and {Curtis-Lake}, Emma and {Carniani}, Stefano and {Baker}, William and {de Graaff}, Anna and {Tacchella}, Sandro and {{\"U}bler}, Hannah and {D'Eugenio}, Francesco and {Witstok}, Joris and {Curti}, Mirko and {Arribas}, Santiago and {Bunker}, Andrew J. and {Charlot}, St{\'e}phane and {Chevallard}, Jacopo and {Eisenstein}, Daniel J. and {Egami}, Eiichi and {Ji}, Zhiyuan and {Jones}, Gareth C. and {Lyu}, Jianwei and {Rawle}, Tim and {Robertson}, Brant and {Rujopakarn}, Wiphu and {Perna}, Michele and {Sun}, Fengwu and {Venturi}, Giacomo and {Williams}, Christina C. and {Willott}, Chris},
        title = "{JADES: The diverse population of infant black holes at 4 < z < 11: Merging, tiny, poor, but mighty}",
      journal = {\aap},
         year = 2024,
        month = nov,
       volume = {691},
          eid = {A145},
        pages = {A145},
          doi = {10.1051/0004-6361/202347640},
archivePrefix = {arXiv},
       eprint = {2308.01230},
 primaryClass = {astro-ph.GA},
       adsurl = {https://ui.adsabs.harvard.edu/abs/2024A&A...691A.145M}
}

@ARTICLE{2025ApJ...978..104G,
       author = {{Guo}, Xiao and {Yu}, Qingjuan and {Lu}, Youjun},
        title = "{Constraining the Binarity of Massive Black Holes in the Galactic Center and Some Nearby Galaxies via Pulsar Timing Array Observations of Gravitational Waves}",
      journal = {\apj},
         year = 2025,
        month = jan,
       volume = {978},
       number = {1},
          eid = {104},
        pages = {104},
          doi = {10.3847/1538-4357/ad94ec},
archivePrefix = {arXiv},
       eprint = {2411.14150},
 primaryClass = {astro-ph.HE},
       adsurl = {https://ui.adsabs.harvard.edu/abs/2025ApJ...978..104G}
}

@INPROCEEDINGS{2016ASPC..502...19L,
       author = {{Lee}, K.~J.},
        title = "{Prospects of Gravitational Wave Detection Using Pulsar Timing Array for Chinese Future Telescopes}",
    booktitle = {Frontiers in Radio Astronomy and FAST Early Sciences Symposium 2015},
         year = 2016,
       editor = {{Qain}, L. and {Li}, D.},
       series = {Astronomical Society of the Pacific Conference Series},
       volume = {502},
        month = feb,
        pages = {19},
       adsurl = {https://ui.adsabs.harvard.edu/abs/2016ASPC..502...19L}
}

@ARTICLE{2010CQGra..27h4016S,
       author = {{Sesana}, A. and {Vecchio}, A.},
        title = "{Gravitational waves and pulsar timing: stochastic background, individual sources and parameter estimation}",
      journal = {Classical and Quantum Gravity},
         year = 2010,
        month = apr,
       volume = {27},
       number = {8},
          eid = {084016},
        pages = {084016},
          doi = {10.1088/0264-9381/27/8/084016},
archivePrefix = {arXiv},
       eprint = {1001.3161},
 primaryClass = {astro-ph.CO},
       adsurl = {https://ui.adsabs.harvard.edu/abs/2010CQGra..27h4016S}
}

@ARTICLE{2025arXiv250610846X,
       author = {{Xin}, Chengcheng and {Isi}, Maximiliano and {Farr}, Will M. and {Haiman}, Zolt{\'a}n},
        title = "{Identifying Compact Chirping SMBHBs in LSST using Bayesian Analysis}",
      journal = {arXiv e-prints},
         year = 2025,
        month = jun,
          eid = {arXiv:2506.10846},
        pages = {arXiv:2506.10846},
          doi = {10.48550/arXiv.2506.10846},
archivePrefix = {arXiv},
       eprint = {2506.10846},
 primaryClass = {astro-ph.HE},
       adsurl = {https://ui.adsabs.harvard.edu/abs/2025arXiv250610846X}
}

@ARTICLE{2025arXiv250505322I,
       author = {{Inayoshi}, Kohei and {Shangguan}, Jinyi and {Chen}, Xian and {Ho}, Luis C. and {Haiman}, Zoltan},
        title = "{The Emergence of Little Red Dots from Binary Massive Black Holes}",
      journal = {arXiv e-prints},
         year = 2025,
        month = may,
          eid = {arXiv:2505.05322},
        pages = {arXiv:2505.05322},
          doi = {10.48550/arXiv.2505.05322},
archivePrefix = {arXiv},
       eprint = {2505.05322},
 primaryClass = {astro-ph.HE},
       adsurl = {https://ui.adsabs.harvard.edu/abs/2025arXiv250505322I}
}

@ARTICLE{2023A&A...677A.184W,
       author = {{Weaver}, J.~R. and {Davidzon}, I. and {Toft}, S. and {Ilbert}, O. and {McCracken}, H.~J. and {Gould}, K.~M.~L. and {Jespersen}, C.~K. and {Steinhardt}, C. and {Lagos}, C.~D.~P. and {Capak}, P.~L. and {Casey}, C.~M. and {Chartab}, N. and {Faisst}, A.~L. and {Hayward}, C.~C. and {Kartaltepe}, J.~S. and {Kauffmann}, O.~B. and {Koekemoer}, A.~M. and {Kokorev}, V. and {Laigle}, C. and {Liu}, D. and {Long}, A. and {Magdis}, G.~E. and {McPartland}, C.~J.~R. and {Milvang-Jensen}, B. and {Mobasher}, B. and {Moneti}, A. and {Peng}, Y. and {Sanders}, D.~B. and {Shuntov}, M. and {Sneppen}, A. and {Valentino}, F. and {Zalesky}, L. and {Zamorani}, G.},
        title = "{COSMOS2020: The galaxy stellar mass function. The assembly and star formation cessation of galaxies at 0.2< z {\ensuremath{\leq}} 7.5}",
      journal = {\aap},
         year = 2023,
        month = sep,
       volume = {677},
          eid = {A184},
        pages = {A184},
          doi = {10.1051/0004-6361/202245581},
archivePrefix = {arXiv},
       eprint = {2212.02512},
 primaryClass = {astro-ph.GA},
       adsurl = {https://ui.adsabs.harvard.edu/abs/2023A&A...677A.184W}
}

@ARTICLE{2025arXiv250821510C,
       author = {{Chiesa}, Alfredo and {Izquierdo-Villalba}, David and {Sesana}, Alberto and {Cocchiararo}, Fabiola and {Franchini}, Alessia and {Lupi}, Alessandro and {Spinoso}, Daniele and {Bonoli}, Silvia},
        title = "{Identifying massive black hole binaries via light curve variability in optical time-domain surveys}",
      journal = {arXiv e-prints},
         year = 2025,
        month = aug,
          eid = {arXiv:2508.21510},
        pages = {arXiv:2508.21510},
          doi = {10.48550/arXiv.2508.21510},
archivePrefix = {arXiv},
       eprint = {2508.21510},
 primaryClass = {astro-ph.HE},
       adsurl = {https://ui.adsabs.harvard.edu/abs/2025arXiv250821510C}
}

@ARTICLE{2025arXiv250910601E,
       author = {{El-Badry}, Kareem and {Hogg}, David W. and {Rix}, Hans-Walter},
        title = "{Active galactic nuclei do not exhibit stably periodic brightness variations}",
      journal = {arXiv e-prints},
         year = 2025,
        month = sep,
          eid = {arXiv:2509.10601},
        pages = {arXiv:2509.10601},
          doi = {10.48550/arXiv.2509.10601},
archivePrefix = {arXiv},
       eprint = {2509.10601},
 primaryClass = {astro-ph.GA},
       adsurl = {https://ui.adsabs.harvard.edu/abs/2025arXiv250910601E}
}

@ARTICLE{2006ApJ...648..523W,
       author = {{Watarai}, Ken-ya},
        title = "{New Analytical Formulae for Supercritical Accretion Flows}",
      journal = {\apj},
         year = 2006,
        month = sep,
       volume = {648},
       number = {1},
        pages = {523-533},
          doi = {10.1086/505854},
archivePrefix = {arXiv},
       eprint = {astro-ph/0605248},
 primaryClass = {astro-ph},
       adsurl = {https://ui.adsabs.harvard.edu/abs/2006ApJ...648..523W}
}

@ARTICLE{2017MNRAS.467..898Q,
       author = {{Qiao}, Erlin and {Liu}, B.~F.},
        title = "{The condensation of the corona for the correlation between the hard X-ray photon index {\ensuremath{\Gamma}} and the reflection scaling factor ℜ in active galactic nuclei}",
      journal = {\mnras},
         year = 2017,
        month = may,
       volume = {467},
       number = {1},
        pages = {898-905},
          doi = {10.1093/mnras/stx121},
archivePrefix = {arXiv},
       eprint = {1701.04211},
 primaryClass = {astro-ph.HE},
       adsurl = {https://ui.adsabs.harvard.edu/abs/2017MNRAS.467..898Q}
}

@ARTICLE{2000A&A...361..175M,
       author = {{Meyer}, F. and {Liu}, B.~F. and {Meyer-Hofmeister}, E.},
        title = "{Evaporation: The change from accretion via a thin disk to a coronal flow}",
      journal = {\aap},
         year = 2000,
        month = sep,
       volume = {361},
        pages = {175-188},
          doi = {10.48550/arXiv.astro-ph/0007091},
archivePrefix = {arXiv},
       eprint = {astro-ph/0007091},
 primaryClass = {astro-ph},
       adsurl = {https://ui.adsabs.harvard.edu/abs/2000A&A...361..175M}
}

@ARTICLE{2022iSci...25j3544L,
       author = {{Liu}, B.~F. and {Qiao}, Erlin},
        title = "{Accretion around black holes: The geometry and spectra}",
      journal = {iScience},
         year = 2022,
        month = jan,
       volume = {25},
       number = {1},
        pages = {103544},
          doi = {10.1016/j.isci.2021.103544},
archivePrefix = {arXiv},
       eprint = {2201.06198},
 primaryClass = {astro-ph.HE},
       adsurl = {https://ui.adsabs.harvard.edu/abs/2022iSci...25j3544L}
}

@ARTICLE{2009MNRAS.394..207C,
       author = {{Cao}, Xinwu},
        title = "{An accretion disc-corona model for X-ray spectra of active galactic nuclei}",
      journal = {\mnras},
         year = 2009,
        month = mar,
       volume = {394},
       number = {1},
        pages = {207-213},
          doi = {10.1111/j.1365-2966.2008.14347.x},
archivePrefix = {arXiv},
       eprint = {0812.1828},
 primaryClass = {astro-ph},
       adsurl = {https://ui.adsabs.harvard.edu/abs/2009MNRAS.394..207C}
}

@ARTICLE{2001MNRAS.324..119Y,
       author = {{Yuan}, Feng},
        title = "{Luminous hot accretion discs}",
      journal = {\mnras},
         year = 2001,
        month = jun,
       volume = {324},
       number = {1},
        pages = {119-127},
          doi = {10.1046/j.1365-8711.2001.04258.x},
archivePrefix = {arXiv},
       eprint = {astro-ph/0009207},
 primaryClass = {astro-ph},
       adsurl = {https://ui.adsabs.harvard.edu/abs/2001MNRAS.324..119Y}
}

@ARTICLE{2008MmSAI..79..134M,
       author = {{Malzac}, J.},
        title = "{Accretion discs, coronae and jets in black hole binaries: prospects for Simbol-X.}",
      journal = {\memsai},
         year = 2008,
        month = jan,
       volume = {79},
        pages = {134},
          doi = {10.48550/arXiv.0801.3767},
archivePrefix = {arXiv},
       eprint = {0801.3767},
 primaryClass = {astro-ph},
       adsurl = {https://ui.adsabs.harvard.edu/abs/2008MmSAI..79..134M}
}

@ARTICLE{2000ApJ...533..682C,
       author = {{Calzetti}, Daniela and {Armus}, Lee and {Bohlin}, Ralph C. and {Kinney}, Anne L. and {Koornneef}, Jan and {Storchi-Bergmann}, Thaisa},
        title = "{The Dust Content and Opacity of Actively Star-forming Galaxies}",
      journal = {\apj},
         year = 2000,
        month = apr,
       volume = {533},
       number = {2},
        pages = {682-695},
          doi = {10.1086/308692},
archivePrefix = {arXiv},
       eprint = {astro-ph/9911459},
 primaryClass = {astro-ph},
       adsurl = {https://ui.adsabs.harvard.edu/abs/2000ApJ...533..682C}
}

@ARTICLE{2026arXiv260102288V,
       author = {{Varniere}, Peggy and {Casse}, Fabien and {Dodu}, Fabrice},
        title = "{Looking for observational signatures of early binary black hole systems}",
      journal = {arXiv e-prints},
         year = 2026,
        month = jan,
          eid = {arXiv:2601.02288},
        pages = {arXiv:2601.02288},
          doi = {10.48550/arXiv.2601.02288},
archivePrefix = {arXiv},
       eprint = {2601.02288},
 primaryClass = {astro-ph.HE},
       adsurl = {https://ui.adsabs.harvard.edu/abs/2026arXiv260102288V}
}

@ARTICLE{2017ApJ...835..199R,
       author = {{Ryan}, Geoffrey and {MacFadyen}, Andrew},
        title = "{Minidisks in Binary Black Hole Accretion}",
      journal = {\apj},
         year = 2017,
        month = feb,
       volume = {835},
       number = {2},
          eid = {199},
        pages = {199},
          doi = {10.3847/1538-4357/835/2/199},
archivePrefix = {arXiv},
       eprint = {1611.00341},
 primaryClass = {astro-ph.HE},
       adsurl = {https://ui.adsabs.harvard.edu/abs/2017ApJ...835..199R}
}

@ARTICLE{2018MNRAS.476.2249T,
       author = {{Tang}, Yike and {Haiman}, Zolt{\'a}n and {MacFadyen}, Andrew},
        title = "{The late inspiral of supermassive black hole binaries with circumbinary gas discs in the LISA band}",
      journal = {\mnras},
         year = 2018,
        month = may,
       volume = {476},
       number = {2},
        pages = {2249-2257},
          doi = {10.1093/mnras/sty423},
archivePrefix = {arXiv},
       eprint = {1801.02266},
 primaryClass = {astro-ph.HE},
       adsurl = {https://ui.adsabs.harvard.edu/abs/2018MNRAS.476.2249T}
}

@ARTICLE{2025ApJ...995...68T,
       author = {{Tiede}, Christopher and {D'Orazio}, Daniel J.},
        title = "{Hot, Cold, and Multicomponent Accretion Flows around Supermassive Black Hole Binaries}",
      journal = {\apj},
         year = 2025,
        month = dec,
       volume = {995},
       number = {1},
          eid = {68},
        pages = {68},
          doi = {10.3847/1538-4357/ae17ba},
archivePrefix = {arXiv},
       eprint = {2508.11748},
 primaryClass = {astro-ph.HE},
       adsurl = {https://ui.adsabs.harvard.edu/abs/2025ApJ...995...68T}
}

@ARTICLE{2025arXiv250614141B,
       author = {{Bang}, Saemi and {Okazaki}, Atsuo T. and {Hayasaki}, Kimitake},
        title = "{Spectral Properties of Irradiated Circumbinary Disks around Binary Black Holes Governed by Hydrogen Opacities Dependent on Temperature and Density}",
      journal = {arXiv e-prints},
         year = 2025,
        month = jun,
          eid = {arXiv:2506.14141},
        pages = {arXiv:2506.14141},
          doi = {10.48550/arXiv.2506.14141},
archivePrefix = {arXiv},
       eprint = {2506.14141},
 primaryClass = {astro-ph.HE},
       adsurl = {https://ui.adsabs.harvard.edu/abs/2025arXiv250614141B}
}

@ARTICLE{2024ApJ...975..141L,
       author = {{Lee}, Yunewoo and {Okazaki}, Atsuo T. and {Hayasaki}, Kimitake},
        title = "{Circumbinary Disk Spectra Irradiated by Two Central Accretion Disks in a Binary Black Hole System}",
      journal = {\apj},
         year = 2024,
        month = nov,
       volume = {975},
       number = {1},
          eid = {141},
        pages = {141},
          doi = {10.3847/1538-4357/ad794a},
archivePrefix = {arXiv},
       eprint = {2407.13366},
 primaryClass = {astro-ph.HE},
       adsurl = {https://ui.adsabs.harvard.edu/abs/2024ApJ...975..141L}
}

@ARTICLE{2025MNRAS.543.2670K,
       author = {{Krauth}, Luke Major and {Davelaar}, Jordy and {Haiman}, Zolt{\'a}n and {Westernacher-Schneider}, John Ryan and {Zrake}, Jonathan and {MacFadyen}, Andrew},
        title = "{Thermal X-ray signatures in late-stage unequal-mass massive black hole binary mergers}",
      journal = {\mnras},
         year = 2025,
        month = nov,
       volume = {543},
       number = {3},
        pages = {2670-2685},
          doi = {10.1093/mnras/staf1583},
archivePrefix = {arXiv},
       eprint = {2503.01494},
 primaryClass = {astro-ph.HE},
       adsurl = {https://ui.adsabs.harvard.edu/abs/2025MNRAS.543.2670K}
}

@ARTICLE{2018ApJ...865..140D,
       author = {{d'Ascoli}, St{\'e}phane and {Noble}, Scott C. and {Bowen}, Dennis B. and {Campanelli}, Manuela and {Krolik}, Julian H. and {Mewes}, Vassilios},
        title = "{Electromagnetic Emission from Supermassive Binary Black Holes Approaching Merger}",
      journal = {\apj},
         year = 2018,
        month = oct,
       volume = {865},
       number = {2},
          eid = {140},
        pages = {140},
          doi = {10.3847/1538-4357/aad8b4},
archivePrefix = {arXiv},
       eprint = {1806.05697},
 primaryClass = {astro-ph.HE},
       adsurl = {https://ui.adsabs.harvard.edu/abs/2018ApJ...865..140D}
}

@ARTICLE{2020ApJ...900..148S,
       author = {{Saade}, M. Lynne and {Stern}, Daniel and {Brightman}, Murray and {Haiman}, Zolt{\'a}n and {Djorgovski}, S.~G. and {D'Orazio}, Daniel and {Ford}, K.~E.~S. and {Graham}, Matthew J. and {Jun}, Hyunsung D. and {Kraft}, Ralph P. and {McKernan}, Barry and {Vikhlinin}, Alexei and {Walton}, Dominic J.},
        title = "{Chandra Observations of Candidate Subparsec Binary Supermassive Black Holes}",
      journal = {\apj},
         year = 2020,
        month = sep,
       volume = {900},
       number = {2},
          eid = {148},
        pages = {148},
          doi = {10.3847/1538-4357/abad31},
archivePrefix = {arXiv},
       eprint = {2001.08870},
 primaryClass = {astro-ph.HE},
       adsurl = {https://ui.adsabs.harvard.edu/abs/2020ApJ...900..148S}
}

@ARTICLE{2025ApJ...989..190M,
       author = {{Malewicz}, Julie and {Ballantyne}, David R. and {Bogdanovi{\'c}}, Tamara and {Brenneman}, Laura and {Dauser}, Thomas},
        title = "{X-Ray Reflection Signatures of Supermassive Black Hole Binaries}",
      journal = {\apj},
         year = 2025,
        month = aug,
       volume = {989},
       number = {2},
          eid = {190},
        pages = {190},
          doi = {10.3847/1538-4357/adea75},
archivePrefix = {arXiv},
       eprint = {2504.14018},
 primaryClass = {astro-ph.HE},
       adsurl = {https://ui.adsabs.harvard.edu/abs/2025ApJ...989..190M}
}

@ARTICLE{2020A&A...636A..73D,
       author = {{Duras}, F. and {Bongiorno}, A. and {Ricci}, F. and {Piconcelli}, E. and {Shankar}, F. and {Lusso}, E. and {Bianchi}, S. and {Fiore}, F. and {Maiolino}, R. and {Marconi}, A. and {Onori}, F. and {Sani}, E. and {Schneider}, R. and {Vignali}, C. and {La Franca}, F.},
        title = "{Universal bolometric corrections for active galactic nuclei over seven luminosity decades}",
      journal = {\aap},
         year = 2020,
        month = apr,
       volume = {636},
          eid = {A73},
        pages = {A73},
          doi = {10.1051/0004-6361/201936817},
archivePrefix = {arXiv},
       eprint = {2001.09984},
 primaryClass = {astro-ph.GA},
       adsurl = {https://ui.adsabs.harvard.edu/abs/2020A&A...636A..73D}
}

@ARTICLE{2020ARA&A..58..257G,
       author = {{Greene}, Jenny E. and {Strader}, Jay and {Ho}, Luis C.},
        title = "{Intermediate-Mass Black Holes}",
      journal = {\araa},
         year = 2020,
        month = aug,
       volume = {58},
        pages = {257-312},
          doi = {10.1146/annurev-astro-032620-021835},
archivePrefix = {arXiv},
       eprint = {1911.09678},
 primaryClass = {astro-ph.GA},
       adsurl = {https://ui.adsabs.harvard.edu/abs/2020ARA&A..58..257G}
}

@ARTICLE{2019ApJ...887..245S,
       author = {{Schutte}, Zachary and {Reines}, Amy E. and {Greene}, Jenny E.},
        title = "{The Black Hole-Bulge Mass Relation Including Dwarf Galaxies Hosting Active Galactic Nuclei}",
      journal = {\apj},
         year = 2019,
        month = dec,
       volume = {887},
       number = {2},
          eid = {245},
        pages = {245},
          doi = {10.3847/1538-4357/ab35dd},
archivePrefix = {arXiv},
       eprint = {1908.00020},
 primaryClass = {astro-ph.GA},
       adsurl = {https://ui.adsabs.harvard.edu/abs/2019ApJ...887..245S}
}

@ARTICLE{2023NatAs...7.1282R,
       author = {{Ricci}, Claudio and {Trakhtenbrot}, Benny},
        title = "{Changing-look active galactic nuclei}",
      journal = {Nature Astronomy},
         year = 2023,
        month = nov,
       volume = {7},
        pages = {1282-1294},
          doi = {10.1038/s41550-023-02108-4},
archivePrefix = {arXiv},
       eprint = {2211.05132},
 primaryClass = {astro-ph.GA},
       adsurl = {https://ui.adsabs.harvard.edu/abs/2023NatAs...7.1282R}
}

@ARTICLE{2018MNRAS.478.1660G,
       author = {{Gaskell}, C. Martin and {Harrington}, P.~Z.},
        title = "{Partial dust obscuration in active galactic nuclei as a cause of broad-line profile and lag variability, and apparent accretion disc inhomogeneities}",
      journal = {\mnras},
         year = 2018,
        month = aug,
       volume = {478},
       number = {2},
        pages = {1660-1669},
          doi = {10.1093/mnras/sty848},
archivePrefix = {arXiv},
       eprint = {1704.06455},
 primaryClass = {astro-ph.HE},
       adsurl = {https://ui.adsabs.harvard.edu/abs/2018MNRAS.478.1660G}
}

@ARTICLE{2016ApJ...825...42V,
       author = {{Veilleux}, S. and {Mel{\'e}ndez}, M. and {Tripp}, T.~M. and {Hamann}, F. and {Rupke}, D.~S.~N.},
        title = "{The Complete Ultraviolet Spectrum of the Archetypal ``Wind-dominated'' Quasar Mrk 231: Absorption and Emission from a High-speed Dusty Nuclear Outflow}",
      journal = {\apj},
         year = 2016,
        month = jul,
       volume = {825},
       number = {1},
          eid = {42},
        pages = {42},
          doi = {10.3847/0004-637X/825/1/42},
archivePrefix = {arXiv},
       eprint = {1605.00665},
 primaryClass = {astro-ph.GA},
       adsurl = {https://ui.adsabs.harvard.edu/abs/2016ApJ...825...42V}
}




\bsp	
\label{lastpage}
\end{document}